\documentclass[aps,floatfix,nofootinbib,preprintnumbers,showpacs,superscriptaddress,twocolumn]{revtex4-2}

\usepackage{caption}
\usepackage[english]{babel}
\usepackage[linesnumbered,lined,boxed,commentsnumbered,ruled,vlined]{algorithm2e}
\usepackage{algpseudocode}
\usepackage{amsfonts}
\usepackage{amsmath}
\usepackage{amssymb}
\usepackage{bm,bbm}
\usepackage{caption}
\usepackage{comment}
\usepackage{dcolumn}
\usepackage{epstopdf}
\usepackage{float}
\usepackage{graphicx}
\usepackage{hyperref}
\usepackage{cleveref}
\usepackage{mathrsfs}
\usepackage{mathtools}
\usepackage{multirow}
\usepackage{physics}
\usepackage{ragged2e}
\usepackage{scalerel}
\usepackage{subfigure}
\usepackage{tabularx}
\usepackage{tikz}
\usepackage{url}
\usepackage{utfsym}
\usepackage{verbatim}
\usepackage{tikz}
\usepackage{ctable}
\usepackage{xcolor}
\usepackage{xkcdcolors}

\newcommand{\vect}{\boldsymbol}

\DeclareCaptionJustification{justified}{\justifying}
\hypersetup{colorlinks=true, citecolor=orange, urlcolor=blue, linkcolor=magenta}

\definecolor{cel}{rgb}{0.0,0.53,0.74}
\definecolor{green}{rgb}{0.0,0.5,0.0}

\definecolor{colSDRG}{RGB}{51,117,56}
\definecolor{colSDCM}{RGB}{194,106,119}
\definecolor{colRec}{RGB}{230,158,0}
\definecolor{colpp}{RGB}{87,181,232}
\definecolor{colnn}{RGB}{0,158,115}
\definecolor{colpn}{RGB}{204,120,166}

\begin{document}

\title{Missing links prediction: comparing machine learning\\with physics-rooted approaches}

\author{Francesca Santucci}
\affiliation{IMT School for Advanced Studies, P.zza San Francesco 19, 55100 Lucca (Italy)}
\author{Giulio Cimini}
\affiliation{`Enrico Fermi' Research Center (CREF), Via Panisperna 89A, 00184 Rome (Italy)}
\affiliation{Department of Physics and INFN, `Tor Vergata' University of Rome, 00133 Rome (Italy)}
\author{Tiziano Squartini}
\affiliation{IMT School for Advanced Studies, P.zza San Francesco 19, 55100 Lucca (Italy)}
\affiliation{Scuola Normale Superiore, P.zza dei Cavalieri 7, 56126 Pisa (Italy)}
\affiliation{INdAM-GNAMPA Istituto Nazionale di Alta Matematica `Francesco Severi', P.le Aldo Moro 5, 00185 Rome (Italy)}

\date{\today}

\begin{abstract}
An active research line within the broader field of network science is the one concerning link prediction. Close in scope to network reconstruction, link prediction targets specific connections with the aim of uncovering the missing ones, as well as predicting those most likely to emerge in the future, from the available information. In this paper, we consider two families of methods, i.e. those rooted in statistical physics and those based upon machine learning: the members of the first family identify missing links as the most probable non-observed ones, the probability coefficients being determined by solving maximum-entropy benchmarks over the accessible network structure; the members of the second family, instead, associate the presence of single edges to explanatory node-specific variables. Running likelihood-based models such as the Configuration Model, or one of its many fitness-based variants, in parallel with the Gradient Boosting Decision Tree algorithm reveals that the accuracy of the former is comparable to the accuracy of the latter. Such a result confirms that white-box algorithms are viable competitors to the currently available black-box ones, being more interpretable and computationally faster.
\end{abstract}


\maketitle

\section{Introduction}
\label{sec:intro}

Link prediction is an active research line within the broader field of network science. Close in scope to network reconstruction~\cite{squartini_reconstruction_2018}, link prediction targets specific connections, aiming to uncover missing ones and predict those most likely to emerge in the future~\cite{lu_link_2011}. Generally speaking, the link prediction problem can be stated by asking the following question: \emph{given a supposedly incomplete snapshot of a network, can the most-likely `yet-to-come' edges be predicted?} Such an issue is relevant in many research areas, such as those concerning socio-economic and financial networks~\cite{liben-nowell_linkprediction_2007,berlusconi_link_2016,jalili_link_2017,parisi_entropybased_2018,mungo_reconstructing_2023}: knowing the structure of the commercial partnerships between firms or of the financial exchanges between banks is, in fact, relevant for a number of reasons, such as quantifying the risk associated with the propagation of a shock~\cite{bardoscia_physics_2021,ialongo_reconstructing_2022,mungo_reconstructing_2023}.\\

Link prediction algorithms rank unconnected node pairs on the basis of a \emph{score}: while some methods rely on purely structural information, others admit external information such as node- and edge-specific features.

The simplest framework to carry out link prediction includes the so-called \emph{similarity-based} algorithms: scores, here, are induced by some measure of similarity between nodes; to this aim, \emph{local}~\cite{barabasi_emergence_1999,zhou_predicting_2009,liben-nowell_linkprediction_2007}, \emph{quasi-local}~\cite{zhou_predicting_2009} or \emph{global}~\cite{katz_new_1953,brin_anatomy_1998,chebotarev_matrixforest_1997,jeh_simrank_2002} information - such as the degree, the degree of common neighbours or the length of paths connecting any two nodes - has been employed~\cite{liben-nowell_linkprediction_2007,zhou_predicting_2009,liu_link_2010,lu_link_2011}.

A more refined framework includes the so-called \emph{likelihood-based} algorithms, defined by a likelihood function whose maximisation provides the probability that any two nodes are connected; this is usually achieved by assuming that a certain amount of information is accessible, hence treating it as a constraint to account for: examples are provided by the methods relying on a graph modular structure~\cite{clauset_hierarchical_2008,guimera_missing_2009} or those induced by entropy maximisation~\cite{parisi_entropybased_2018,adriaens_blockapproximated_2020}.

\begin{figure*}[t!]
\begin{tikzpicture}
\node[anchor=south east](a) at (0,0) {\textbf{a)}};
\node[anchor=north west](fig1) at (a.south east) {\includegraphics[width=.45\textwidth]{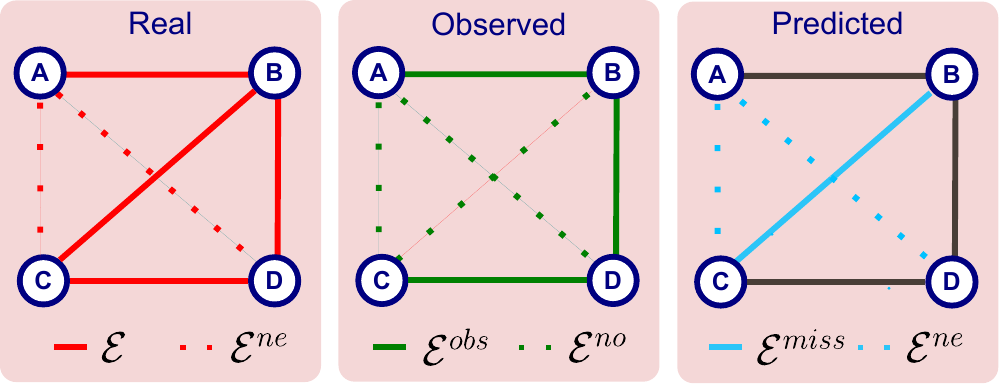}};
\node[anchor=south west](b) at (fig1.north east) {\textbf{b)}};
\node[anchor=north west, shift={(0,.1)}] (fig2) at (b.south east) {\includegraphics[width=.45\linewidth]{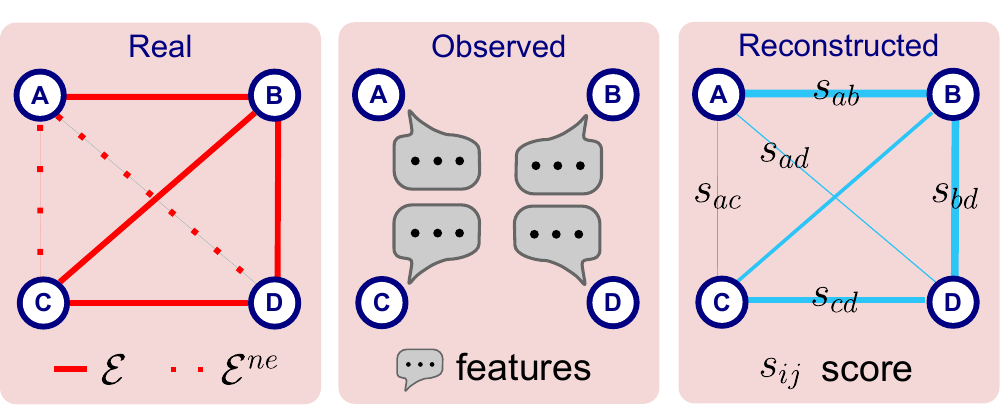}};
\end{tikzpicture}
\caption{Visual representation of the framework adopted to carry out link prediction in~\cite{parisi_entropybased_2018}. Panel \textbf{a}: the red, solid lines indicate the empirical edges while the red, dotted lines indicate the non-existent edges; the green, solid lines indicate the edges we have direct access to and the information about which can be employed to predict the ones indicated by the green, dotted lines; the aim of a link prediction exercise is understanding which of the non-observed links is actually non-existent (indicated by the light blue, dotted lines) and which is genuinely missing (indicated by the light blue, solid lines). Panel \textbf{b}: the three subpanels depict a slightly different framework, within which link prediction is carried out by employing node-specific features only - in a way that is reminiscent of network reconstruction exercises~\cite{ialongo_reconstructing_2022}.}
\label{fig:1}
\end{figure*}

More computer science-oriented approaches are, instead, defined by models whose parameters represent node- and edge-specific features~\cite{friedman_learning_1999,yu_stochastic_2006,lu_link_2011,mungo_reconstructing_2023}: specifying a model, thus, amounts to learning a function that determines the probability for nodes $i$ and $j$ to be connected while taking as inputs \textcolor{black}{features that may be node-specific (e.g. the sales, the number of employees, the degree of country $i$ and country $j$), edge-specific (e.g. the geographic distance between countries $i$ and $j$, the volume of sector-specific exchanges) or both}.

In all cases, algorithms are trained on a portion of the whole network, i.e. the \emph{training set}, while the prediction is carried out on the links of the unobserved part, i.e. the \emph{test set}.\\

The present work is devoted to comparing the performance of different link prediction methods in recovering the missing portion of the edge set of the (binary, undirected representation of the) World Trade Web and of the electronic Market for Interbank Deposits. The framework is the same as in~\cite{parisi_entropybased_2018}, where the training set coincides with the whole network apart from a percentage of deleted connections and the proper prediction exercise solely concerns these missing edges (see also figure~\ref{fig:1}). Hereby, we consider algorithms belonging to each of the three, aforementioned classes, such as the Gravity Model (taking as inputs `exogenous' features like each country's Gross Domestic Product and their geographic distances), the Configuration Model (taking as inputs `endogenous' features like the degrees) and the Gradient Boosting Decision Tree (taking as inputs both `exogenous' and `endogenous' features\footnote{\textcolor{black}{The terms `endogenous' and `exogenous' are borrowed from economics to distinguish between features that are structural in nature, like the degrees (`endogenous', i.e. `inherent to the structure') and features that, instead, are not, like the Gross Domestic Products and the geographic distances (`exogenous', i.e. `not inherent to the structure')}}).

\section{Setting up the framework}
\label{sec:framework}

Let us indicate with $\mathbf{A}$ the $N\times N$ adjacency matrix of the network we are considering, with $\mathcal{U}$ the set of all node pairs and with $\mathcal{E}$ the edge-set: as a consequence, $\mathcal{E}^{ne}=\mathcal{U}\setminus\mathcal{E}$ indicates the set of non-existent links~\cite{lu_link_2011,parisi_entropybased_2018}. The edge-set is, then, partitioned into the set of observed, or accessible, links $\mathcal{E}^{obs}$ and the set of missing links $\mathcal{E}^{miss}=\mathcal{E}\setminus\mathcal{E}^{obs}$; lastly, let us denote the union of the set of non-existent links and the set of missing links - collecting the node pairs appearing as unconnected either because the link does not exist or because the information about its presence is missing - as $\mathcal{E}^{no}=\mathcal{E}^{ne}\cup\mathcal{E}^{miss}=\mathcal{U}\setminus\mathcal{E}^{obs}$~\cite{lu_link_2011,parisi_entropybased_2018} (see also the first panel of figure \ref{fig:1}\textbf{a}). Link prediction algorithms output $|\mathcal{E}^{no}|=|\mathcal{E}^{ne}|+|\mathcal{E}^{miss}|$ scores, \textcolor{black}{hereby indicated as $\{s_{ij}\}_{ij\in\mathcal{E}^{no}}$}, to be assigned to the non-observed links: while the $|\mathcal{E}^{miss}|$ largest scores identify the predicted missing links, the remaining $|\mathcal{E}^{ne}|$ scores identify the predicted non-existent ones; let us indicate the two sets as $\overline{\mathcal{E}^{miss}}$ and $\overline{\mathcal{E}^{ne}}$.

A bit more compactly, upon considering that $\mathcal{U}=\mathcal{E}^{obs}\cup\mathcal{E}^{miss}\cup\mathcal{E}^{ne}$, one can define an $N\times N$ adjacency matrix $\mathbf{A}^{obs}=\{a_{ij}^{obs}\}_{ij\in\mathcal{U}}$, where

\begin{equation}
a_{ij}^{obs}=
\begin{cases}
1\text{ if }ij\in\mathcal{E}^{obs}\\
0\text{ if }ij\in\mathcal{E}^{miss}\\
0\text{ if }ij\in\mathcal{E}^{ne}
\end{cases}
\end{equation}
- in practice, the original matrix $\mathbf{A}$ where $|\mathcal{E}^{miss}|$ connections have been deleted; the \emph{training set}, then, becomes $\mathbf{D}^{tr}=\{\mathbf{X},\mathbf{A}^{obs}\}$, \textcolor{black}{with $\mathbf{X}=\{\mathbf{x}_{ij}\}_{ij\in\mathcal{U}}$ and $\mathbf{x}_{ij}$ being a vector of features, which may refer to nodes $i$ and $j$, to the pair $ij$ or both} (see also the second panel of figure \ref{fig:1}\textbf{a}). Let us explicitly notice that \emph{i)} `endogenous' features are evaluated on the accessible portion of the dataset $\mathbf{A}^{obs}$; \emph{ii)} the structural portion of the training set consists of both connected and unconnected pairs of nodes - as we will see, the main difference between the black-box and white-box algorithms considered here lies precisely in the way such information is processed to make predictions.

After having trained a given model on $\mathbf{D}^{tr}$, prediction is performed on the \emph{test set}, individuated by the node pairs with no observed connection: in symbols, $\mathbf{D}^{te}=\{a_{ij}\}_{ij\in\mathcal{E}^{no}}$ (see also the third panel of figure \ref{fig:1}\textbf{a}).

The goodness of a given prediction performance can be evaluated according to a number of different statistical indicators. Let us indicate the number of true positives (i.e. the correctly identified missing links) with

\begin{equation}
\text{TP}=|\overline{\mathcal{E}^{miss}}\cap\mathcal{E}^{miss}|,
\end{equation}
the number of true negatives with

\begin{equation}
\text{TN}=|\overline{\mathcal{E}^{ne}}\cap\mathcal{E}^{ne}|
\end{equation}
and the number of false positives with

\begin{equation}
\text{FP}=|\overline{\mathcal{E}^{miss}}\cap\mathcal{E}^{ne}|;
\end{equation}
upon doing so, we can define the \emph{True Positives Rate} (TPR) as

\begin{equation}
\text{TPR}=\frac{|\overline{\mathcal{E}^{miss}}\cap\mathcal{E}^{miss}|}{|\mathcal{E}^{miss}|},
\end{equation}
the \emph{False Positives Rate} (FPR) as

\begin{equation}
\text{FPR}=\frac{|\overline{\mathcal{E}^{miss}}\cap\mathcal{E}^{ne}|}{|\mathcal{E}^{ne}|},
\end{equation}
the \emph{Jaccard Index} (JI) as

\begin{equation}
\text{JI}=\frac{|\overline{\mathcal{E}^{miss}}\cap\mathcal{E}^{miss}|}{|\overline{\mathcal{E}^{miss}}\cup\mathcal{E}^{miss}|}
\end{equation}
and the \emph{Accuracy} (ACC) as

\begin{equation}
\text{ACC}=\frac{\text{TP}+\text{TN}}{|\mathcal{E}^{no}|}.
\end{equation}

In words, the TPR accounts for the percentage of missing links that are correctly identified as such, the JI enriches the picture provided by the TPR by accounting for the number of false positives too and the ACC accounts for the percentage of correctly identified links, be they missing or non-existent. 

Lastly, the \emph{Area Under the Receiver Operating Curve} (AUROC) is defined as the area under the curve obtained upon scattering the TPR versus the FPR as the list of links ranked in decreasing order of the chosen score is gone through and quantifies the extent to which a given link prediction algorithm performs better than a random one - that flips a coin to classify each non-observed link either as non-existent or missing. {\color{black} A complementary index is the \emph{Area Under the Precision-Recall Curve} (AUPRC) \cite{polanco_drugdisease_2025}, defined as the area under the curve obtained upon scattering the \emph{Positive Predictive Value} (PPV), defined as

\begin{equation}
\text{PPV}=\frac{|\overline{\mathcal{E}^{miss}}\cap\mathcal{E}^{miss}|}{|\overline{\mathcal{E}^{miss}}|},
\end{equation}
versus the TPR as the list of links ranked in decreasing order of the chosen score is gone through and quantifying the likelihood that a predictive positive is, indeed, a true one.}

\section{Alternative sampling schemes}
\label{sec:alternatives}

\textcolor{black}{As our link prediction exercise needs the explicit definition of the aforementioned sets and there are many ways of selecting $|\mathcal{E}^{miss}|$ links out of $|\mathcal{E}|$ possible ones, we have sampled uniformly at random the $10\%$, $20\%$ and $30\%$ of links a system-specific number of times (see below) to populate $\mathcal{E}^{miss}$ and generate $\mathcal{E}^{obs}=\mathcal{E}\setminus\mathcal{E}^{miss}$; each of the statistical indicators above has, then, been averaged over such samples.}

\textcolor{black}{Such a sampling scheme, however, `favours' well-connected nodes; in order to test alternative ones, we have also considered a scheme that samples links from $\mathcal{E}$ with a probability that is inversely proportional to the product of the degrees of the involved vertices and the one named \texttt{Random Node-Edge Sampler}~\cite{krishnamurthy_reducing_2005}, sampling nodes and edges uniformly at random in two, subsequent steps.}

\section{Missing links prediction algorithms}
\label{sec:algorithms}

\textcolor{black}{The present section is devoted to illustrate the missing links prediction algorithms considered in the present paper, noticing that each algorithm is characterised by a specific functional form of the prediction scores $\{s_{ij}\}_{ij\in\mathcal{E}^{no}}$: given a node pair $ij$, each algorithm outputs a score $s_{ij}$ accounting for the likelihood that the involved nodes are connected by a link, with higher values of $s_{ij}$ indicating a higher chance that $ij\in\mathcal{E}$.}

\subsection{Econometric approaches}

A simple, yet successful, recipe to model the World Trade Web (WTW) is the one provided by the Gravity Model (GM)~\cite{isard_location_1954,tinbergen_shaping_1962}. According to it, the prediction score reads

\begin{equation}
s_{ij}^\text{GM-I}=\text{GDP}_i\text{GDP}_j,
\end{equation}
where $\text{GDP}_i$ ($\text{GDP}_j$) is the Gross Domestic Product of country $i$ (country $j$). A more refined GM-based link prediction recipe reads

\begin{align}\label{eq:score_Gravity}
s_{ij}^\text{GM-II}&=\ln\left[\epsilon\frac{(\text{GDP}_i\text{GDP}_j)^\beta}{d_{ij}^\gamma}\right]\nonumber\\
&=\alpha+\beta\ln(\text{GDP}_i\text{GDP}_j)-\gamma\ln d_{ij},
\end{align}
where $d_{ij}$ is the geographic distance between countries $i$ and $j$ and $\alpha=\ln\epsilon$. To ease numerical manipulations, the GM features have been normalised, thus reading $\omega_i=\text{GDP}_i/\overline{\text{GDP}}$ and $\delta_{ij}=d_{ij}/\overline{d}$, where $\overline{\text{GDP}}=\sum_{i=1}^N\text{GDP}_i/N$ and $\overline{d}=2\sum_{i=1}^N\sum_{j(>i)}d_{ij}/N(N-1)$.

Such a model can be fully determined upon maximising the log-likelihood function

\begin{equation}
\mathcal{L}(\mathbf{A}^{obs}|\alpha,\beta,\gamma)=\sum_{i=1}^N\sum_{j(>i)}[a_{ij}^{obs}\ln p_{ij}+(1-a_{ij}^{obs})\ln(1-p_{ij})],
\end{equation}
where

\begin{equation}
p_{ij}=\frac{1}{1+e^{-s_{ij}}},
\end{equation}
with $s_{ij}=s_{ij}^\text{GM-II}$. In order to evaluate the amount of information required by the recipe above, let us explicitly compute the gradient of the corresponding log-likelihood function. Since

\begin{align}
\mathcal{L}(\mathbf{A}^{obs}|\alpha,\beta,\gamma)&=\sum_{i=1}^N\sum_{j(>i)}[a_{ij}^{obs}\ln(e^{s_{ij}})-\ln(1+e^{s_{ij}})]\nonumber\\
&=\sum_{i=1}^N\sum_{j(>i)}[a_{ij}^{obs}s_{ij}-\ln\left(1+e^{s_{ij}}\right)],
\end{align}
one finds

\begin{align}
\frac{\partial\mathcal{L}}{\partial\bm{\theta}}&=\sum_{i=1}^N\sum_{j(>i)}\left[a_{ij}^{obs}-\frac{e^{s_{ij}}}{1+e^{s_{ij}}}\right]\frac{\partial s_{ij}}{\partial\bm{\theta}}\nonumber\\
&=\sum_{i=1}^N\sum_{j(>i)}\left[a_{ij}^{obs}-p_{ij}\right]\frac{\partial s_{ij}}{\partial\bm{\theta}},
\end{align}
where $\bm{\theta}=\{\alpha,\beta,\gamma\}$, an expression showing that the required amount of information coincides with the entire set of observed connections.

The inferred coefficients, obtained by running the scikit-learn Python package~\cite{pedregosa_scikitlearn_2011}, are shown in table~\ref{tab:coeffGravity}.

\begin{table}[t!]
\centering
\begin{tabular}{c|c|c|c}
\hline
\hline
year & $\alpha$ & $\beta$ & $\gamma$ \\
\hline 
\hline
$1990$ & $2.102\pm 0.020$ & $0.436\pm 0.004$ & $0.526\pm 0.009$ \\
\hline
$1991$ & $1.643\pm 0.159$ & $0.409\pm 0.004$ & $0.492\pm 0.007$ \\
\hline
$1992$ & $1.766\pm 0.191$ & $0.425\pm 0.004$ & $0.482\pm 0.009$ \\
\hline
$1993$ & $1.868\pm 0.012$ & $0.443\pm 0.004$ & $0.538\pm 0.013$ \\
\hline
$1994$ & $1.955\pm 0.195$ & $0.442\pm 0.004$ & $0.583\pm 0.028$ \\
\hline
$1995$ & $2.00\pm 0.027$ & $0.442\pm 0.004$ & $0.579\pm 0.015$ \\
\hline
$1996$ & $2.162\pm 0.017$ & $0.449\pm 0.005$ & $0.580\pm 0.016$ \\
\hline
$1997$ & $2.346\pm 0.035$ & $0.472\pm 0.007$ & $0.584\pm 0.016$ \\
\hline
$1998$ & $2.346\pm 0.014$ & $0.474\pm 0.004$ & $0.595\pm 0.019$ \\
\hline
$1999$ & $3.361\pm 0.017$ & $0.477\pm 0.004$ & $0.575\pm 0.015$ \\
\hline
$2000$ & $2.410\pm 0.030$ & $0.486\pm 0.005$ & $0.578\pm 0.017$ \\
\hline
\hline
\end{tabular}
\caption{Coefficients of the GM-based link prediction recipe defined by eq.~\ref{eq:score_Gravity}, trained on the years listed in the first column. The WTW has been considered in its binary, undirected representation.}
\label{tab:coeffGravity}
\end{table}

\subsection{Machine learning approaches}

The model we consider here is named Gradient Boosting Decision Tree (GBDT) and belongs to the family of \emph{ensemble methods}, that combine several base-learners, trained in sequence, to achieve a better performance than the one that would be obtained by employing each of them singularly~\cite{friedman_greedy_2001,ke_lightgbm_2017,mungo_reconstructing_2023}: for binary classification tasks, the GBDT outputs a raw score which is, then, mapped to a probability coefficient. More in detail, the raw score at the $K$-th step reads

\begin{equation}
\label{eq:GB1}
\textcolor{black}{s_{ij}^{(K)}=\alpha+\sum_{k=1}^K\rho_{(k)}\phi(\mathbf{x}_{ij}|\bm{\theta}_{(k)}),}
\end{equation}
i.e. is a summation over the aforementioned base-learners, \textcolor{black}{taking as input the vector of features $\mathbf{x}_{ij}$} and parametrised by $\bm{\theta}$; the parameters $\rho$ and $\bm{\theta}$ are, then, learnt sequentially. More explicitly, at the zero-th step, one has

\begin{equation}
s_{ij}^{(0)}=\alpha,
\end{equation}
where the value of $\alpha$ is chosen in order to minimise the loss function (equivalently, maximise the corresponding log-likelihood function)

\begin{equation}
\ell(\mathbf{A}^{obs}|\mathbf{s}^{(0)})=-\sum_{i=1}^N\sum_{j(>i)}[a_{ij}^{obs}\ln p_{ij}+(1-a_{ij}^{obs})\ln(1-p_{ij})],
\end{equation}
where $p_{ij}=1/(1+e^{-s_{ij}^{(0)}})$. At the first step one has

\begin{equation}
\textcolor{black}{s_{ij}^{(1)}=\alpha+\rho_{(1)}\phi(\mathbf{x}_{ij}|\bm{\theta}_{(1)}),}
\end{equation}
where $\bm{\theta}_{(1)}$ is determined as

\textcolor{black}{
\begin{align}
&\bm\theta_{(1)}=\\
&\arg\min_{\bm{\theta},\eta}\left\{\sum_{i=1}^N\sum_{j(>i)}\left[-\frac{\partial\ell(a_{ij}^{obs}|s_{ij})}{\partial s_{ij}}\bigg|_{s_{ij}=s_{ij}^{(0)}}-\eta\phi(\bm x_{ij}|\vect\theta)\right]^2\right\},\nonumber
\end{align}
i.e. $\bm{\theta}$ is chosen to minimise the distance from the gradient of the loss function, evaluated in $s_{ij}=s_{ij}^{(0)}$~\cite{friedman_greedy_2001,mungo_reconstructing_2023}; subsequently, $\rho_{(1)}$ is chosen to minimise the `updated' loss function
\begin{equation}
\rho_{(1)}=\arg\min_\rho\left\{\sum_{i=1}^N\sum_{j(>i)}\ell(a_{ij}^{obs}|\alpha+\rho\phi(\bm{x}_{ij}|\bm{\theta}_{(1)}))\right\}.
\end{equation}}

By iterating the steps above $K$ times, it is possible to build the predictor $s^{(K)}$~\cite{friedman_greedy_2001}.

The base-learners are usually chosen to be \emph{decision trees}~\cite{friedman_greedy_2001}, hence returning a prediction by performing a sequence of data splits according to the decision rules embodied by the nodes of the trees, with the root corresponding to the initial split and the leaves corresponding to the prediction itself.

The GBDT, trained on a number of different combinations of features - i.e. each country's GDP, each country's GDP and their geographic distances, each country's degree, each country's degree and their geographic distances - has been implemented by running the LightGBM Python package~\cite{ke_lightgbm_2017}; \textcolor{black}{all `endogenous' features - both those concerning the individual nodes $i$ and $j$ and those concerning the pair $ij$ - are computed on $\mathbf{A}^{obs}$}.

\subsection{Entropy-based approaches}
\label{sec:linear ERG}

Maximum-entropy models have been widely employed, either to detect the empirical patterns of a networked configuration that cannot be interpreted as a mere by-product of the enforced constraints or to individuate the most likely networked configuration(s) that are compatible with the available information: also named \emph{Exponential Random Graphs} (ERGs)~\cite{park_statistical_2004, bianconi_entropy_2007,squartini_analytical_2011,fronczak_statistical_2012,squartini_unbiased_2015,saracco_randomizing_2015,cimini_statistical_2019}, the models of this class embody a physics-rooted approach to the study of network science~\cite{dorogovtsev_principles_2003,park_origin_2003,park_statistical_2004}, ensuring that the imposed constraints are preserved as ensemble averages.

More formally, entropy-based approaches define a probability distribution over a properly-defined ensemble of graphs that is as random as possible, given a set of $M$ constraints $\mathbf{C}(\mathbf{G})=\{C_i(\mathbf{G})\}_{i=1}^M$ on the expected value of a number of observables: in symbols,

\begin{equation}\label{eq:ERG}
P(\mathbf{G}|\bm{\theta})=\frac{e^{-H(\mathbf{G},\bm{\theta})}}{Z(\bm{\theta})}=\frac{e^{-\mathbf{C}(\mathbf{G})\cdot\bm{\theta}}}{Z(\bm{\theta})},
\end{equation}
where the parameters $\bm{\theta}$ are the Lagrange multipliers associated with the constraints themselves. They can be numerically determined by maximising the log-likelihood $\ln P(\mathbf{G}^*|\bm{\theta})$ of observing the empirical configuration $\mathbf{G}^*$. It is easy to prove that such a recipe leads to solve the system of non-linear, coupled equations $\langle\mathbf{C}\rangle=\mathbf{C}(\mathbf{G}^*)$, i.e. $\langle C_i\rangle=C_i(\mathbf{G}^*)$, $i=1\dots M$~\cite{garlaschelli_maximum_2008,vallarano_fast_2021,divece_deterministic_2023}.

In what follows, we will consider instances of the \emph{separable} Hamiltonian~\cite{marzi_reproducing_2025}

\begin{equation}\label{eq:separableH}
H(\mathbf{A},\bm{\theta})=\sum_{i=1}^N\sum_{j(>i)}H_{ij}(\mathbf{A},\theta_{ij})=\sum_{i=1}^N\sum_{j(>i)}a_{ij}\cdot f(\mathbf{A},\theta_{ij})
\end{equation}
that induces a probability distribution reading

\begin{equation}
P(\mathbf{A}|\bm{\theta})=\prod_{i=1}^N\prod_{j(>i)}p_{ij}^{a_{ij}}(1-p_{ij})^{1-a_{ij}}
\end{equation}
with

\begin{equation}
p_{ij}=\frac{e^{-f(\mathbf{A},\theta_{ij})}}{1+e^{-f(\mathbf{A},\theta_{ij})}};
\label{eq:Pfact}
\end{equation}
notice that \emph{linear} Hamiltonians can be recovered as special cases upon posing $f(\mathbf{A},\theta_{ij})=f(\theta_{ij})$.

Within such a framework, the $|\mathcal{E}^{miss}|$ largest coefficients of the set $\{p_{ij}\}_{(i,j)\in\mathcal{E}^{no}}$ identify a subgraph $\mathbf{\Sigma}^*$ whose probability reads

\begin{equation}
P(\mathbf{\Sigma}^*|\bm{\theta})=\prod_i\prod_{j(>i)}p_{ij}^{\sigma_{ij}^*}(1-p_{ij})^{1-\sigma_{ij}^*},
\end{equation}
the product running over the pairs of nodes in $\mathcal{E}^{no}$ and $\sigma_{ij}^*$ being $1$ only for the $(i,j)$s corresponding to the $|\mathcal{E}^{miss}|$ largest probability coefficients - the remaining entries reading zero: among all subgraphs induced by precisely $|\mathcal{E}^{miss}|$ links, $\mathbf{\Sigma}^*$ is the one attaining the highest probability. In words, entropy-based methods lead us to interpret the non-observed links which have been assigned the largest probability coefficients as the ones that are most likely to appear given the chosen constraints~\cite{parisi_entropybased_2018}.

\subsubsection{Link prediction via linear ERGs}

\noindent\textit{Configuration Model.} Let us, now, consider the first specification of the entropy-based framework: it boils down to posing $\mathbf{C}=\mathbf{k}$, with $k_i=\sum_{j(\neq i)}a_{ij}$, an identification that leads to $H_\text{CM}(\mathbf{A})=\sum_{i=1}^N\sum_{j(>i)}a_{ij}\cdot(\theta_i+\theta_j)$, further inducing

\begin{equation}\label{eq:score_ConfigModel}
p_{ij}^\text{CM}=\frac{x_ix_j}{1+x_ix_j},
\end{equation}
where $x_i=e^{-\theta_i}$; the parameters $\{x_i\}_{i=1}^N$ can be determined by solving the system of non-linear, coupled equations

\begin{equation}\label{eq:CM_system}
k_i(\mathbf{A}^{obs})=\sum_{j(\neq i)}p_{ij}^\text{CM}=\sum_{j(\neq i)}\frac{x_ix_j}{1+x_ix_j},\quad\forall\:i,
\end{equation}
a task that can be accomplished by implementing the iterative recipe~\cite{vallarano_fast_2021}

\begin{equation}
x_i^{t+1}=\frac{k_i(\mathbf{A}^{obs})}{\sum_{j(\neq i)}\left(\frac{x_j^t}{1+x_i^tx_j^t}\right)}.
\end{equation}

The prescription above, characterising the so-called \emph{Configuration Model} (CM)~\cite{park_statistical_2004,squartini_analytical_2011,fronczak_statistical_2012,squartini_unbiased_2015,cimini_statistical_2019}, matches the one considered in~\cite{parisi_entropybased_2018}, where the accessible portion of a graph was employed to estimate its degree sequence and, on this basis, obtain the probability for any two nodes to be connected - to be later employed as a score for the existence of a link between any two unconnected vertices. Such a procedure rests upon the assumption that if the accessible portion of a graph is reproduced with a certain accuracy by a certain topological information, the inaccessible portion will be reproduced with the same degree of accuracy by the same kind of information\footnote{In a sense, the link prediction problem is interpreted as an instance of the network reconstruction one~\cite{squartini_reconstruction_2018,cimini_reconstructing_2021}, the aim of which is precisely that of inferring a graph structure from a set of observed quantities - although network reconstruction usually deals with \emph{less information} to predict \emph{more aggregate} properties (see also figure \ref{fig:1}\textbf{b}).}.

A simpler prescription ensuring that the expected degree sequence matches the observed one is provided by the Chung-Lu model (CL):

\begin{equation}\label{eq:score_ChungLu}
p_{ij}^\text{CL}=\frac{k_ik_j}{2L},
\end{equation}
where $k_i$ and $k_j$ must be interpreted as the degrees of the nodes evaluated on the accessible portion of the dataset $\mathbf{A}^{obs}$, i.e. $k_i(\mathbf{A}^{obs})$ and $k_j(\mathbf{A}^{obs})$; as it can be recovered from the CM by posing $p_{ij}^\text{CM}\simeq x_ix_j$, the validity of this model is limited to the case in which the network is sparse and the degree distribution is not too broad~\cite{vallarano_fast_2021}.\\

\noindent\textit{Configuration Model with Distances.} Let us, now, enrich the CM by adding information about the geographic distances between countries: this choice boils down to posing $\mathbf{C}=\{\mathbf{k},d\}$, with $k_i=\sum_{j(\neq i)}a_{ij}$ and $d=\sum_{i=1}^N\sum_{j(>i)}a_{ij}d_{ij}$~\cite{picciolo_role_2012,bianconi_information_2022}. Such an identification leads to $H_\text{CMD}(\mathbf{A})=\sum_{i=1}^N\sum_{j(>i)}a_{ij}\cdot(\theta_i+\theta_j+\gamma d_{ij})$, an expression inducing

\begin{equation}\label{eq:score_ConfigDist}
p_{ij}^\text{CMD}=\frac{x_ix_jw^{d_{ij}}}{1+x_ix_jw^{d_{ij}}},
\end{equation}
where $x_i=e^{-\theta_i}$ and $w=e^{-\gamma}$. The parameters $\{x_i\}_{i=1}^N$ and $w$ can be determined by solving the system of non-linear, coupled equations

\begin{equation}\label{eq:CMD_sys}
\left\{
\begin{split}
k_i(\mathbf{A}^{obs})&=\sum_{j(\neq i)}p_{ij}^\text{CMD},\quad\forall\:i\\
d(\mathbf{A}^{obs})&=\sum_{i=1}^N\sum_{j(>i)}p_{ij}^\text{CMD}d_{ij}
\end{split}
\right.
\end{equation}
a task that can be accomplished by implementing the iterative recipe

\begin{equation}
\left\{
\begin{split}
x_i^{t+1}&=\frac{k_i(\mathbf{A}^{obs})}{\sum_{j(\neq i)}\left(\frac{x_j^tw_t^{d_{ij}}}{1+x_i^tx_j^tw_t^{d_{ij}}}\right)},\quad\forall\:i\\
w_{t+1}&=\frac{d(\mathbf{A}^{obs})}{\sum_{i=1}^N\sum_{j(>i)}\left(\frac{x_i^tx_j^tw_t^{d_{ij}-1}}{1+x_i^tx_j^tw_t^{d_{ij}}}\right)d_{ij}}
\end{split}
\right.
\end{equation}
notice how the \emph{Configuration Model with Distances} (CMD) relies upon both `endogenous' (i.e. topological) and `exogenous' (i.e. non-topological) features. To ease numerical manipulations, the distances have been normalised, thus reading $\delta_{ij}=d_{ij}/\overline{d}$, where $\overline{d}=2\sum_{i=1}^N\sum_{j(>i)}d_{ij}/N(N-1)$.\\

\noindent\textit{Fitness Model.} Firstly introduced in~\cite{caldarelli_scalefree_2002} as a generative mechanism alternative to the `rich-gets-richer' one, the \emph{Fitness Model} (FM) prescribes to associate each node with a hidden variable: following~\cite{garlaschelli_fitnessdependent_2004}, here we pose $x_i=\sqrt{z}\omega_i$, with $\omega_i=\text{GDP}_i/\overline{\text{GDP}}$ and $\overline{\text{GDP}}=\sum_{i=1}^N\text{GDP}_i/N$, thus obtaining

\begin{equation}\label{eq:score_fitness}
p_{ij}^\text{FM}=\frac{z\omega_i\omega_j}{1+z\omega_i\omega_j}.
\end{equation}

The parameter $z$ can be inferred by imposing that the expected number of links matches the observed one $L(\mathbf{A}^{obs})=|\mathcal{E}^{obs}|$, i.e. that

\begin{equation}
L(\mathbf{A}^{obs})=\sum_{i=1}^N\sum_{j(>i)}p_{ij}^\text{FM},
\end{equation}
a task that can be accomplished by implementing the iterative recipe

\begin{equation}
z_{t+1}=\frac{L(\mathbf{A}^{obs})}{\sum_{i=1}^N\sum_{j(>i)}\left(\frac{\omega_i\omega_j}{1+z_t\omega_i\omega_j}\right)}.
\end{equation}

A different position reads $x_i=\sqrt{z}s_i$, where, now, $s_i=\sum_{j(\neq i)}w_{ij}$ indicates the strength of node $i$. This second choice is a popular one when analysing financial networks, the available data about which usually concern lending and borrowing relationships. This second specification is also known with the name of \emph{density-corrected Gravity Model} (dcGM)~\cite{cimini_systemic_2015,cimini_reconstructing_2021}.\\

\noindent\textit{Fitness Model with Distances.} As for the CM, also the CMD can be turned into a \emph{Fitness Model with Distances} (FMD). By operating the same replacement as above, one gets 

\begin{equation}\label{eq:score_fitnessDist}
p_{ij}^\text{FMD}=\frac{z\omega_i\omega_jw^{d_{ij}}}{1+z\omega_i\omega_jw^{d_{ij}}},
\end{equation}
the system to be solved, now, reading

\begin{equation}\label{eq:FMD_sys}
\left\{
\begin{split}
L(\mathbf{A}^{obs})&=\sum_{i=1}^N\sum_{j(>i)}p_{ij}^\text{FMD}\\
d(\mathbf{A}^{obs})&=\sum_{i=1}^N\sum_{j(>i)}p_{ij}^\text{FMD}d_{ij}
\end{split}
\right.
\end{equation}
and the iterative recipe becoming

\begin{equation}
\left\{
\begin{split}
z_{t+1}&=\frac{L(\mathbf{A}^{obs})}{\sum_{i=1}^N\sum_{j(>i)}\left(\frac{\omega_i\omega_jw^{d_{ij}}}{1+z_t\omega_i\omega_jw^{d_{ij}}}\right)}\\
w_{t+1}&=\frac{d(\mathbf{A}^{obs})}{\sum_{i=1}^N\sum_{j(>i)}\left(\frac{z_t\omega_i\omega_jw_t^{d_{ij}-1}}{1+z_t\omega_i\omega_jw_t^{d_{ij}}}\right)d_{ij}}
\end{split}
\right.
\end{equation}

To ease numerical manipulations, the distances have been normalised, thus reading $\delta_{ij}=d_{ij}/\overline{d}$, where $\overline{d}=2\sum_{i=1}^N\sum_{j(>i)}d_{ij}/N(N-1)$.

\subsubsection{Link prediction via non-linear ERGs}

So far, we have focused on linear ERGs; let us, now, move to considering non-linear ERGs.\\

\noindent\textit{Fitness-induced 2-Star Model.} The simplest model of this kind is the so-called \emph{2-Star Model} (2SM)~\cite{park_statistical_2004,park_solution_2004,bolfe_analytic_2021}, induced by posing $\mathbf{C}=\{L,S\}$, with $L=\sum_{i=1}^N\sum_{j(>i)}a_{ij}=\sum_{i=1}^Nk_i/2$ and $S=\sum_{i=1}^N\sum_{j(>i)}\sum_{m(\neq i,j)}a_{im}a_{mj}=\sum_{i=1}^Nk_i^2/2-L$. Such an identification leads to $H_\text{2SM}(\mathbf{A})=\sum_{i=1}^N\sum_{j(>i)}a_{ij}\cdot[\alpha+\beta(k_i+k_j)]$, an expression inducing

\begin{equation}\label{eq:score_TwoStar}
p_{ij}^\text{2SM}=\frac{xy^{\langle k_i\rangle+\langle k_j\rangle}}{1+xy^{\langle k_i\rangle+\langle k_j\rangle}},
\end{equation}
where $x=e^{-\alpha}$ and $y=e^{-\beta}$. One would be tempted to identify the expected value of the degrees with the observed one but consistency forbids such a match: as the only available information concerns two, global quantities, degrees are not directly accessible. In order to make the model fully consistent, a more refined formulation is, thus, needed.

First, the model we are looking for should be heterogeneous, thus ensuring that node-specific differences are preserved. More quantitatively, one would be looking for an expression like

\begin{equation}\label{eq:score_dcTwoStar}
p_{ij}^\text{dc2SM}=\frac{x_ix_jy^{k_i+k_j}}{1+x_ix_jy^{k_i+k_j}},
\end{equation}
where $k_i$ and $k_j$ must be interpreted as the degrees of the nodes evaluated on $\mathbf{A}^{obs}$, i.e. $k_i(\mathbf{A}^{obs})$ and $k_j(\mathbf{A}^{obs})$; as proven in~\cite{marzi_reproducing_2025}, however, the model above, named \emph{degree-corrected 2-Star Model} (dc2SM), suffers from limitations inherited by the mean-field approach inducing it. One is, thus, forced to consider a fitness-based variant of it, reading either

\begin{equation}\label{eq:score_fitTwoStarI}
p_{ij}^\text{fit2SM-I}=\frac{z\omega_i\omega_jy^{\langle k_i\rangle+\langle k_j\rangle}}{1+z\omega_i\omega_jy^{\langle k_i\rangle+\langle k_j\rangle}},
\end{equation}
where $\langle k_i\rangle=\sum_{j(\neq i)}p_{ij}^\text{fit2SM-I}$, or

\begin{equation}
p_{ij}^\text{fit2SM-II}=\frac{zs_is_jy^{\langle k_i\rangle+\langle k_j\rangle}}{1+zs_is_jy^{\langle k_i\rangle+\langle k_j\rangle}},
\end{equation}
where $\langle k_i\rangle=\sum_{j(\neq i)}p_{ij}^\text{fit2SM-II}$. A third variant can be defined as

\begin{equation}\label{eq:score_fitTwoStar}
p_{ij}^\text{fit2SM-III}=\frac{z\tilde{x}_i\tilde{x}_jy^{\langle k_i\rangle+\langle k_j\rangle}}{1+z\tilde{x}_i\tilde{x}_jy^{\langle k_i\rangle+\langle k_j\rangle}},
\end{equation}
where $\langle k_i\rangle=\sum_{j(\neq i)}\tilde{x}_i\tilde{x}_j/(1+\tilde{x}_i\tilde{x}_j)=k_i$; the major difference with respect to the first recipe is the disentangled resolution of the formula above: in other words, one, first, determines the parameters $\{\tilde{x}_i\}_{i=1}^N$ by solving the CM and, then, insert them above to determine $z$ and $y$.

The difference with respect to the 2SM is evident: the three variants above are heterogenous, hence likely reproducing the observed degrees better than the 2SM. The parameters $z$ and $y$ can be determined by solving the system of non-linear, coupled equations

\begin{equation}
\left\{
\begin{split}
L(\mathbf{A}^{obs})&=\sum_{i=1}^N\sum_{j(>i)}p_{ij}^\text{fit2SM}\\
S(\mathbf{A}^{obs})&=\sum_{i=1}^N\sum_{j(>i)}\sum_{m(\neq i,j)}p_{im}^\text{fit2SM}p_{mj}^\text{fit2SM}
\end{split}
\right.
\end{equation}
where we have dropped the reference to the model specification as the recipe is the same in all cases.

Let us, now, face the problem of determining the expected degrees; to this aim, let us focus on the third recipe: as the parameters $\{\tilde{x}_i\}_{i=1}^N$ have been determined by solving the CM on $\mathbf{A}^{obs}$, we can insert the observed degrees directly into the recipe above. As a consequence, the two parameters $z$ and $y$ can, now, be obtained by implementing the iterative recipe

\textcolor{black}{
\begin{equation}
\left\{
\begin{split}
&z_{t+1}=\frac{L(\mathbf{A}^{obs})}{\sum_{i=1}^N\sum_{j(>i)}\frac{\tilde{x}_i\tilde{x}_jy_t^{k_i+k_j}}{1+z_t\tilde{x}_i\tilde{x}_jy^{k_i+k_j}_t}}\\
&y_{t+1}=\\
&\frac{S(\mathbf{A}^{obs})}{\sum_{i=1}^N\sum_{j(>i)}\sum_{m(\neq i,j)}\frac{z_t^2\tilde{x}_i\tilde{x}_j\tilde{x}_m^2y_t^{k_i+2k_m+k_j-1}}{\left(1+z\tilde{x}_i\tilde{x}_my^{k_i+k_m}_t\right)\left(1+z\tilde{x}_m\tilde{x}_jy^{k_m+k_j}_t\right)}}
\end{split}
\right.
\end{equation}}
where $k_i$, $k_m$ and $k_j$ must be interpreted as the degrees of the nodes evaluated on the accessible portion of the dataset $\mathbf{A}^{obs}$, i.e. $k_i(\mathbf{A}^{obs})$, $k_m(\mathbf{A}^{obs})$ and $k_j(\mathbf{A}^{obs})$.

\section{Data description}
\label{sec:data}

The aforementioned link prediction algorithms have been run and compared on a number of instances of economic and financial datasets.

As a first dataset, we have considered the WTW curated by Gleditsch~\cite{gleditsch_expanded_2002} and including yearly, bilateral, aggregated data on trade (the generic entry $\text{exp}_{ij}(y)$ is the sum of the single, commodity-specific exports from $i$ to $j$, during the year $y$, reported in millions of US dollars), yearly GDP values (reported in millions of US dollars) and the (time-independent) matrix of geographic distances between the capital cities of the countries in the data. We have, then, symmetrised and binarised the trade matrix by posing $a_{ij}=\Theta[w_{ij}]$, where $\Theta$ is the Heaviside step function and $w_{ij}=(\text{exp}_{ij}+\text{exp}_{ji})/2$ - in words, $w_{ij}$ is the bilateral trade volume defined as the arithmetic mean of the export volume from country $i$ to country $j$ and the export volume from country $j$ to country $i$ - for each of the eleven years going from $1990$ to $2000$.

As a second dataset, we have considered the electronic Market for Interbank Deposit (eMID)~\cite{marzi_reproducing_2025}, represented as a weighted, directed network whose nodes are anonymised banks and weights represent exposures in millions of euros. Reported data cover the period January $1999$-December $2014$ on a daily frequency: the presence of a weight $w_{ij}$ at time $t$ indicates the existence of a total exposure $w_{ij}\geq 50.000$ euros, registered at the end of the particular period $t$ and directed from bank $i$ to bank $j$. Considering that $\simeq98\%$ of banks are Italian and that the volume of their transactions covers $\simeq85\%$ of the total volume (as of $2011$), our analysis solely focuses on the subgraph induced by such a subset of nodes. We have, then, symmetrised and binarised the exposure matrix by posing $a_{ij}=\Theta[w_{ij}+w_{ji}]$ - in words, $w_{ij}+w_{ij}$ is the sum of the exposures of the involved banks - for each aggregation period ranging from daily to yearly.

\section{Results}
\label{sec:results}

As in a real-case scenario the choice between different methods would be determined by the available features, it is useful to compare models informed by the same quantities. To this aim, let us recall that \emph{i)} the GM, defined by eq.~\ref{eq:score_Gravity}, takes as inputs the GDPs and the geographic distances; \emph{ii)} the first formulation of the FM, defined by eq.~\ref{eq:score_fitness}, takes as inputs the GDPs; \emph{iii)} the second formulation of the FM, defined by eq.~\ref{eq:score_fitnessDist}, takes as inputs the GDPs and the geographic distances; \emph{iv)} the CM and the CL, respectively defined by eqs.~\ref{eq:score_ConfigModel} and~\ref{eq:score_ChungLu}, take as inputs the degrees; \emph{v)} the CMD, defined by eq.~\ref{eq:score_ConfigDist}, takes as inputs the degrees and the geographic distances; \emph{vi)} the fit2SM, defined by eq.~\ref{eq:score_fitTwoStar}, takes as inputs the degrees (as the total number of links and the total number of two-stars can be deduced from them). Naturally, we have compared the performance of each white-box method with the instance of the GBDT taking as input the same amount of information.\\

Let us start commenting on the WTW. The basic statistics are reported in table~\ref{tab:datasets1}: as it can be appreciated, the number of nodes $N$ ranges between $159$ and $177$ and the total number of links $L$ ranges between $7639$ and $9866$; still, the link density $c=2L/N(N-1)$ does not fluctuate much, ranging between $0.55$ and $0.63$, as well as the average degree $\overline{k}=2L/N$, ranging between $\gtrsim95$ and $\gtrsim111$.

\begin{table}[t!]
\centering
\begin{tabular}{c|c|c|c|c}
\hline
\hline
year & $N$ & $L$ & $c$ & $\overline{k}$ \\
\hline
\hline 
$1990$ & $159$ & $7639$ & $0.608$ & $96.08$ \\
\hline
$1991$ & $173$ & $8249$ & $0.554$ & $95.36$ \\
\hline
$1992$ & $175$ & $8552$ & $0.562$ & $97.74$ \\
\hline 
$1993$ & $177$ & $8837$ & $0.567$ & $99.85$ \\
\hline 
$1994$ & $177$ & $9052$ & $0.591$ & $103.94$ \\
\hline
$1995$ & $177$ & $9199$ & $0.591$ & $103.94$ \\
\hline
$1996$ & $177$ & $9623$ & $0.618$ & $108.73$ \\
\hline
$1997$ & $177$ & $9864$ & $0.633$ & $111.46$ \\
\hline
$1998$ & $177$ & $9866$ & $0.633$ & $111.48$ \\
\hline
$1999$ & $177$ & $9864$ & $0.633$ & $111.46$ \\
\hline
$2000$ & $177$ & $9865$ & $0.633$ & $111.47$ \\
\hline
\hline
\end{tabular}
\caption{Basic statistics concerning the WTW dataset~\protect\cite{gleditsch_expanded_2002}, i.e. the total number of countries $N$, the total number of links $L$, the network density $c$ and the average degree $\overline{k}$.}
\label{tab:datasets1}
\end{table}

\begin{figure*}[t!]
\includegraphics[trim= 0 0 0 0, clip,width=\textwidth]{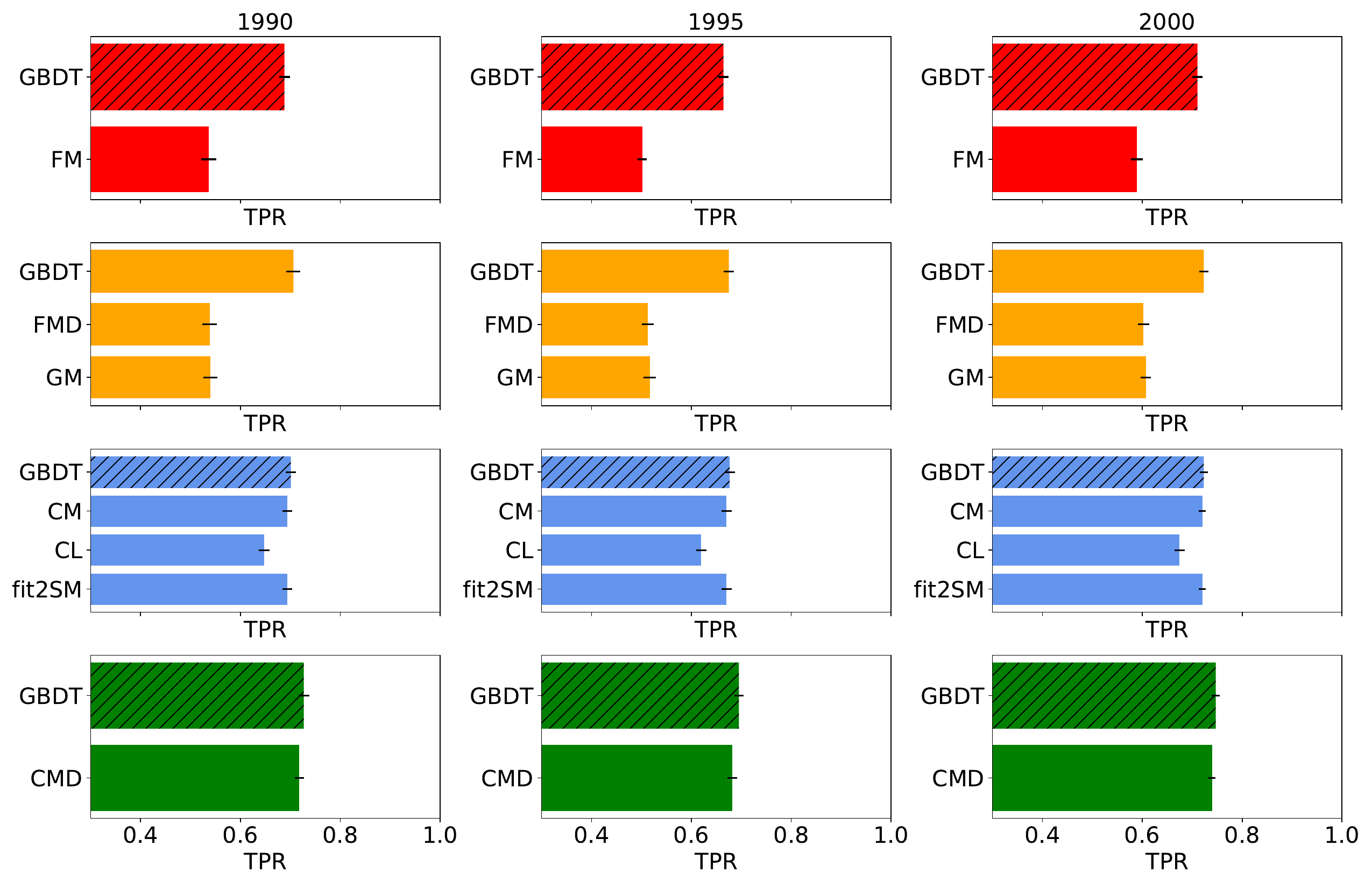}
{\includegraphics[trim= 300 10 100 15, clip,width=.55\textwidth]{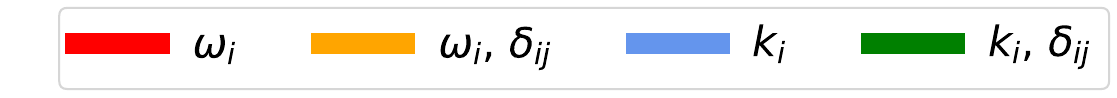}}
\caption{Performance of the models, measured in terms of TPR, for the years $1990$, $1995$ and $2000$ of the WTW. Each panel collects methods that rely on the same set of features (see the colour legend). The upper bar in each panel, marked with a diagonal pattern, corresponds to the black-box GBDT. \textcolor{black}{We have randomly selected the $10\%$ of links $40$ times to populate $\mathcal{E}^{miss}$ and generate $\mathcal{E}^{obs}=\mathcal{E}\setminus\mathcal{E}^{miss}$;} each statistical indicator has, then, been averaged over such samples, the standard deviation being represented by a horizontal, black bar. When considering `exogenous' features, each instance of the GBDT outperforms its white-box counterpart. The two classes of models, instead, perform in a comparable way when coming to consider `endogenous' features such as the degrees.}
\label{fig:2}
\end{figure*}

Figure~\ref{fig:2} depicts the TPR for each of the aforementioned models, on the years $1990$, $1995$ and $2000$. Let us recall that it is defined as $\text{TPR}=|\overline{\mathcal{E}^{miss}}\cap\mathcal{E}^{miss}|/|\mathcal{E}^{miss}|$, hence accounting for the percentage of missing links that are correctly identified as such. First, let us notice that when considering `exogenous' features, each instance of the GBDT outperforms its white-box counterpart. The two classes of models, instead, perform in a comparable way when coming to consider `endogenous' features such as the degrees and the number of two-stars: quite interestingly, the CL is the method performing worst, while the CM and the fit2SM perform in a very similar way - a result indicating that the present application does not necessarily require to consider non-linear ERGs.

\begin{figure*}[t!]
\includegraphics[trim= 0 0 0 0, clip,width=\textwidth]{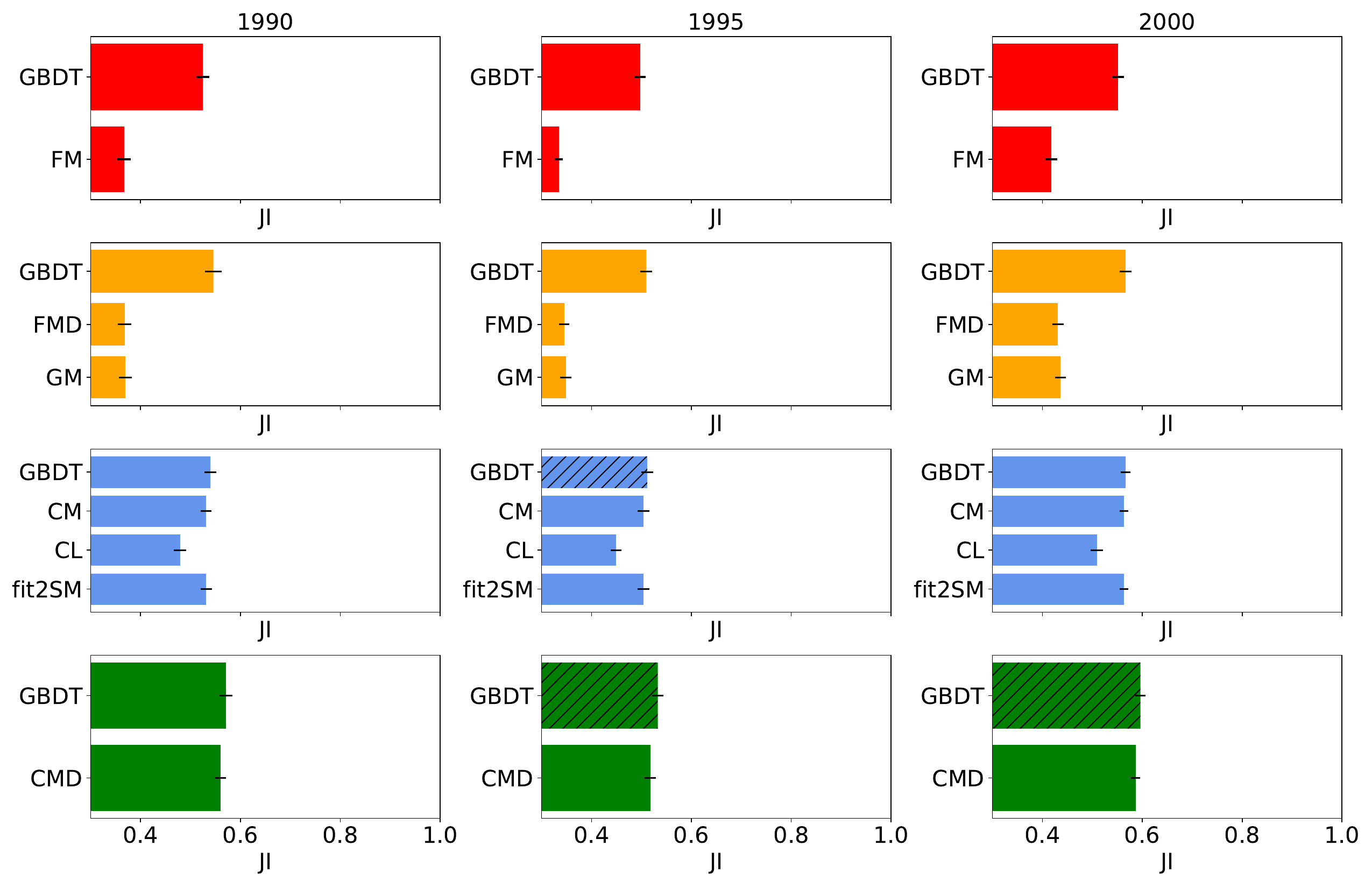}
{\includegraphics[trim= 300 10 100 15, clip,width=.55\textwidth]{allscores_legend1.pdf}}
\caption{Performance of the models, measured in terms of JI, for the years $1990$, $1995$ and $2000$ of the WTW. Each panel collects methods that rely on the same set of features (see the colour legend). The upper bar in each panel, marked with a diagonal pattern, corresponds to the black-box GBDT. \textcolor{black}{We have randomly selected the $10\%$ of links $40$ times to populate $\mathcal{E}^{miss}$ and generate $\mathcal{E}^{obs}=\mathcal{E}\setminus\mathcal{E}^{miss}$;} each statistical indicator has, then, been averaged over such samples, the standard deviation being represented by a horizontal, black bar. When considering `exogenous' features, each instance of the GBDT outperforms its white-box counterpart. The two classes of models, instead, perform in a comparable way when coming to consider `endogenous' features such as the degrees.}
\label{fig:JI_hbars}
\end{figure*}

\begin{figure*}[t!]
\includegraphics[trim= 0 0 0 0, clip,width=\textwidth]{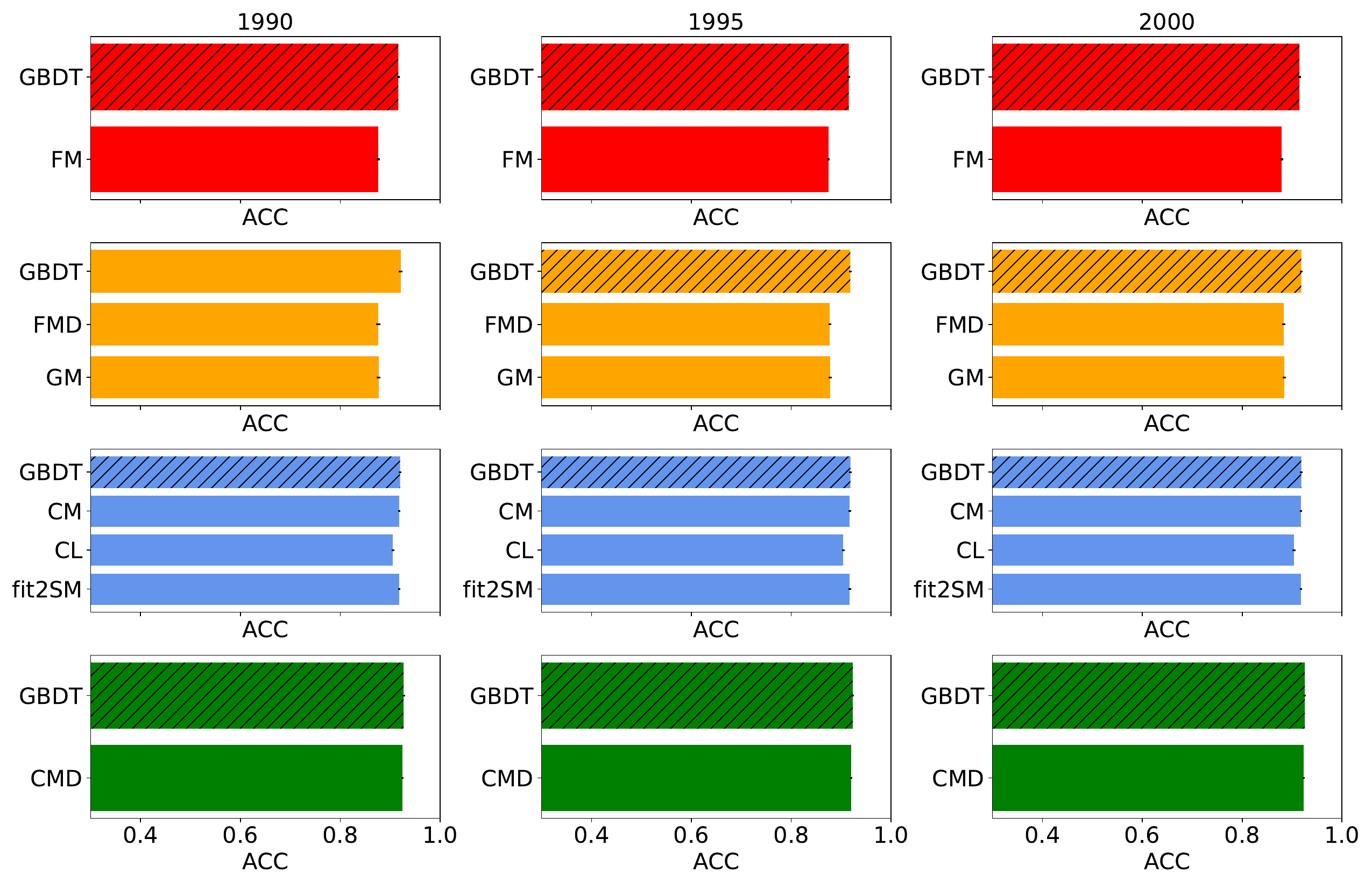}
{\includegraphics[trim= 300 10 100 15, clip,width=.55\textwidth]{allscores_legend1.pdf}}
\caption{Performance of the models, measured in terms of ACC, for the years $1990$, $1995$ and $2000$ of the WTW. Each panel collects methods that rely on the same set of features (see the colour legend). The upper bar in each panel, marked with a diagonal pattern, corresponds to the black-box GBDT. \textcolor{black}{We have randomly selected the $10\%$ of links $40$ times to populate $\mathcal{E}^{miss}$ and generate $\mathcal{E}^{obs}=\mathcal{E}\setminus\mathcal{E}^{miss}$;} each statistical indicator has, then, been averaged over such samples, the standard deviation being represented by a horizontal, black bar. When considering `exogenous' features, each instance of the GBDT outperforms its white-box counterpart. The two classes of models, instead, perform in a comparable way when coming to consider `endogenous' features such as the degrees.}
\label{fig:ACC_hbars}
\end{figure*}

Figure~\ref{fig:JI_hbars} depicts $\text{JI}=|\overline{\mathcal{E}^{miss}}\cap\mathcal{E}^{miss}|/|\overline{\mathcal{E}^{miss}}\cup\mathcal{E}^{miss}|$, hence enriching the picture provided by the TPR by accounting for the number of false positives too; figure~\ref{fig:ACC_hbars} depicts $\text{ACC}=(\text{TP}+\text{TN})/|\mathcal{E}^{no}|$ for each of the aforementioned models, hence accounting for the percentage of correctly identified links, be they missing or non-existent: both return a picture that is consistent with the one returned by the TPR.

\begin{figure*}[t!]
\begin{tikzpicture}
\node[anchor=south west,inner sep=0] (fig1) at (0,0) {\includegraphics[trim= 0 0 0 0, clip,width=\textwidth]{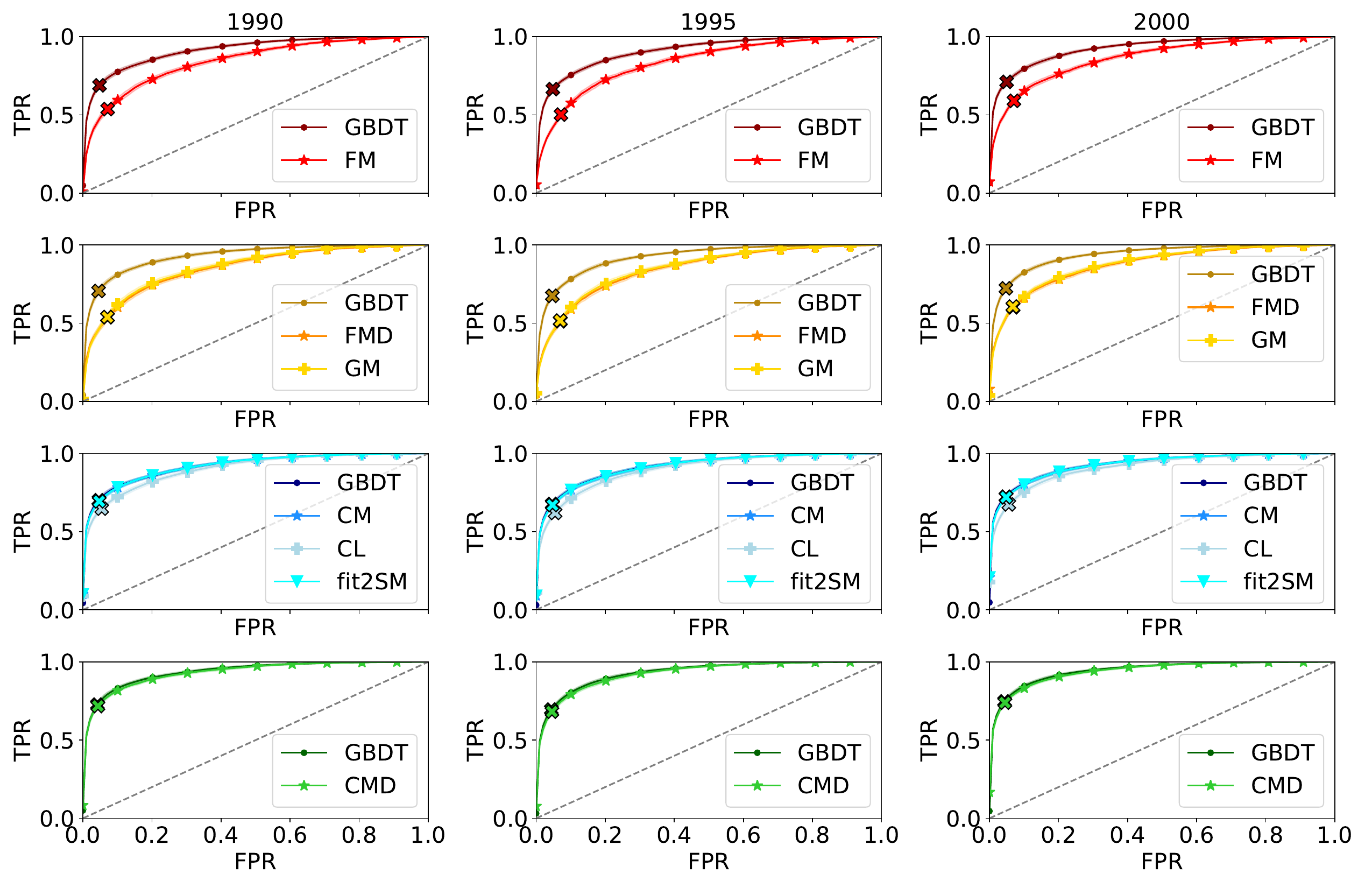}};
\node[anchor=north west, shift={(-5pt,3pt)}] at (fig1.north west) {\textbf{a)}};
\node[anchor=north west,inner sep=0] (fig2) at (fig1.south west) {\includegraphics[trim= 0 0 0 0 0, clip,width=\textwidth]{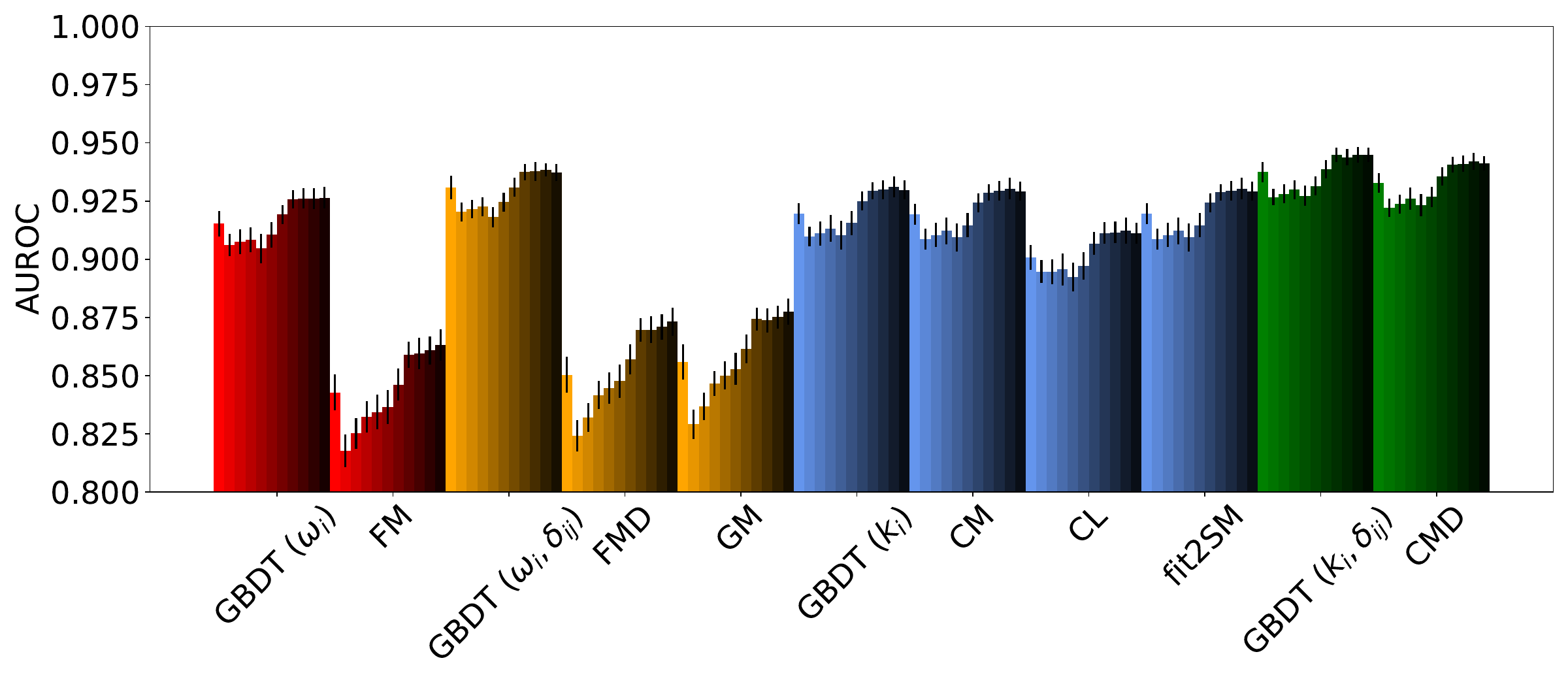}};
\node[anchor=north west, shift={(-5pt,3pt)}] at (fig2.north west) {\textbf{b)}};
\end{tikzpicture}
\caption{Panel \textbf{a}: performance of the models, measured in terms of ROC curves, for the years $1990$, $1995$ and $2000$ of the WTW. Each panel collects methods that rely on the same set of features. \textcolor{black}{We have randomly selected the $10\%$ of links $40$ times to populate $\mathcal{E}^{miss}$ and generate $\mathcal{E}^{obs}=\mathcal{E}\setminus\mathcal{E}^{miss}$ (the curves relative to each indicator are, in fact, $40$ partially overlapping curves corresponding to each realization).} The `X' markers indicate the TPR and FPR values obtained by selecting a number of missing links equal to $|\mathcal{E}^{miss}|$. Panel \textbf{b}: \textcolor{black}{each statistical indicator has been averaged over the $40$ realizations, the standard deviation being represented by a vertical, black bar.} The AUROC, depicted for all years (darker shades correspond to more recent years), enlarges when moving from less structured models, like the GM, to more structured ones, like the CMD.}
\label{fig:ROCcurves}
\end{figure*}

A complementary picture is provided by figure~\ref{fig:ROCcurves}, showing the ROC curves, obtained upon scattering the $\text{TPR}=|\overline{\mathcal{E}^{miss}}\cap\mathcal{E}^{miss}|/|\mathcal{E}^{miss}|$ versus the $\text{FPR}=|\overline{\mathcal{E}^{miss}}\cap\mathcal{E}^{ne}|/|\mathcal{E}^{ne}|$ as the list of links ranked in decreasing order of the chosen score is gone through, for each of the aforementioned models, on the years $1990$, $1995$ and $2000$. As evident upon looking at the picture, the AUROC of white-box models enlarges when moving from less structured models like the GM to more structured ones like the CMD.

\textcolor{black}{All the aforementioned exercises have been carried out by randomly selecting the $10\%$ of links $40$ times to populate $\mathcal{E}^{miss}$ and generate $\mathcal{E}^{obs}=\mathcal{E}\setminus\mathcal{E}^{miss}$} - each statistical indicator has, then, been averaged over such samples: \textcolor{black}{the results we have obtained for a different percentage of randomly selected links, as well as for a different indicator, namely AUPRC, are shown in Appendix~\hyperlink{AppA}{A}.} As it can be appreciated, the performance of the CMD remains high up to a large percentage of removed connections - equivalently, of missing links; still, the performance of the GBDT deteriorates to a lesser extent than the one of white-box models.\\

\begin{table}[t!]
\centering
\begin{tabular}{c|c|c|c|c}
\hline
\hline
year & $N$ & $L$ & $c$ & $\overline{k}$ \\
\hline
\hline 
$1999$ & $215$ & $9770$ & $0.424$ & $90.88$ \\
\hline
$2000$ & $195$ & $8172$ & $0.432$ & $83.81$ \\
\hline
$2001$ & $182$ & $6669$ & $0.401$ & $73.29$ \\
\hline 
$2002$ & $176$ & $5528$ & $0.359$ & $62.82$ \\
\hline 
$2003$ & $176$ & $4892$ & $0.318$ & $55.59$ \\
\hline
$2004$ & $175$ & $4327$ & $0.284$ & $49.45$ \\
\hline
$2005$ & $172$ & $4119$ & $0.280$ & $47.90$ \\
\hline
$2006$ & $172$ & $3910$ & $0.266$ & $45.47$ \\
\hline
$2007$ & $170$ & $3886$ & $0.270$ & $45.72$ \\
\hline
$2008$ & $157$ & $3161$ & $0.258$ & $40.27$ \\
\hline
$2009$ & $137$ & $2381$ & $0.256$ & $34.76$ \\
\hline
$2010$ & $127$ & $2192$ & $0.274$ & $34.52$ \\
\hline
$2011$ & $112$ & $2036$ & $0.328$ & $36.36$ \\
\hline
$2012$ & $97$ & $1421$ & $0.306$ & $29.30$ \\
\hline
\hline
\end{tabular}
\caption{Basic statistics concerning the eMID dataset~\protect\cite{marzi_reproducing_2025}, aggregated at the yearly level. The table shows the total number of banks $N$, the total number of links $L$, the network density $c$ and the average degree $\overline{k}$.}
\label{tab:datasets2}
\end{table}

Let us, now, move to commenting on eMID, aggregated at the yearly level as well. The basic statistics are reported in table~\ref{tab:datasets2}: as it can be appreciated, the number of nodes $N$ ranges between $97$ and $215$ and the total number of links $L$ ranges between $1421$ and $9770$; the link density $c=2L/N(N-1)$ fluctuates appreciably between $0.26$ and $0.43$, as well as the average degree $k=2L/N$, ranging between $\gtrsim29$ and $\lesssim91$. As, in this case, no external feature is available a priori, we have excluded comparisons of the kind.

\begin{figure*}[t!]
\begin{tikzpicture}
\node[anchor=south west,inner sep=0] (fig1) at (0,0) {\includegraphics[trim= 0 0 0 0, clip,width=\textwidth]{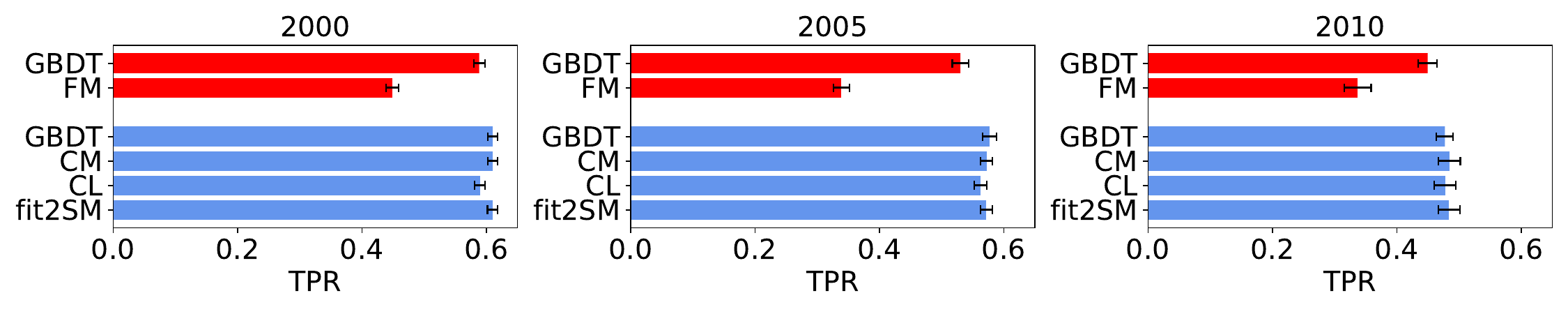}};
\node[anchor=north west, shift={(-5pt,3pt)}] at (fig1.north west) {\textbf{a)}};
\node[anchor=north west,inner sep=0] (fig2) at (fig1.south west) {\includegraphics[trim= 0 0 0 0, clip,width=\textwidth]{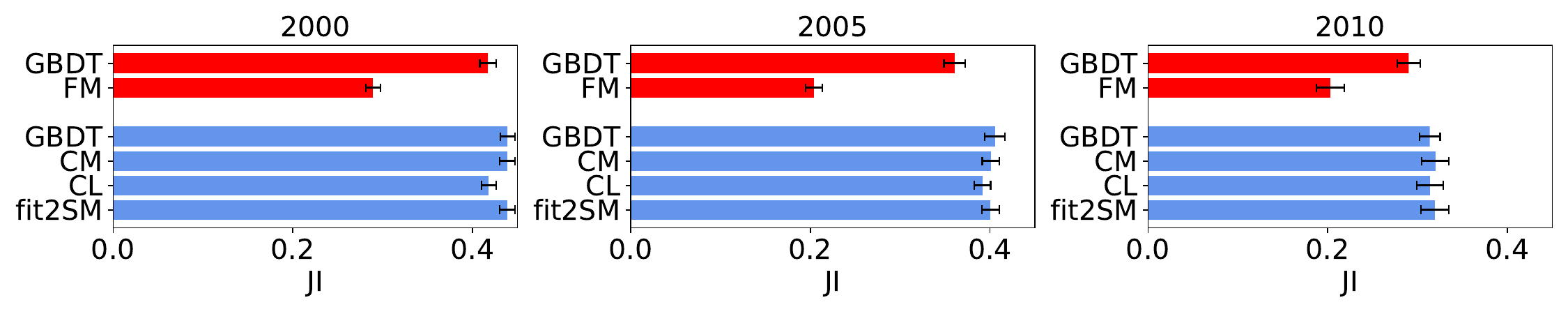}};
\node[anchor=north west, shift={(-5pt,3pt)}] at (fig2.north west) {\textbf{b)}};
\node[anchor=north west,inner sep=0] (fig3) at (fig2.south west){\includegraphics[trim= 0 0 0 0, clip,width=\textwidth]{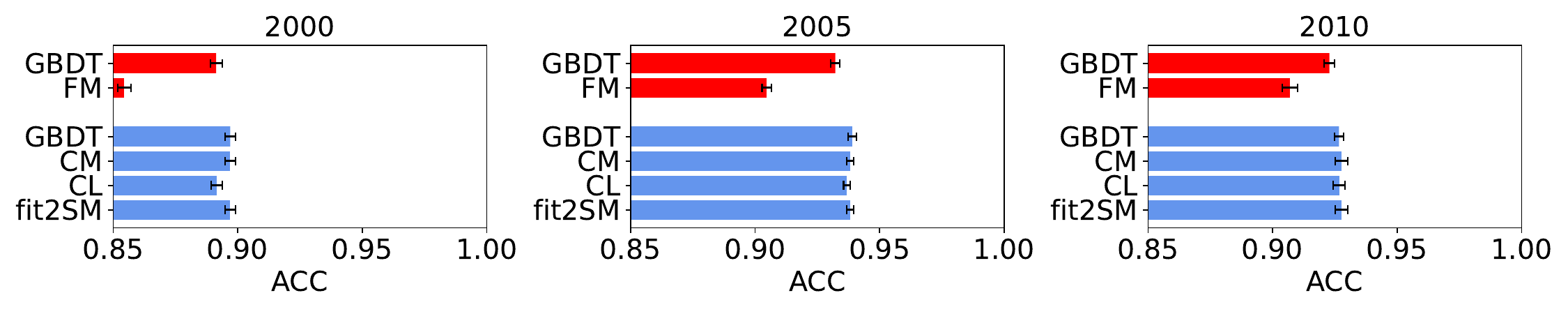}};
\node[anchor=north west, shift={(-5pt,3pt)}] at (fig3.north west) {\textbf{c)}};
\node[anchor=north west,inner sep=0] (fig4) at (fig3.south west){\includegraphics[trim= 0 0 0 0, clip,width=\textwidth]{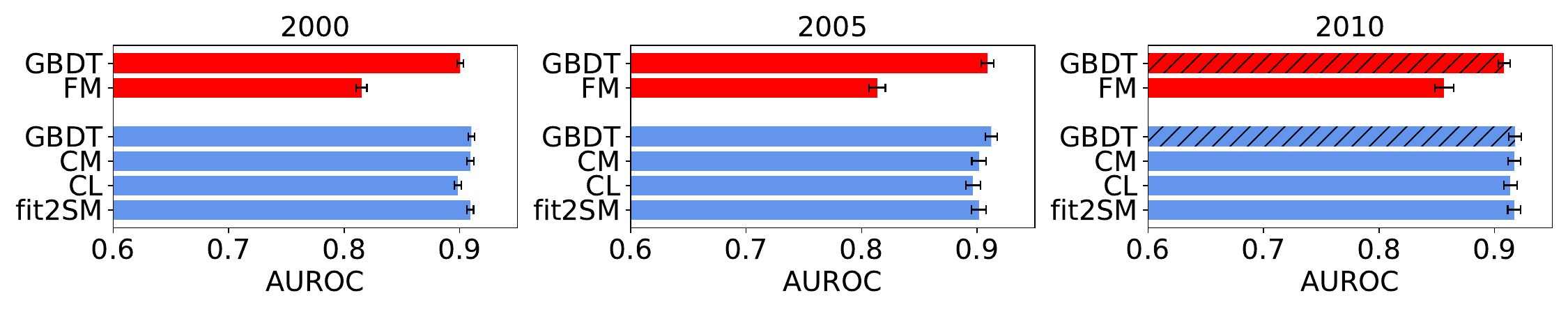}};
\node[anchor=north west, shift={(-5pt,3pt)}] at (fig4.north west) {\textbf{d)}};
\node[anchor=north,inner sep=0] at (fig4.south){{\includegraphics[trim= 40 10 30 10, clip,width=.30\textwidth]{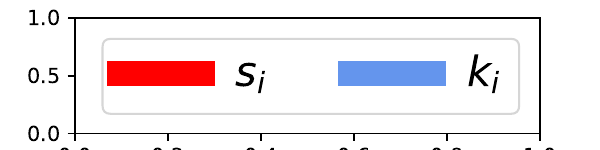}}};
\end{tikzpicture}
\caption{Performance of the models, measured in terms of TPR (row \textbf{a}), JI (row \textbf{b}), ACC (row \textbf{c}) and AUROC (row \textbf{d}), for the years $2000$, $2005$ and $2010$ of eMID. The upper bar in each panel, marked with a diagonal pattern, corresponds to the black-box GBDT. \textcolor{black}{We have randomly selected the $20\%$ of links $30$ times to populate $\mathcal{E}^{miss}$ and generate $\mathcal{E}^{obs}=\mathcal{E}\setminus\mathcal{E}^{miss}$;} each statistical indicator has, then, been averaged over such samples, the standard deviation being represented by a horizontal, black bar. The FM has been implemented by replacing $\omega_i$ with $s_i$ in eq.~\ref{eq:score_fitness} (see the colour legend). As for the WTW, each `endogenous' instance of the GBDT performs in a way that is comparable to that of its purely structural, white-box counterparts - especially for what concerns the CM and the fit2SM; overall, however, the higher sparsity of the system under consideration leads to (much) lower TPR and JI scores than the ones obtained when analysing the WTW.}
\label{fig:EMID_hbars}
\end{figure*}

All results are shown in figure~\ref{fig:EMID_hbars}. Overall, the larger sparsity of the system under consideration leads to TPR and JI scores that are lower than the ones obtained when analysing the WTW; in the case of eMID, in fact, reproducing the $1$s is (much) more difficult than reproducing the $0$s: as a consequence, a (much) worse performance on the number of true positives than on the number of true negatives is expected. As for the WTW, however, each `endogenous' instance of the GBDT performs in a way that is comparable to the one of its purely structural, white-box counterpart, especially for what concerns the CM and the fit2SM - although the sparsity of eMID lets the CL perform as well as the CM and the fit2SM.

\textcolor{black}{All the aforementioned exercises on eMID have been carried out by randomly selecting the $20\%$ of links $30$ times to populate $\mathcal{E}^{miss}$ and generate $\mathcal{E}^{obs}=\mathcal{E}\setminus\mathcal{E}^{miss}$} - each statistical indicator has, then, been averaged over such samples: the results we have obtained for a different aggregation level\textcolor{black}{, as well as for a different indicator, namely AUPRC,} are shown in Appendix~\hyperlink{AppB}{B}. As can be appreciated, very similar results to the ones characterising the yearly aggregation level are obtained, \textcolor{black}{with the only exception of the AUPRC, which appears to be sensitive to the configuration density.}

\section{Discussion}

\textcolor{black}{The aim of the present contribution is that of comparing the performance of several algorithms in carrying out a link prediction exercise, one of the selection criteria concerning data requirements: the algorithms implemented here have been instantiated with the same set of either `endogenous' or `exogenous' features - in other words, have been compared on the same ground.}

In order to do so, we have posed ourselves in the same setting of~\cite{lu_link_2011,parisi_entropybased_2018}, where the variables of interest are computed on the observed portion of the graph (i.e. $\mathbf{A}^{obs}$) and the prediction is carried out on the pairs of nodes in $\mathcal{E}^{no}$, which correspond to the entries $a_{ij}^{obs}=0$, $\forall\:ij\in\mathcal{E}^{no}$, hence appearing as not connected. Within such a framework, the best performance is obtained when the purely structural information is combined with the available, external information: for what concerns the WTW, the latter consists of two sets of properties, i.e. the GDPs of countries and the geographic distances between them; while, however, the FMD does not perform better than the FM and the GM, enriching the CM with the information concerning the geographic distances leads to a better performance than the one characterising the GBDT taking as inputs the same sets of quantities.\\

Such a result leads us to the conclusion that the structural information plays a fundamental role in shaping a network topology: it is, in fact, so relevant to make the CM alone capable of competing with the instance of the GBDT taking as inputs the degrees; in other words, although machine learning algorithms appear as among those performing best, white-box models taking as inputs purely `endogenous' features offer a comparable performance. This is even more relevant when considering that the GBDT needs a very fine-grained kind of information, i.e. the exact position of the connections populating the training set, while the CM takes as input a much coarser one, i.e. (just) the number of connections that are incident to each node (its degree) - see also below.

While the first consideration is in agreement with the results of~\cite{mungo_reconstructing_2023}, the second is not. Such a discrepancy may be due to several reasons: \emph{i)} there, the considered systems are supply chains, represented as binary, directed networks, while, here, the binary, undirected representation of the world trade and of a financial system has been considered. While these are not big differences \emph{per se}, the chosen representation affects the way models are defined, as well as the number and type of `exogenous' features (e.g. industry, location, size) that can be employed - the FM implemented in~\cite{mungo_reconstructing_2023} is undirected in nature, hence particularly unsuited to model directed configurations characterised by a small value of reciprocity such as production networks; \emph{ii)} there, local, topological information such as the one encoded into the degree sequence(s) has been apparently ignored. Here, instead, it represents the backbone of many of the models taking as inputs `endogenous' features; \emph{iii)} there, algorithms are trained on subsets of the original dataset, identified to contain both connected and unconnected pairs of nodes - more specifically, $\mathcal{U}$ is split into training and test set, i.e. $\mathcal{U}=\mathcal{U}^{tr}\cup\mathcal{U}^{te}$, and, then, $\mathcal{U}^{tr}$ is sampled, the collected information consisting of one observation for each pair of firms that has been drawn, i.e. whether the two firms are connected or not, plus a vector of features. Here, instead, all algorithms are trained on the same (incomplete) portion of the dataset of interest, to let all of them start on equal footing. Still, we have carried out an additional set of comparisons, more in line with the approach pursued in~\cite{mungo_reconstructing_2023}; the results are shown in Appendix~\hyperlink{AppC}{C}: as evident upon inspecting the figures there, very similar considerations to the ones above hold true.\\

\textcolor{black}{As we have already noticed, the usual sampling scheme `favours' well connected nodes. Appendix~\hyperlink{AppD}{D} has been, now, dedicated to compare it with a couple of alternative ones: the first one prescribes to remove links by selecting edges with a probability that is inversely proportional to the product of the degrees of the involved vertices; the second one is the so-called \texttt{Random Node-Edge Sampler}~\cite{krishnamurthy_reducing_2005}, first sampling nodes uniformly at random and, then, sampling edges uniformly at random from the ones that are incident to the sampled nodes - although both alternatives aim at contrasting the aforementioned bias, giving the less connected nodes a higher chance to be selected, this is achieved only to a partial extent.}

\textcolor{black}{Overall, it should be noticed that the performance of all methods degrades as `targeted' link removal recipes are implemented, whence our conclusion that existing link prediction algorithms (be they black-box or white-box) perform better for better connected nodes. Although negative, such a result offers an additional indication about the identity of the `best' explanatory variables in a link-prediction exercise: as also noticed elsewhere~\cite{cimini_systemic_2015,squartini_reconstruction_2018,cimini_reconstructing_2021} economic, financial (and social) systems are constituted by nodes the `activity' of which is proportional to (if not driven by) their connectedness; as a consequence, controlling for the degrees implies being capable of explaining (most of) each node's `activity' - as well as most of each node's structural properties.}\\

\textcolor{black}{Beside incurring into the additional cost required by tuning hyperparameters (see Appendix~\hyperlink{AppE}{E}), machine learning algorithms need to be fed with (much) more information than that required by white-box algorithms} - what is required to make predictions on the `test' portion of the dataset is, in fact, the exact knowledge of the `training' portion of the dataset - in order to achieve a performance that is not necessarily better. While such a need may even represent an advantage when considering a specific dataset, it turns out to constitute a limitation when coming to generalise, as machine learning approaches either perform quite poorly (e.g. whenever trained on a dataset to predict the missing links of another) or cannot perform at all (e.g. whenever in presence of missing information). White-box models, on the contrary, perform robustly: examples are provided by the network reconstruction tasks carried out in~\cite{cimini_systemic_2015,ialongo_reconstructing_2022}, where the considered systems have been fully reconstructed by (solely) employing the strengths of nodes and (an estimation of) the link density.\\

\textcolor{black}{Finally, a statistical analysis about the differences characterising the performances of our algorithms (see Appendix~\hyperlink{AppF}{F}) reveals that a clear winner cannot be always unambiguously identified, as the presence of a winner depends on both score and the algorithm. When considering the AUPRC, for instance, \emph{i)} machine learning approaches perform better than the white-box models not taking structural information as input; \emph{ii)} the former and the latter perform in a statistically undistinguishable way when `mixed' structural information is taken as input; \emph{iii)} white-box models (typically) perform better than machine learning approaches when either the plain degree sequence or non-linear structural information is taken as input.}\\

\textcolor{black}{Let us conclude by stressing once again that our motivation, here, is not that of challenging the performance of machine learning approaches, rather checking under which conditions they can be said to outperform white-box models in link prediction tasks: what emerges from our analysis is that answering such a question crucially depends on the way these exercises are carried out. Still, the results shown in Appendices~\hyperlink{AppE}{E} and~\hyperlink{AppF}{F} motivate us to indicate maximum-entropy models as strong competitors of machine learning approaches, often performing in a comparable way at a substantially lower computational cost.}

\section{Acknowledgments}

This work has been supported by the following projects: `RE-Net - Reconstructing economic networks: from physics to machine learning and back' - 2022MTBB22, funded by the European Union Next Generation EU, PNRR Mission 4 Component 2 Investment 1.1, CUP: D53D23002330006; `C2T - From Crises to Theory: towards a science of resilience and recovery for economic and financial systems' - P2022E93B8, funded by the European Union Next Generation EU, PNRR Mission 4 Component 2 Investment 1.1, CUP: D53D23019330001; `SoBigData RI PPP - SoBigData RI Preparatory Phase Project', funded by the European Union under the scheme HORIZON-INFRA-2021-DEV-02-01, preparatory phase of new ESFRI research infrastructure projects, G.A. 101079043; `FAIR - Future Artificial Intelligence Research' - Spoke 1 `Human-centered AI', funded by the European Commission under the Next Generation EU program, PNRR Mission 4 Component 2 Investment 3.1, G.A. PE00000013.

The authors thank Sadamori Kojaku and Mattia Marzi for insightful discussions.

\section{Author contributions}

Study conception and design: FS, GC, TS. Data collection: FS. Analysis and interpretation of results: FS, GC, TS. Draft manuscript preparation: FS, GC, TS. Draft manuscript revision: FS, GC, TS.

\section{Competing interests}

The authors declare no competing interests.

\bibliographystyle{apsrev4-2}

\bibliography{references}

\clearpage

\onecolumngrid

\appendix

\hypertarget{AppA}{}
\section*{APPENDIX A.\\MISSING LINKS PREDICTION ON THE WORLD TRADE WEB}\label{AppA}

\setcounter{section}{0}
\renewcommand{\thefigure}{A\arabic{figure}}
\setcounter{figure}{0}
\renewcommand{\thetable}{A\arabic{table}}
\setcounter{table}{0}

\begin{figure}[t!]
\begin{tikzpicture}
\node[anchor=south west,inner sep=0] (fig1) at (0,0) {\includegraphics[trim= 0 132 0 0, clip,width=\linewidth]{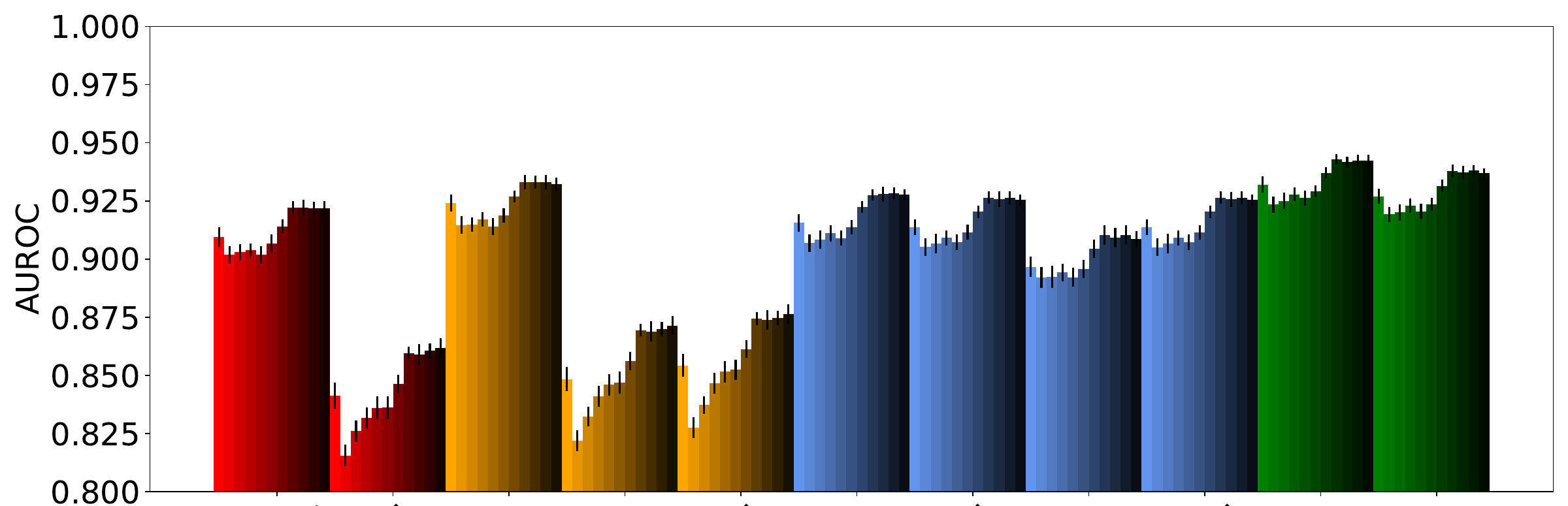}};
\node[anchor=north west, shift={(-5pt,3pt)}] at (fig1.north west) {\textbf{a)}};
\node[anchor=north,inner sep=0] (fig2) at (fig1.south) {\includegraphics[trim= 0 132 0 0, clip,width=\linewidth]{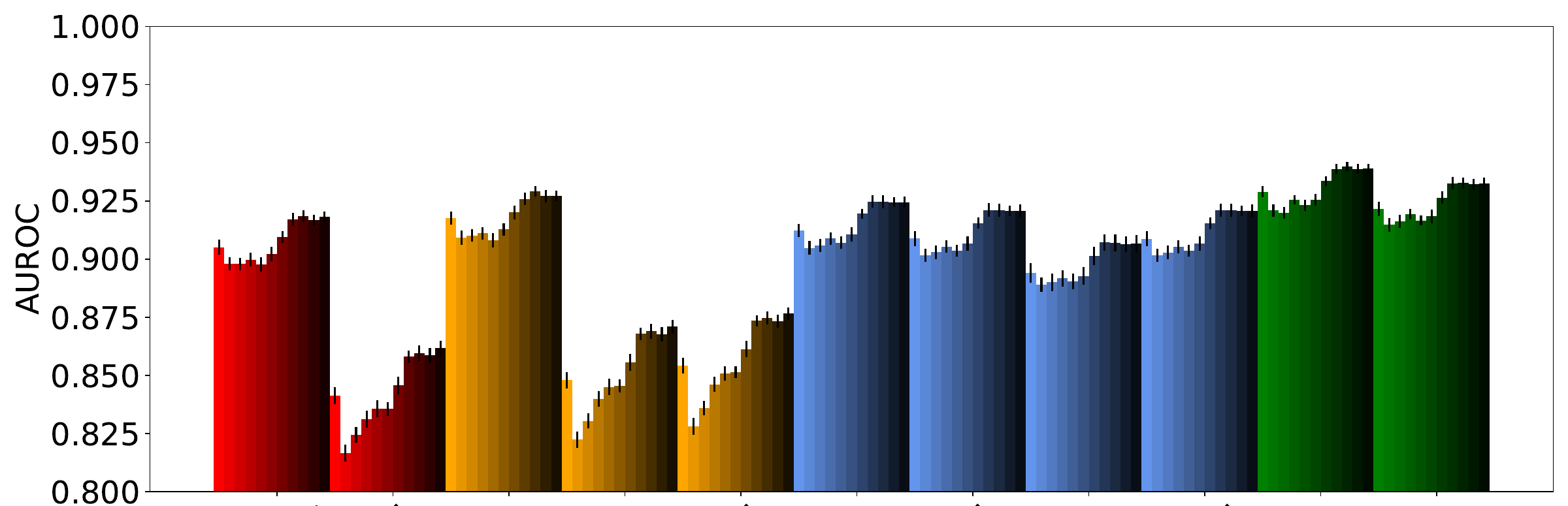}};
\node[anchor=north west, shift={(-5pt,3pt)}] at (fig2.north west) {\textbf{b)}};
\end{tikzpicture}
\caption{\textcolor{black}{Performance of the models, measured in terms of AUROC, for all years of the WTW (darker shades correspond to more recent years). \textcolor{black}{We have randomly selected the $20\%$ (panel \textbf{a}) and the $30\%$ (panel \textbf{b}) of links $40$ times to populate the test set, on which the prediction is performed after having trained the model on the remaining pairs of nodes.} Each statistical indicator has, then, been averaged over such samples, the standard deviation being represented by a vertical, black bar. The results are in agreement with those reported in the main text, as each `endogenous' instance of the GBDT performs in a way that is comparable to that of its purely structural, white-box counterparts - especially for what concerns the CM and the fit2SM.}}
\label{fig:AUROC_shares}
\end{figure}

Figure~\ref{fig:AUROC_shares} depicts the performance of our models for different shares of removed links: as it can be appreciated, the performance of the CMD remains high up to a large percentage of missing links; \textcolor{black}{to be noticed that the performance of the GBDT taking as inputs both `exogenous' and `endogenous' features deteriorates to a lesser extent than the one of the CMD, i.e. its white-box counterpart, as the size of the training set decreases.}

\begin{figure}[t!]
\begin{tikzpicture}
\node[anchor=south west,inner sep=0] (fig1) at (0,0) {\includegraphics[trim= 0 0 0 0, clip,width=\linewidth]{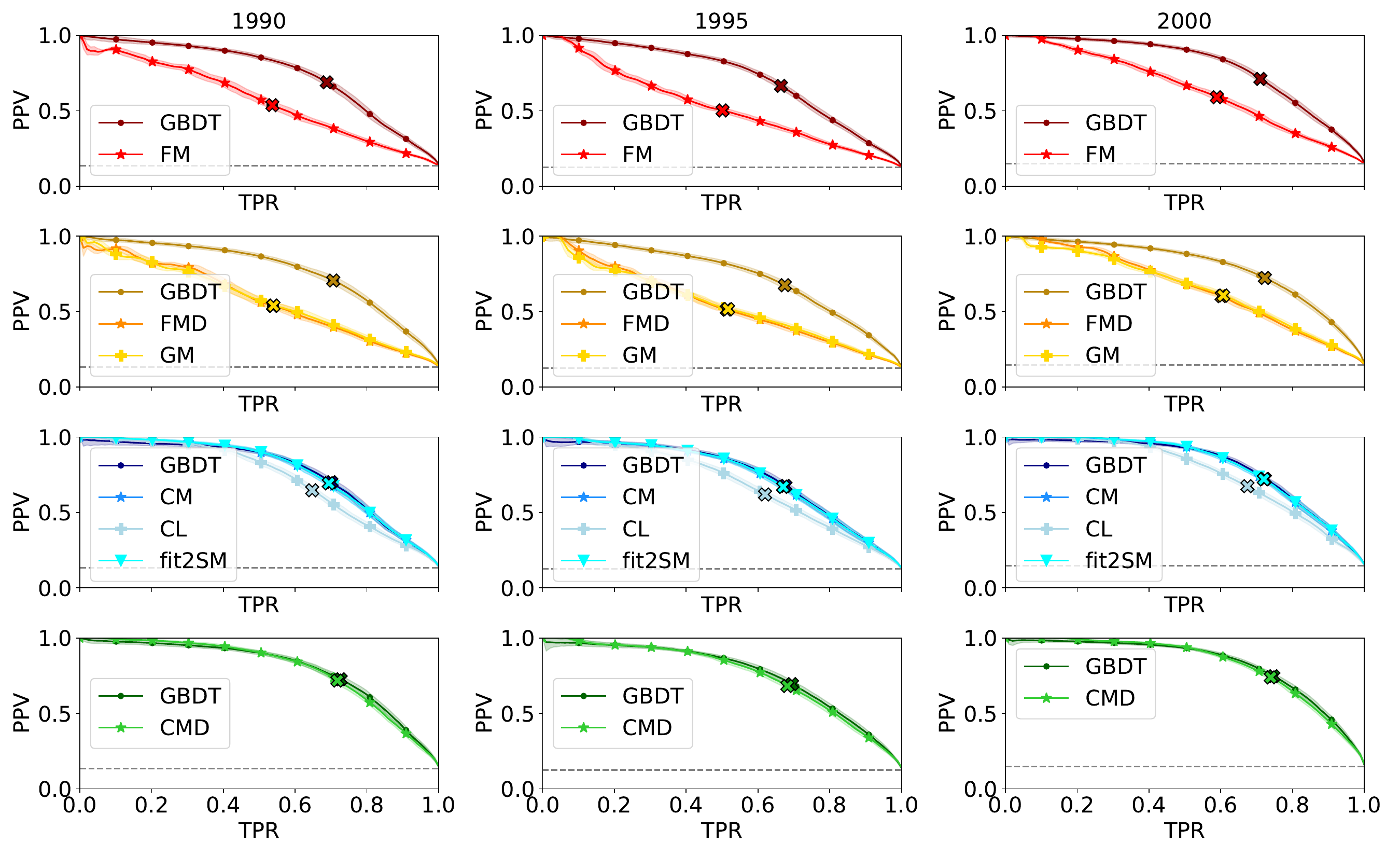}};
\node[anchor=north west, shift={(-5pt,3pt)}] at (fig1.north west) {\textbf{a)}};
\node[anchor=north,inner sep=0] (fig2) at (fig1.south) {\includegraphics[trim= 0 0 0 0, clip,width=\linewidth]{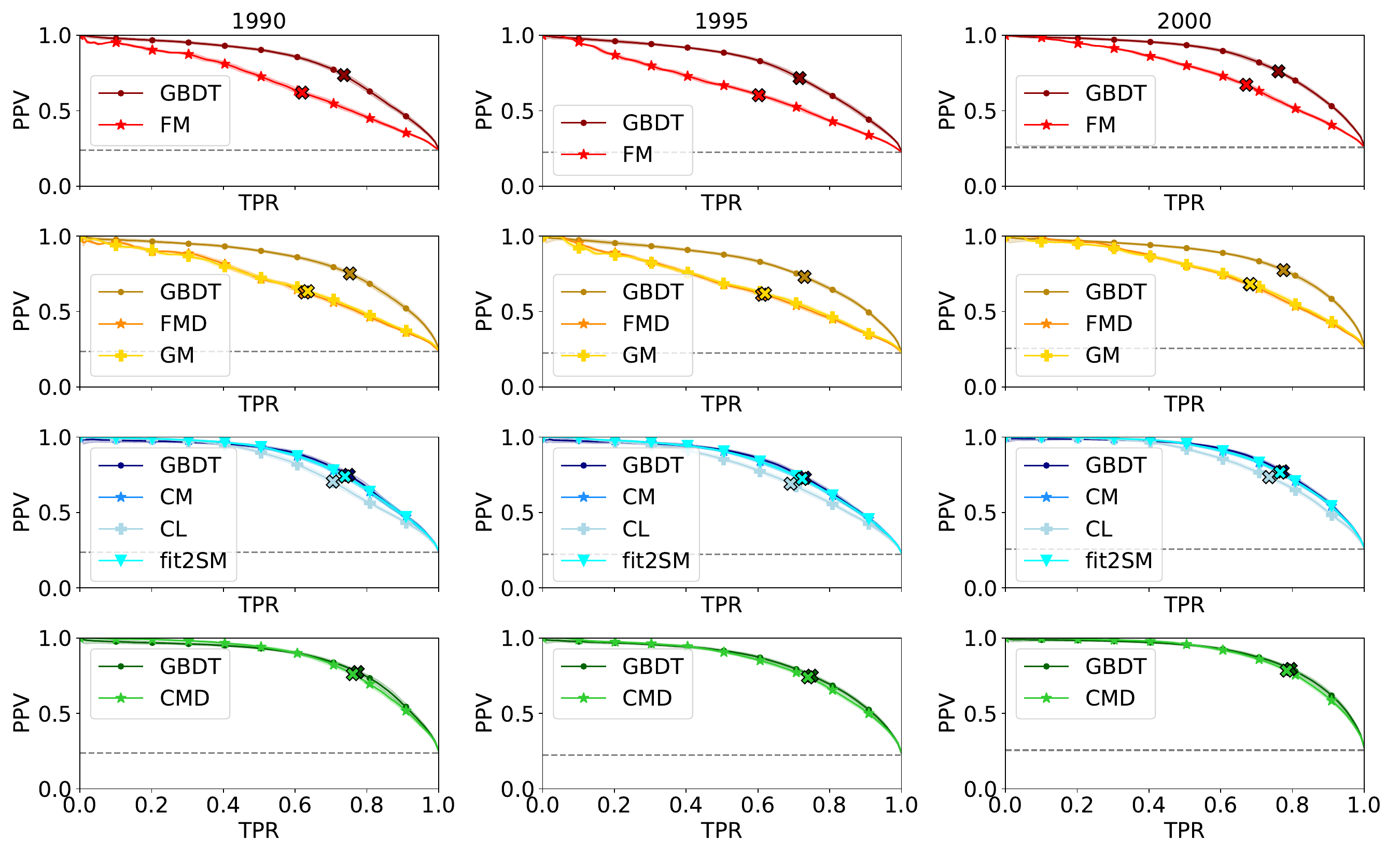}};
\node[anchor=north west, shift={(-5pt,3pt)}] at (fig2.north west) {\textbf{b)}};
\end{tikzpicture}
\caption{\textcolor{black}{Performance of the models, measured in terms of PRC curves, for the years $1990$, $1995$ and $2000$ of the WTW. Each panel collects methods that rely on the same set of features. \textcolor{black}{We have randomly selected the $10\%$ (panel \textbf{a}) and the $20\%$ (panel \textbf{b}) of links $40$ times to populate the test set (the curves relative to each indicator are, in fact, $40$ partially overlapping curves corresponding to each realization), on which the prediction is performed after having trained the model on the remaining pairs of nodes.} The `X' markers indicate the PPV and TPR values obtained by selecting a number of missing links equal to $|\mathcal{E}^{miss}|$. The AUPRC enlarges when moving from less structured models, like the GM, to more structured ones, like the CMD.}}
\label{fig:AUPR}
\end{figure}

\begin{figure}[t!]
\begin{tikzpicture}
\node[anchor=south west,inner sep=0] (fig1) at (0,0) {\includegraphics[trim= 0 132 0 0, clip,width=\linewidth]{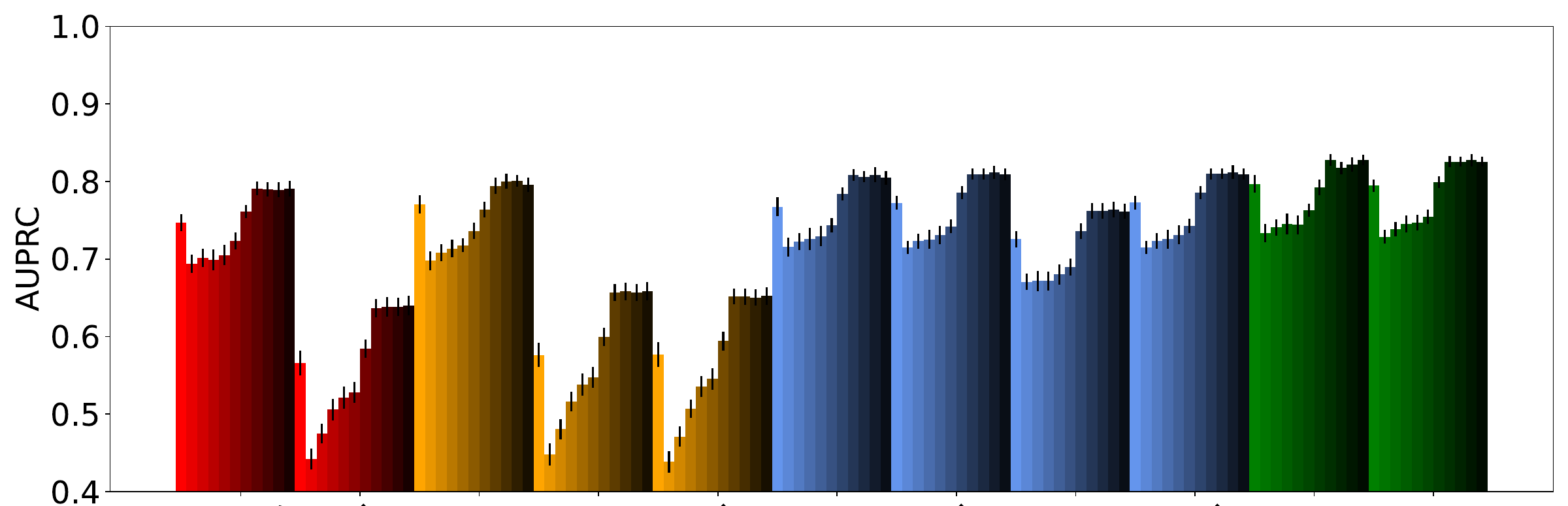}};
\node[anchor=north west, shift={(-5pt,3pt)}] at (fig1.north west) {\textbf{a)}};
\node[anchor=north,inner sep=0] (fig2) at (fig1.south) {\includegraphics[trim= 0 0 0 0, clip,width=\linewidth]{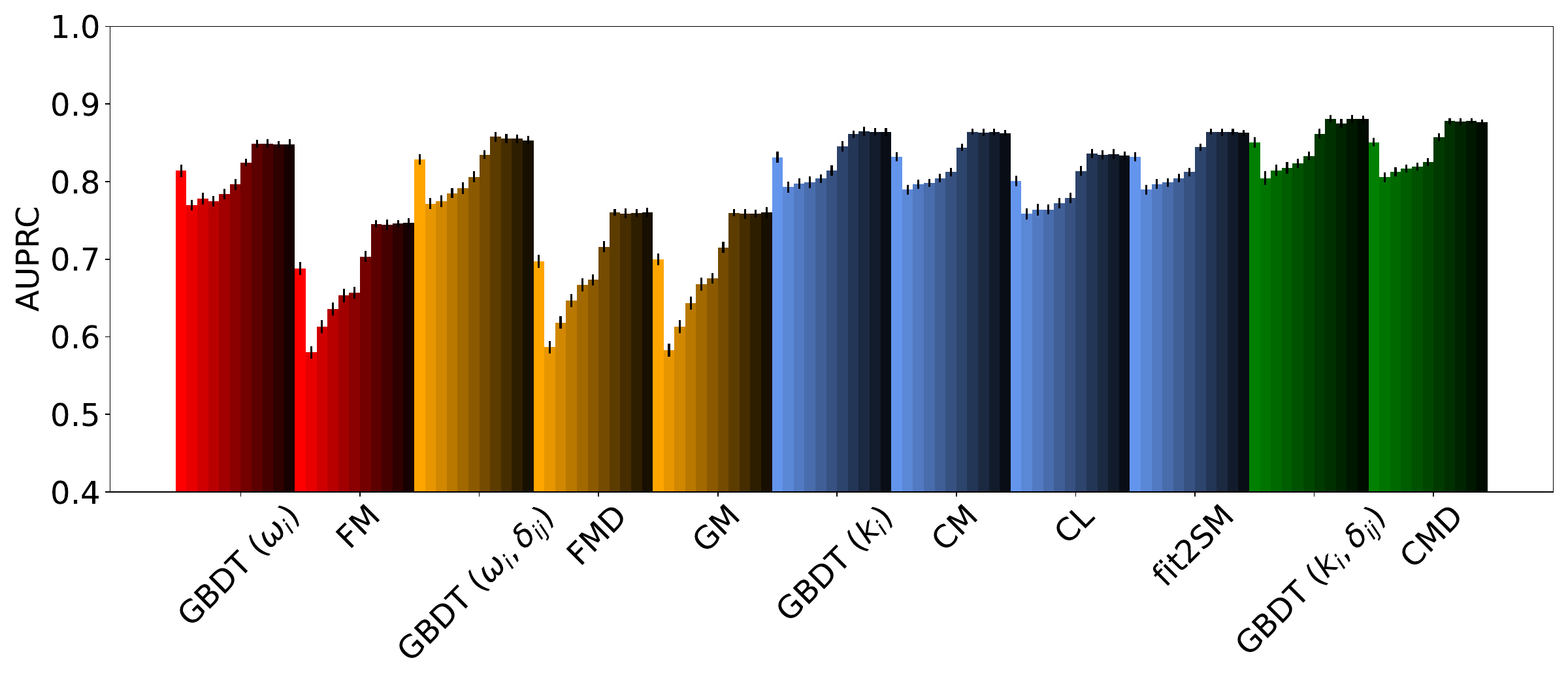}};
\node[anchor=north west, shift={(-5pt,3pt)}] at (fig2.north west) {\textbf{b)}};
\end{tikzpicture}
\caption{\textcolor{black}{Performance of the models, measured in terms of AUPRC, for all years of the WTW (darker shades correspond to more recent years). \textcolor{black}{We have randomly selected the $10\%$ (panel \textbf{a}) and the $20\%$ (panel \textbf{b}) of links $40$ times to populate the test set, on which the prediction is performed after having trained the model on the remaining pairs of nodes.} Each statistical indicator has, then, been averaged over such samples, the standard deviation being represented by a vertical, black bar. The results are in agreement with those reported in the main text, as each `endogenous' instance of the GBDT performs in a way that is comparable to that of its purely structural, white-box counterparts - especially for what concerns the CM and the fit2SM.}}
\label{fig:AUPR2}
\end{figure}

\textcolor{black}{The results concerning the AUPRC, shown in figures~\ref{fig:AUPR} and~\ref{fig:AUPR2}, are in agreement with those reported in the main text as well, as each `endogenous' instance of the GBDT performs in a way that is comparable to that of its purely structural, white-box counterpart - especially for what concerns the CM and the fit2SM.}

\clearpage

\clearpage

\clearpage

\hypertarget{AppB}{}
\section*{APPENDIX B.\\MISSING LINKS PREDICTION ON THE ELECTRONIC MARKET FOR INTERBANK DEPOSITS}\label{AppB}

\setcounter{section}{0}
\renewcommand{\thefigure}{B\arabic{figure}}
\setcounter{figure}{0}
\renewcommand{\thetable}{B\arabic{table}}
\setcounter{table}{0}

\begin{table}[t!]
\centering
\begin{tabular}{l|c|c|c|c}
\hline
\hline
quarter & $N$ & $L$ & $c$ & $\overline{k}$ \\
\hline
\hline 
$2000$ - $Q01$ & $181$ & $5483$ & $0.337$ & $60.59$ \\
\hline
$2000$ - $Q03$ & $178$ & $5205$ & $0.330$ & $58.48$ \\
\hline
$2010$ - $Q01$ & $113$ & $1433$ & $0.226$ & $25.36$ \\
\hline 
$2010$ - $Q03$ & $107$ & $1476$ & $0.260$ & $27.59$ \\
\hline
\hline
month & $N$ & $L$ & $c$ & $\overline{k}$ \\
\hline
\hline 
$2000$ - $M01$ & $177$ & $3634$ & $0.230$ & $41.06$ \\
\hline
$2000$ - $M08$ & $175$ & $3403$ & $0.224$ & $38.89$ \\
\hline
$2010$ - $M01$ & $102$ & $1014$ & $0.197$ & $19.88$ \\
\hline 
$2010$ - $M08$ & $97$ & $920$ & $0.198$ & $18.97$   \\
\hline
\hline
week & $N$ & $L$ & $c$ & $\overline{k}$ \\
\hline
\hline 
$2000$ - $W01$ & $170$ & $1428$ & $0.099$ & $16.80$ \\
\hline
$2000$ - $W30$ & $167$ & $1493$ & $0.108$ & $17.88$ \\
\hline
$2010$ - $W01$ & $82$ & $346$ & $0.104$ & $8.44$    \\
\hline 
$2010$ - $W30$ & $90$ & $423$ & $0.106$ & $9.40$    \\
\hline
\hline
day & $N$ & $L$ & $c$ & $\overline{k}$ \\
\hline
\hline 
$2000$ - $01$ - $03$ & $149$ & $523$ & $0.047$ & $7.02$ \\
\hline
$2005$ - $04$ - $15$ & $108$ & $320$ & $0.055$ & $5.93$ \\
\hline
$2010$ - $08$ - $30$ & $72$  & $166$ & $0.065$ & $4.61$ \\
\hline 
\hline 
\end{tabular}
\caption{Basic statistics concerning the eMID dataset~\protect\cite{marzi_reproducing_2025}, aggregated at the quarterly, monthly, weekly and daily level. The table shows the total number of banks $N$, the total number of links $L$, the network density $c$ and the average degree $\overline{k}$ for each aggregation scale.}
\label{tab:eMIDstats_daily}
\end{table}

\begin{figure*}[t!]
\begin{tikzpicture}
\node[anchor=south west,inner sep=0] (fig1) at (0,0) {\includegraphics[trim= 0 0 0 0, clip,width=\textwidth]{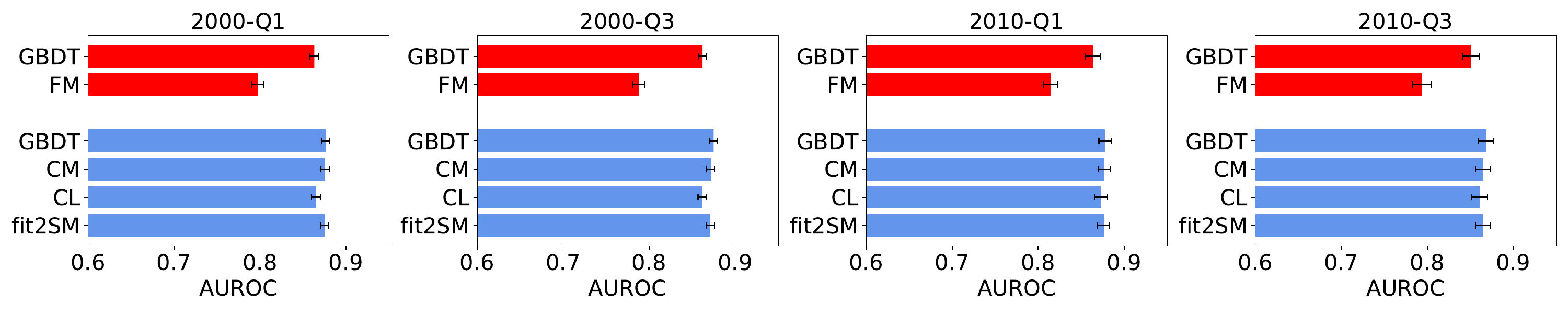}};
\node[anchor=north west, shift={(-5pt,3pt)}] at (fig1.north west) {\textbf{a)}};
\node[anchor=north west,inner sep=0] (fig2) at (fig1.south west) {\includegraphics[trim= 0 0 0 0, clip,width=\textwidth]{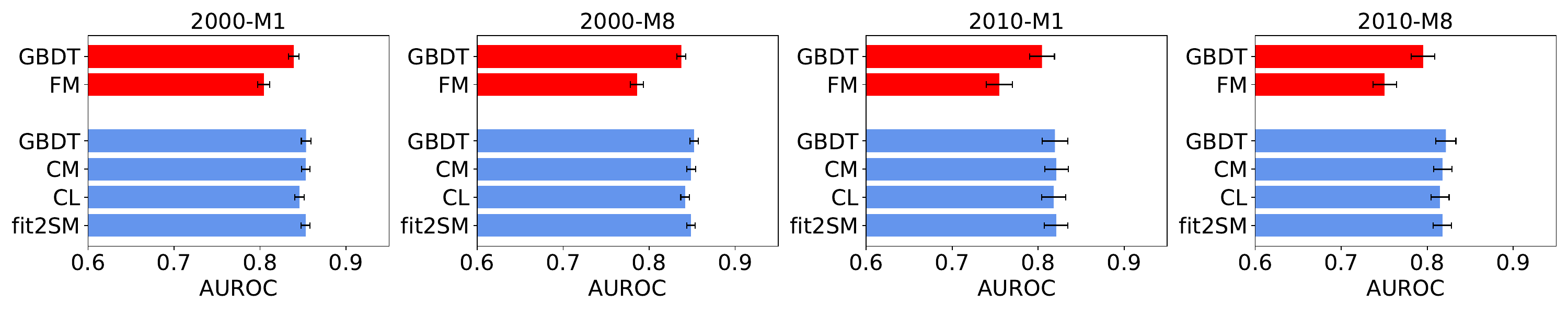}};
\node[anchor=north west, shift={(-5pt,3pt)}] at (fig2.north west) {\textbf{b)}};
\node[anchor=north west,inner sep=0] (fig3) at (fig2.south west) {\includegraphics[trim= 0 0 0 0, clip,width=\textwidth]{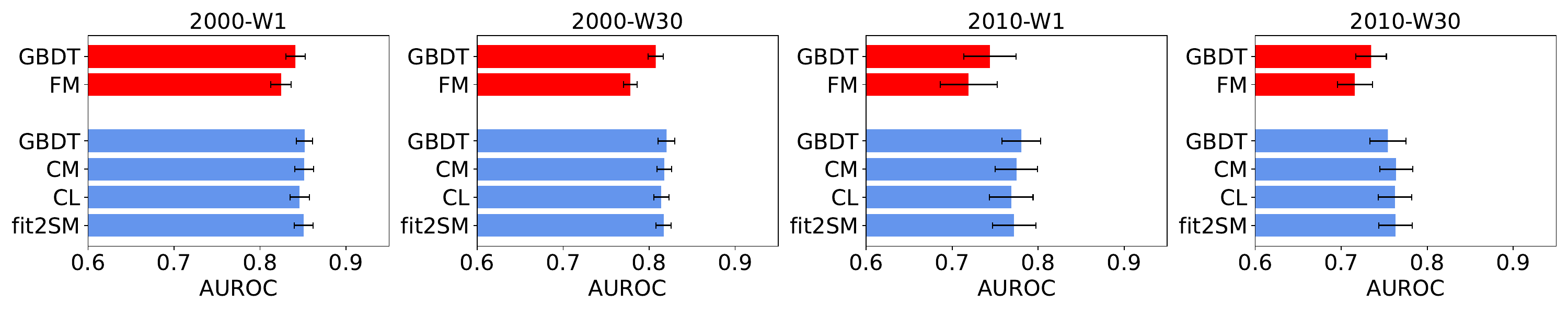}};
\node[anchor=north west, shift={(-5pt,3pt)}] at (fig3.north west) {\textbf{c)}};
\node[anchor=north,inner sep=0] (fig4) at (fig3.south) {\includegraphics[trim= 0 0 0 0, clip,width=.75\textwidth]{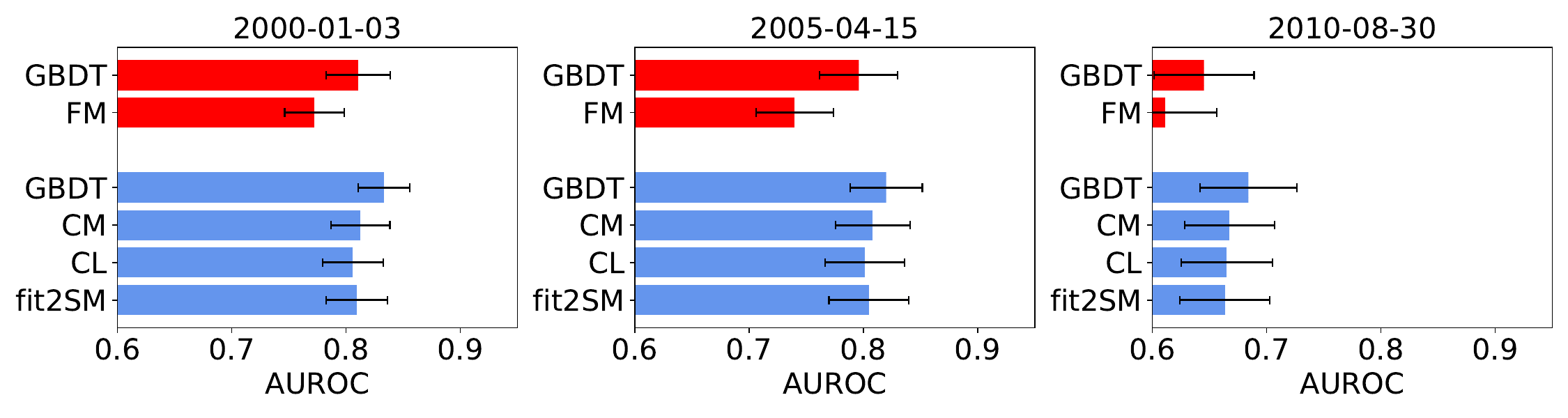}};
\node[anchor=north west, shift={(-5pt,3pt)}] at (fig4.north west) {\textbf{d)}};
\node[anchor=north,inner sep=0] at (fig4.south) {{\includegraphics[trim= 40 10 30 10, clip,width=.25\textwidth]{allscores_legend3.pdf}}};
\end{tikzpicture}
\caption{Performance of the models, measured in terms of AUROC, for snapshots of eMID at different aggregation levels: quarterly (panel \textbf{a}), monthly (panel \textbf{b}), weekly (panel \textbf{c}) and daily (panel \textbf{d}). The upper bar in each panel, marked with a diagonal pattern, corresponds to the black-box GBDT. \textcolor{black}{We have randomly selected the $20\%$ of links $30$ times to populate the test set.} Each statistical indicator has, then, been averaged over such samples, the standard deviation being represented by a horizontal, black bar. The FM has been implemented by replacing $\omega_i$ with $s_i$ in eq.~\ref{eq:score_fitness} (see the colour legend). As for the WTW, each `endogenous' instance of the GBDT performs in a way that is comparable to that of its purely structural, white-box counterpart; the results are overall consistent across aggregation levels.}
\label{fig:EMID_aggregations}
\end{figure*}

\begin{figure*}[t!]
\begin{tikzpicture}
\node[anchor=south west,inner sep=0] (fig0) at (0,0) {\includegraphics[trim= 0 0 0 0, clip,width=.75\textwidth]{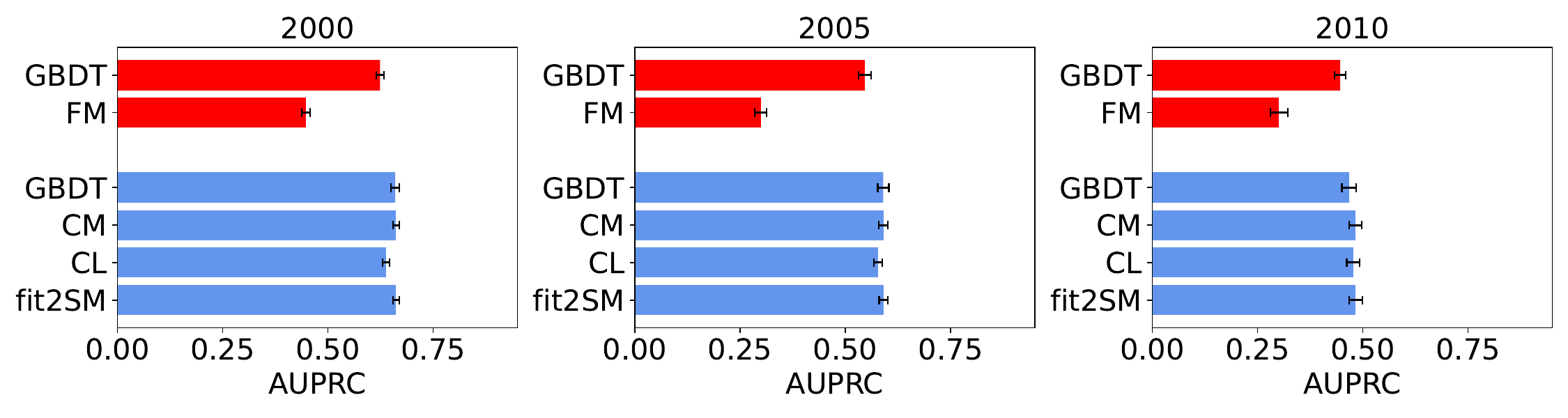}};
\node[anchor=north west, shift={(-5pt,3pt)}] at (fig0.north west) {\textbf{a)}};
\node[anchor=north west,inner sep=0] (fig1) at (fig0.south west) {\includegraphics[trim= 0 0 0 0, clip,width=\textwidth]{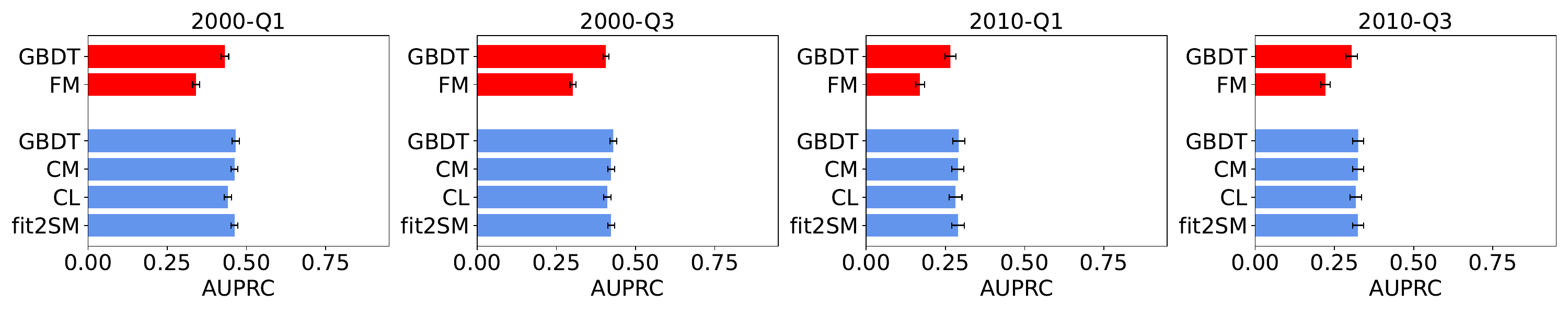}};
\node[anchor=north west, shift={(-5pt,3pt)}] at (fig1.north west) {\textbf{b)}};
\node[anchor=north west,inner sep=0] (fig2) at (fig1.south west) {\includegraphics[trim= 0 0 0 0, clip,width=\textwidth]{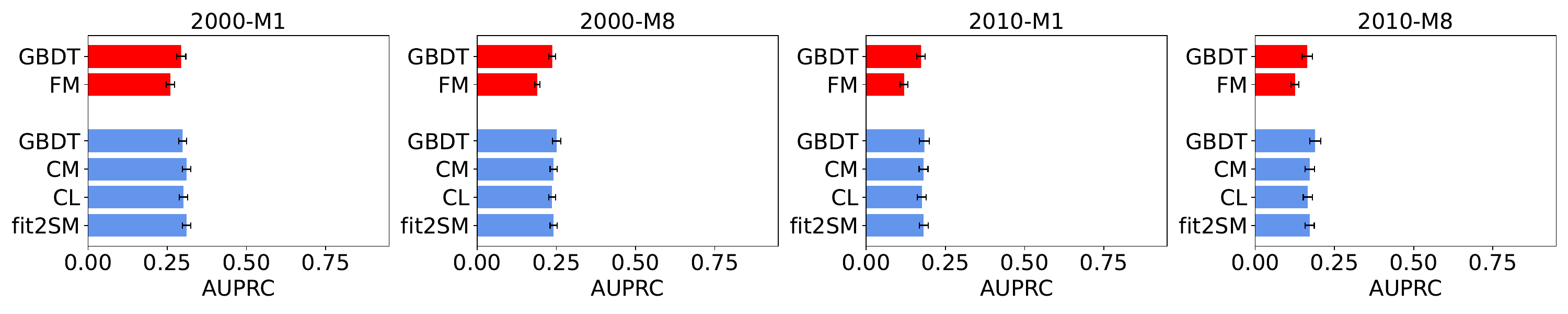}};
\node[anchor=north west, shift={(-5pt,3pt)}] at (fig2.north west) {\textbf{c)}};
\node[anchor=north west,inner sep=0] (fig3) at (fig2.south west) {\includegraphics[trim= 0 0 0 0, clip,width=\textwidth]{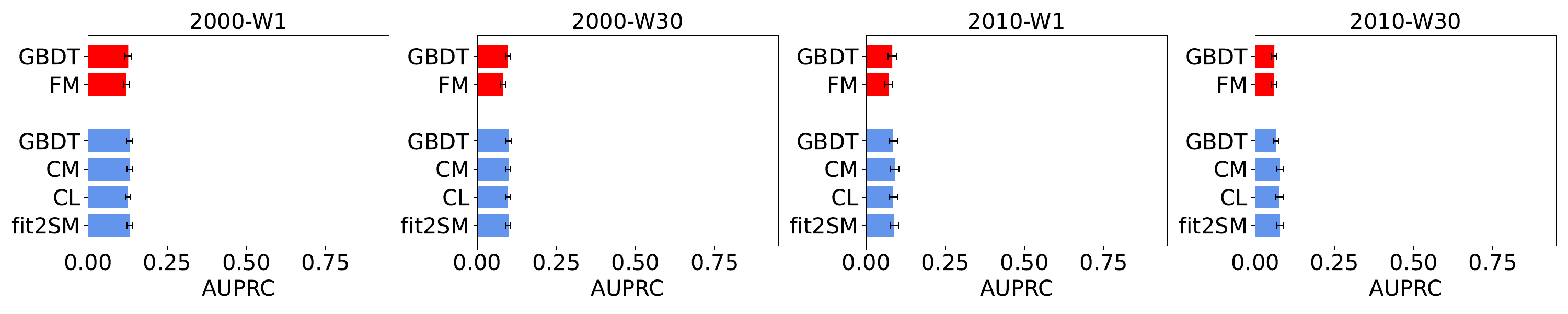}};
\node[anchor=north west, shift={(-5pt,3pt)}] at (fig3.north west) {\textbf{d)}};
\node[anchor=north,inner sep=0] (fig4) at (fig3.south) {\includegraphics[trim= 0 0 0 0, clip,width=.75\textwidth]{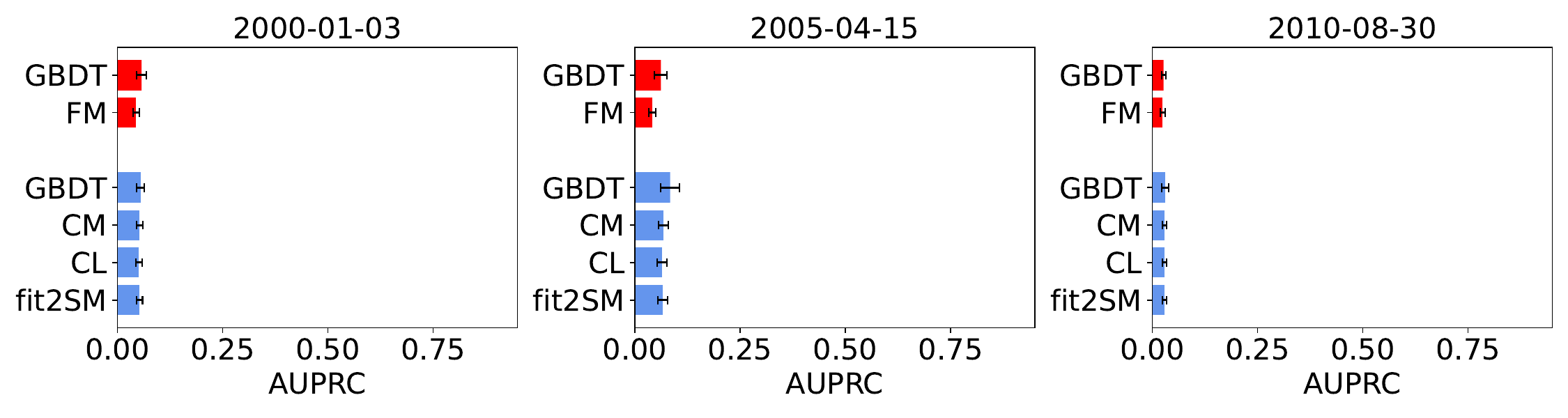}};
\node[anchor=north west, shift={(-5pt,3pt)}] at (fig4.north west) {\textbf{e)}};
\node[anchor=west,inner sep=0,shift={(1.5,0)}] at (fig0.east) {{\includegraphics[trim= 25 50 20 20, clip,width=.10\textwidth]{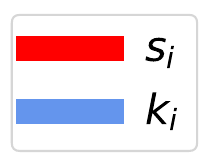}}};
\end{tikzpicture}
\caption{\textcolor{black}{Performance of the models, measured in terms of AUPRC, for snapshots of eMID at different aggregation levels: yearly (panel \textbf{a}), quarterly (panel \textbf{b}), monthly (panel \textbf{c}), weekly (panel \textbf{d}) and daily (panel \textbf{e}). The upper bar in each panel, marked with a diagonal pattern, corresponds to the black-box GBDT. \textcolor{black}{We have randomly selected the $20\%$ of links $30$ times to populate the test set.} Each statistical indicator has, then, been averaged over such samples, the standard deviation being represented by a horizontal, black bar. The FM has been implemented by replacing $\omega_i$ with $s_i$ in eq.~\ref{eq:score_fitness} (see the colour legend). As for the WTW, each `endogenous' instance of the GBDT performs in a way that is comparable to that of its purely structural, white-box counterpart; the results change as finer aggregation levels are considered.}}
\label{fig:EMID_AUPRC}
\end{figure*}

{\color{black}
Table~\ref{tab:eMIDstats_daily} reports the basic statistics for each aggregation scale; figures~\ref{fig:EMID_aggregations} and~\ref{fig:EMID_AUPRC} depict the performance of our models for different aggregation scales. As it can be appreciated, while \emph{i)} the results concerning the AUROC are consistent with the ones characterising the yearly aggregation level, \emph{ii)} the results concerning the AUPRC change as finer aggregation levels are considered: more specifically, the AUPRC shrinks (much) faster than the AUROC as the aggregation level becomes finer; such a result indicates that, as the list of potentially missing links is gone through, \emph{i)} the overall number of truly missing links is (practically, always) recovered quite accurately, although \emph{ii)} the number of false positives rises - in other words, the performance of all algorithms in spotting single connections degrades quite rapidly as we move from yearly to daily snapshots (equivalently, towards sparser configurations).

Upon comparing these results with those in Appendix~\hyperlink{AppA}{A}, we are led to conclude that existing link prediction algorithms (be they black-box or white-box) perform better for denser configurations.}

\clearpage

\label{sec:AppC}
\section*{APPENDIX C.\\MISSING LINKS PREDICTION IN A DIFFERENT SETTING}

\setcounter{section}{0}
\renewcommand{\thefigure}{C\arabic{figure}}
\setcounter{figure}{0}
\renewcommand{\thetable}{C\arabic{table}}
\setcounter{table}{0}

\begin{figure}[t!]
\begin{tikzpicture}
\node[anchor=south west,inner sep=0] (fig1) at (0,0) {\includegraphics[trim= 0 132 0 0, clip,width=\linewidth]{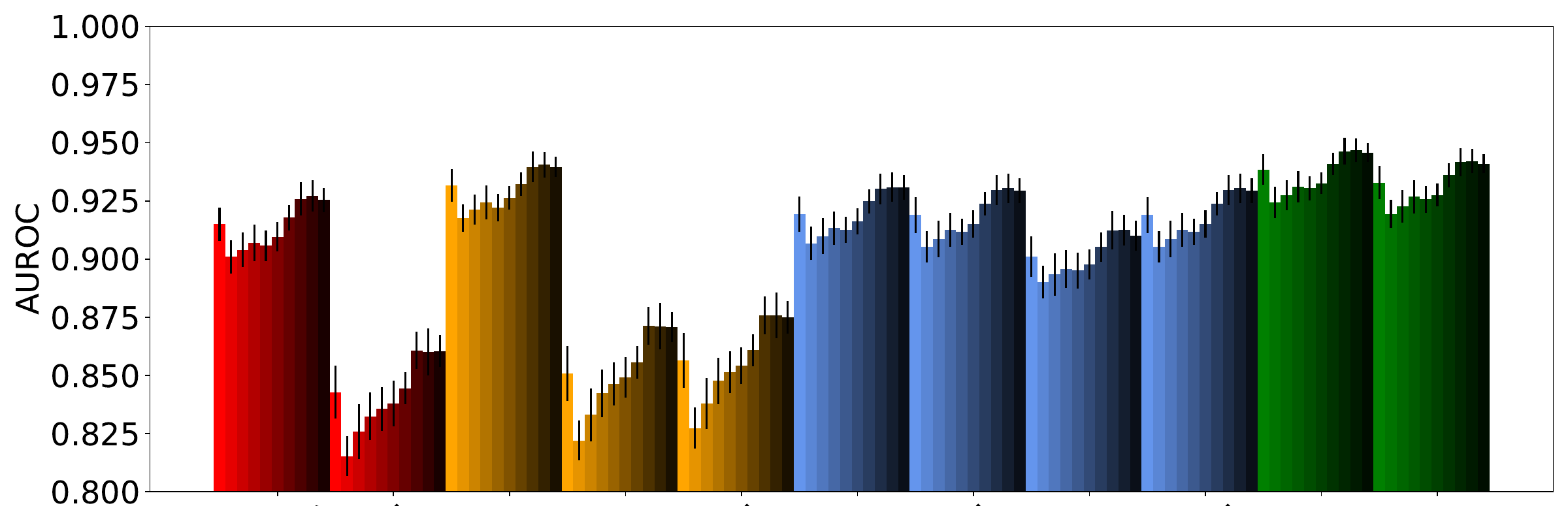}};
\node[anchor=north west, shift={(-5pt,3pt)}] at (fig1.north west) {\textbf{a)}};
\node[anchor=north,inner sep=0] (fig2) at (fig1.south) {\includegraphics[trim= 0 132 0 0, clip,width=\linewidth]{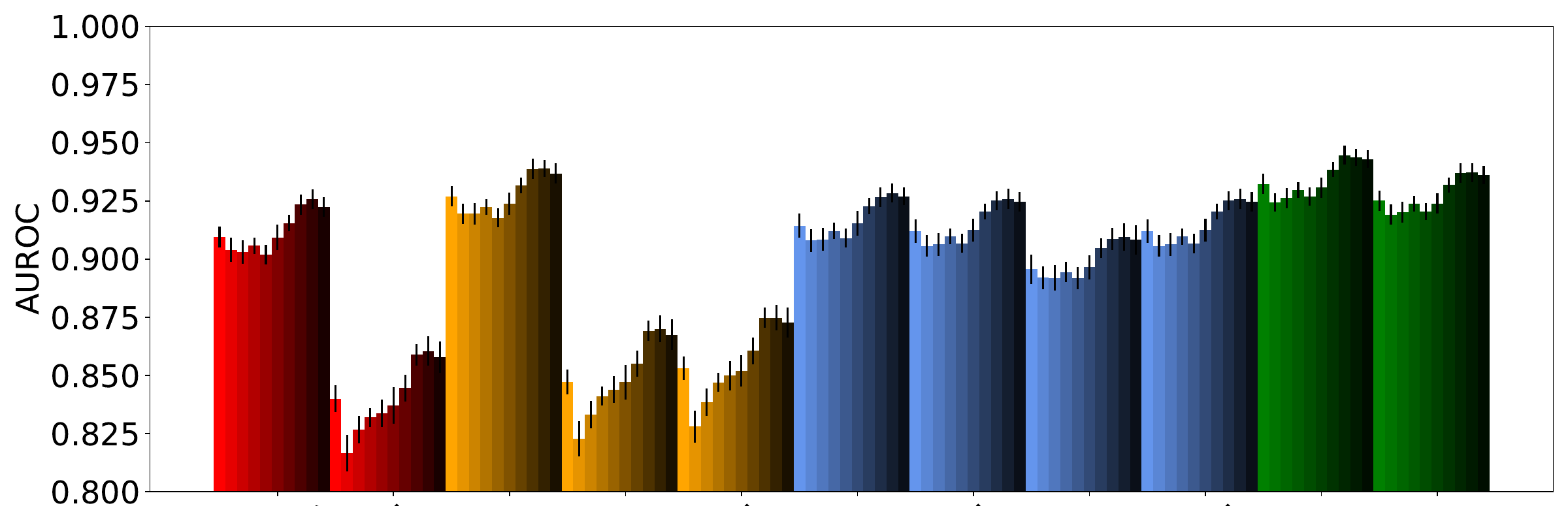}};
\node[anchor=north west, shift={(-5pt,3pt)}] at (fig2.north west) {\textbf{b)}};
\node[anchor=north,inner sep=0] (fig3) at (fig2.south) {\includegraphics[trim= 0 0 0 0, clip,width=\linewidth]{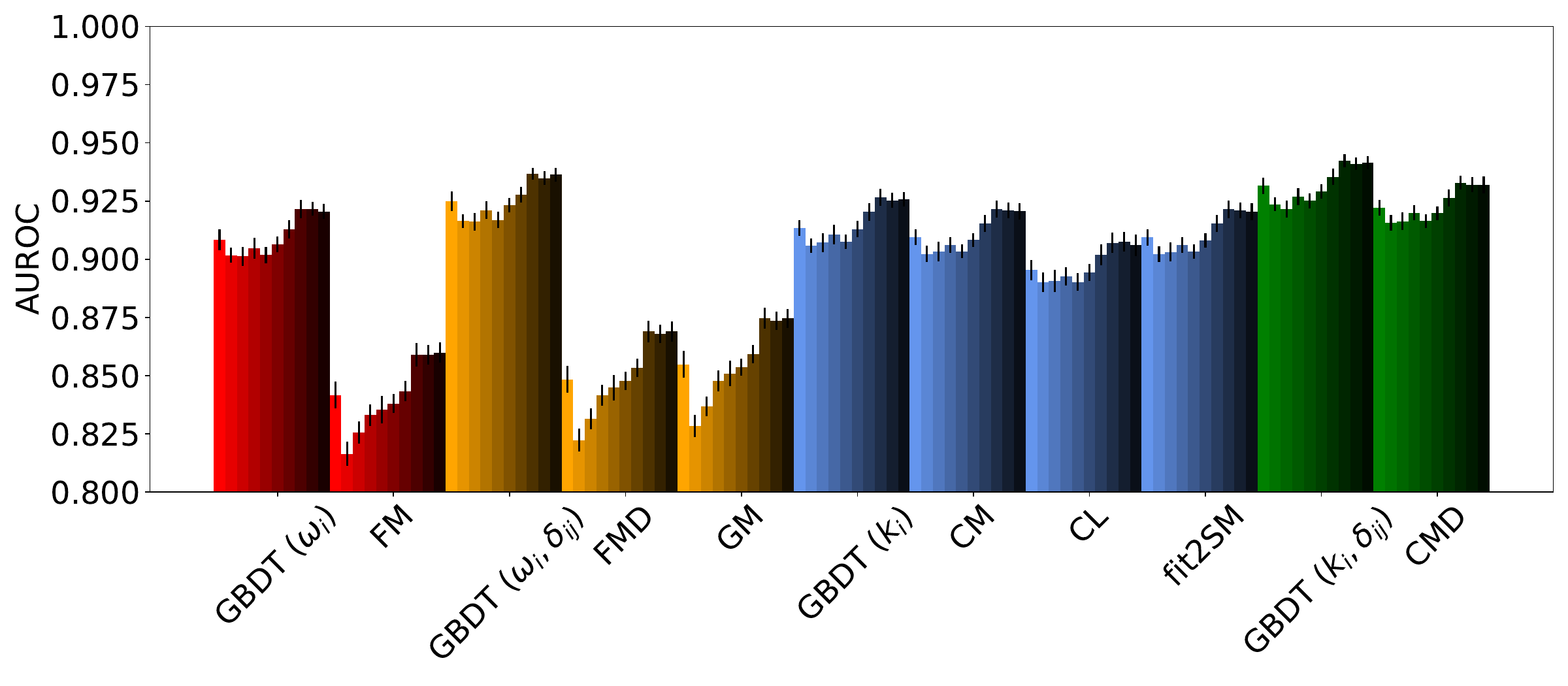}};
\node[anchor=north west, shift={(-5pt,3pt)}] at (fig3.north west) {\textbf{c)}};
\end{tikzpicture}
\caption{Performance of the models, measured in terms of AUROC, for all years of the WTW (darker shades correspond to more recent years), when the alternative prediction setting described in Appendix~\hyperref[sec:AppC]{C} is adopted. \textcolor{black}{We have randomly selected the $10\%$ (panel \textbf{a}), the $20\%$ (panel \textbf{b}) and the $30\%$ (panel \textbf{c}) of node pairs $40$ times to populate the test set, on which the prediction is performed after having trained the model on the remaining pairs of nodes.} Each statistical indicator has, then, been averaged over such samples, the standard deviation being represented by a vertical, black bar. The results are in agreement with those reported in the main text, as each `endogenous' instance of the GBDT performs in a way that is comparable to that of its purely structural, white-box counterparts - especially for what concerns the CM and the fit2SM.}
\label{fig:resMungo}
\end{figure}

{
Another, common setting for prediction - similar to the one adopted in~\cite{mungo_reconstructing_2023} - consists in randomly splitting all node pairs into a training and a test set, i.e. $\mathcal{U}=\mathcal{U}^{tr}\cup\mathcal{U}^{te}$: the training set is, then, identified with $\mathbf{D}^{tr}=\{\vect{x}_{ij},a_{ij}\}_{ij\in\mathcal{U}^{tr}}$ and the prediction is performed on $\mathbf{D}^{te}=\{a_{ij}\}_{ij\in\mathcal{U}^{te}}$. As the evaluation of `endogenous' features is based on the links populating $\mathcal{U}^{tr}$, this is equivalent to treating all pairs of nodes in $\mathcal{U}^{te}$ as unconnected.

The results, shown in figure~\ref{fig:resMungo}, are in agreement with those reported in the main text, as each `endogenous' instance of the GBDT performs in a way that is comparable to that of its purely structural, white-box counterparts - especially for what concerns the CM and the fit2SM. To be noticed that the performance of the GBDT taking as inputs both `exogenous' and `endogenous' features deteriorates to a lesser extent than the one of the CMD, i.e. its white-box counterpart, as the size of the training set decreases.}

\clearpage

\hypertarget{AppD}{}
\section*{APPENDIX D.\\LINK REMOVAL IN A DIFFERENT SETTING}\label{AppD}

\setcounter{section}{0}
\renewcommand{\thefigure}{D\arabic{figure}}
\setcounter{figure}{0}
\renewcommand{\thetable}{D\arabic{table}}
\setcounter{table}{0}

\begin{figure}[t!]
\begin{tikzpicture}
\node[anchor=north west](fig1) at (0,0) {\includegraphics[width=\textwidth]{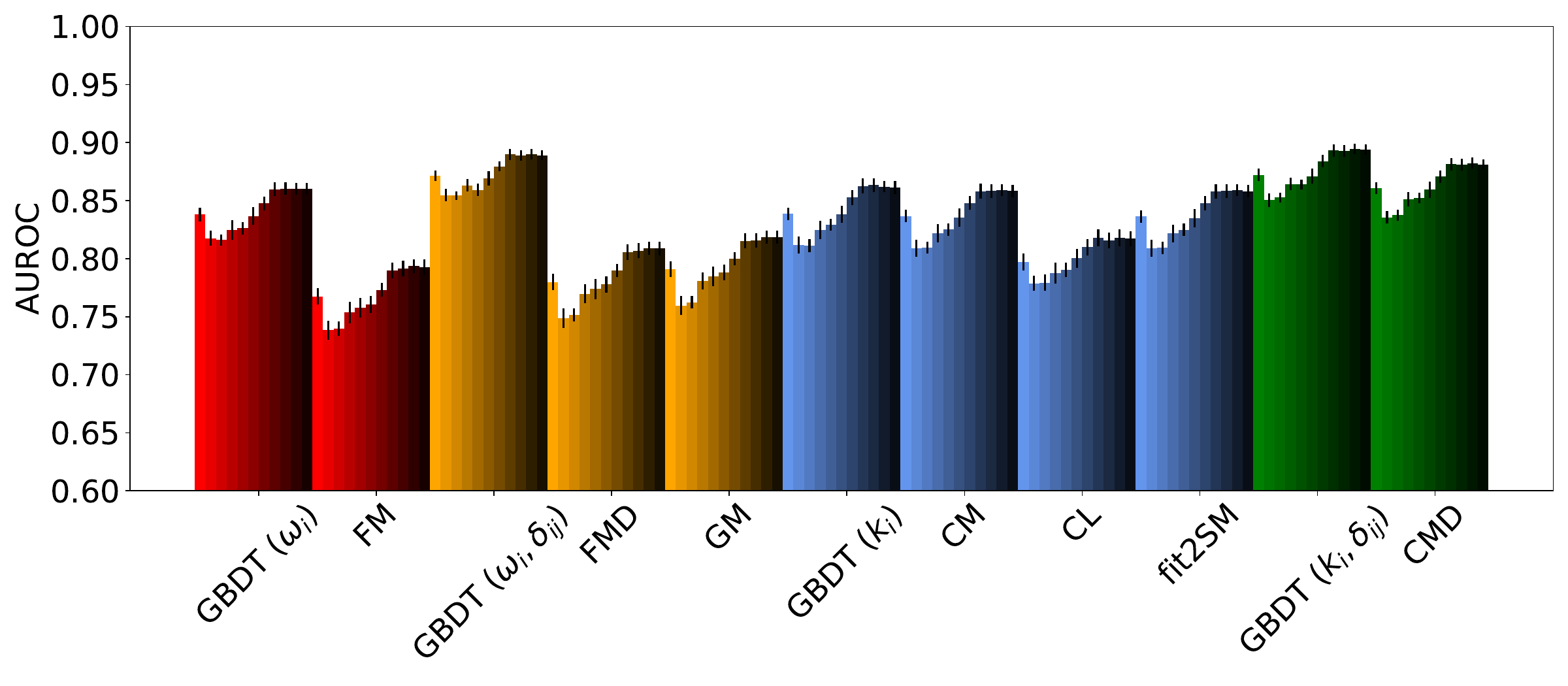}};  
\node[anchor=north west] (a) at (fig1.north west) {\textbf{a)}};
\node[anchor=north west](fig2) at (fig1.south west){\includegraphics[width=0.49\textwidth]{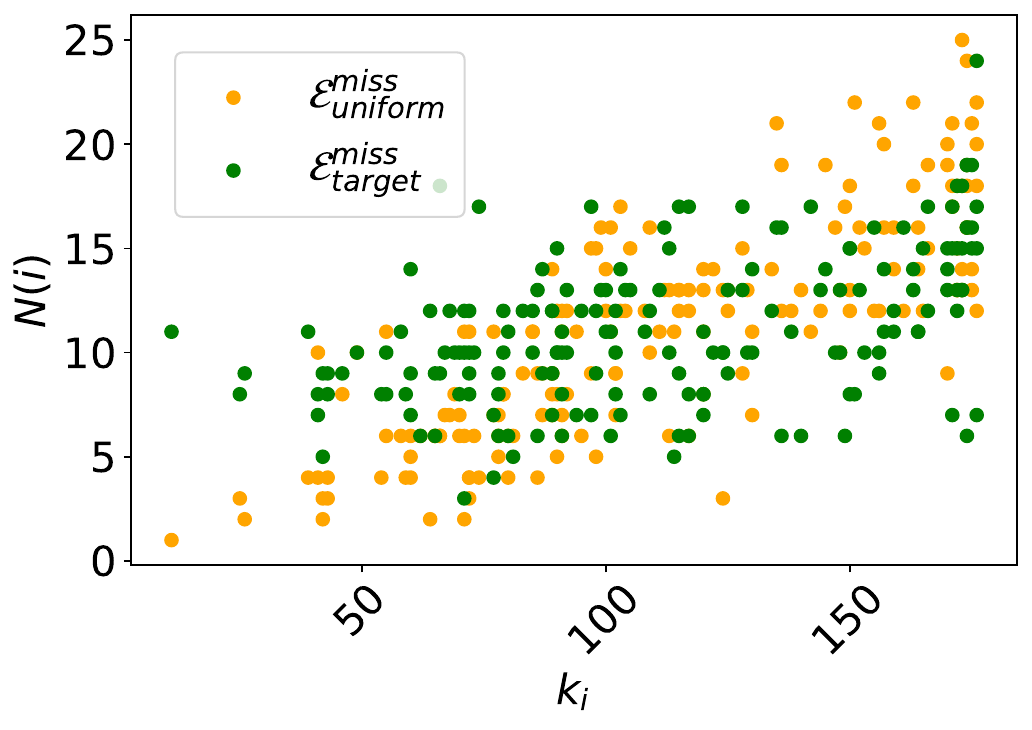}};    
\node[anchor=north west] (b) at (fig2.north west) {\textbf{b)}};
\node[anchor=north west](fig3) at (fig2.north east) {\includegraphics[width=0.49\textwidth]{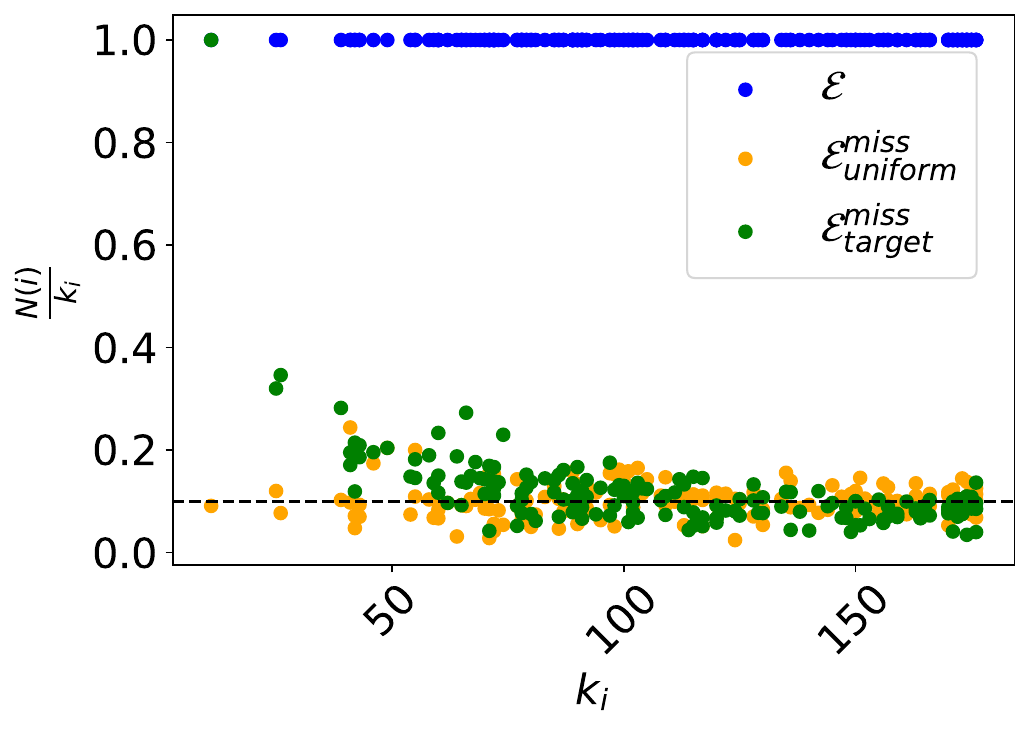}};
\node[anchor=north west] (c) at (fig3.north west) {\textbf{c)}};
\end{tikzpicture}
\caption{\textcolor{black}{Panel \textbf{a}: AUROC scores for `targeted' link removal, for all years of the WTW (darker shades correspond to more recent years). The statistical indicators have been averaged over all samples, the standard deviation being represented by a vertical, black bar. We have selected the $10\%$ of links $20$ times to populate the test set, on which the prediction is performed after having trained the model on the remaining pairs of nodes. Panel \textbf{b}: number of occurrences of node $i$ in $\mathcal{E}^{miss}_{uniform}$ (populated with randomly selected links from the year $2000$ of the WTW) and $\mathcal{E}^{miss}_{target}$ (populated by selecting links with a probability that is inversely proportional to the product of the degrees of the involved vertices from the year $2000$ of the WTW), scattered versus the degree $k_i$. Panel \textbf{c}: number of occurrences of node $i$ in $\mathcal{E}$, $\mathcal{E}^{miss}_{uniform}$ and $\mathcal{E}^{miss}_{target}$, rescaled by the degree, scattered versus the degree $k_i$. The results indicate that the performance of all methods degrades as `targeted' link removal recipes are implemented.}}
\label{fig:targeted_WTW}
\end{figure}

\begin{figure}[t!]
\begin{tikzpicture}
\node[anchor=north west](fig3) at (0,0) {\includegraphics[width=\textwidth]{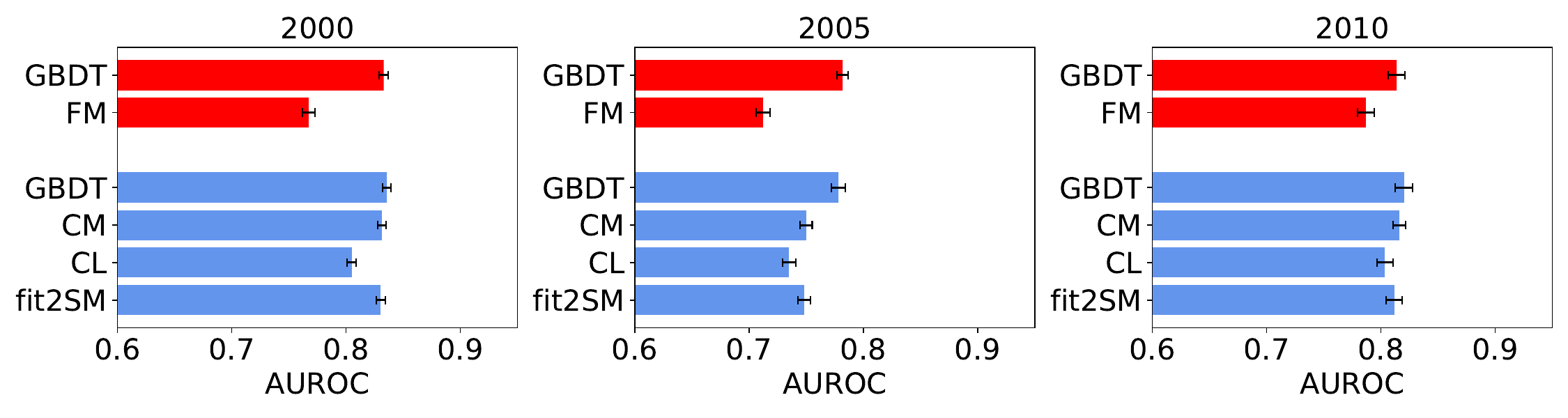}};    
\node[anchor=north west] (a) at (fig3.north west) {\textbf{a)}};
\node[anchor=north west](fig1) at (fig3.south west) {\includegraphics[width=0.49\textwidth]{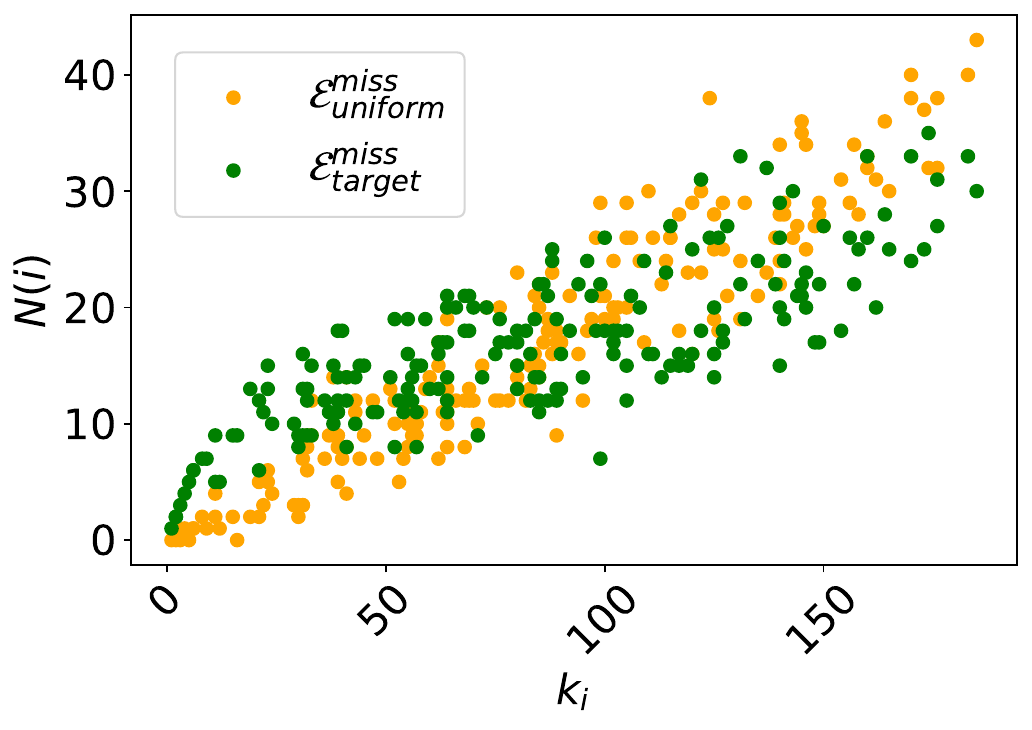}};    
\node[anchor=north west] (b) at (fig1.north west) {\textbf{b)}};
\node[anchor=north west](fig2) at (fig1.north east) {\includegraphics[width=0.49\textwidth]{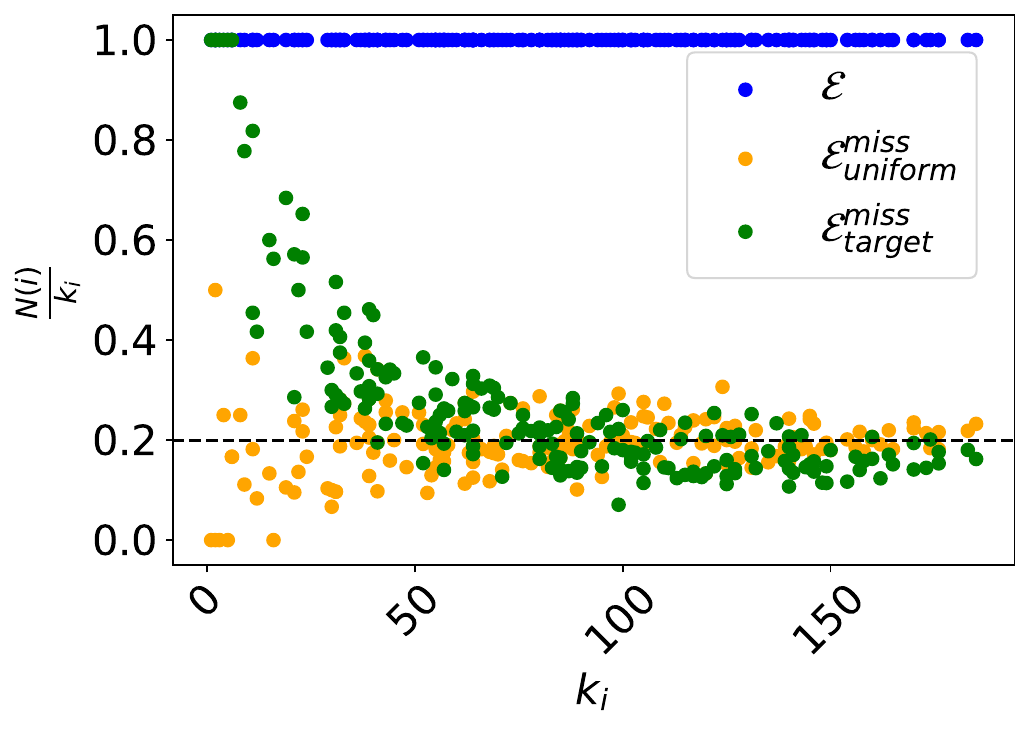}};
\node[anchor=north west] (c) at (fig2.north west) {\textbf{c)}};
\end{tikzpicture}
\caption{\textcolor{black}{Panel \textbf{a}: AUROC scores for targeted link removal, for the years $2000$, $2005$ and $2010$ of eMID. The statistical indicators have been averaged over all samples, the standard deviation being represented by a vertical, black bar. We have selected the $20\%$ of links $30$ times to populate the test set, on which the prediction is performed after having trained the model on the remaining pairs of nodes. Panel \textbf{b}: number of occurrences of node $i$ in $\mathcal{E}^{miss}_{uniform}$ (populated with randomly selected links from the yearly aggregation of the eMID dataset, where year is $2000$) and $\mathcal{E}^{miss}_{target}$ (populated by selecting links with a probability that is inversely proportional to the product of the degrees of the involved vertices from the same dataset snapshot), scattered versus the degree $k_i$. Panel \textbf{c}: number of occurrences of node $i$ in $\mathcal{E}$, $\mathcal{E}^{miss}_{uniform}$ and $\mathcal{E}^{miss}_{target}$ rescaled by the degree, scattered versus the degree $k_i$. The results indicate that the performance of all methods degrades as `targeted' link removal recipes are implemented.}}
\label{fig:targeted_eMID}
\end{figure}

\begin{figure}[t!]
\begin{tikzpicture}
\node[anchor=north west](fig1) at (0,0) {\includegraphics[width=0.49\textwidth]{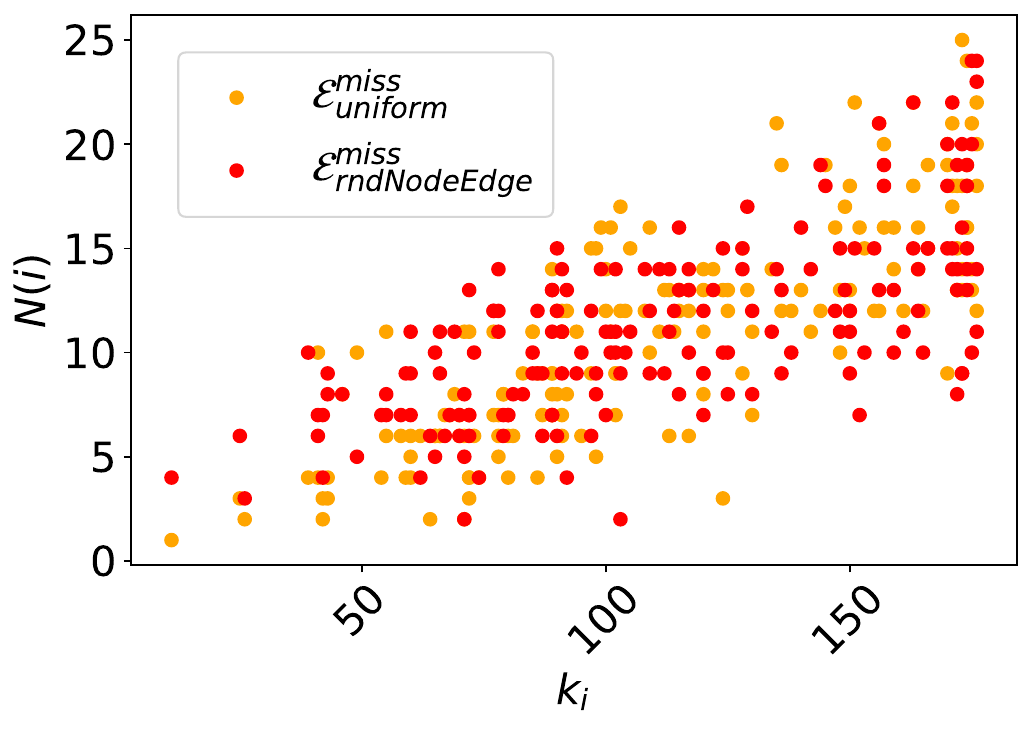}};    
\node[anchor= west] (a) at (fig1.north west) {\textbf{a.1)}};
\node[anchor=north west](fig2) at (fig1.north east) {\includegraphics[width=0.49\textwidth]{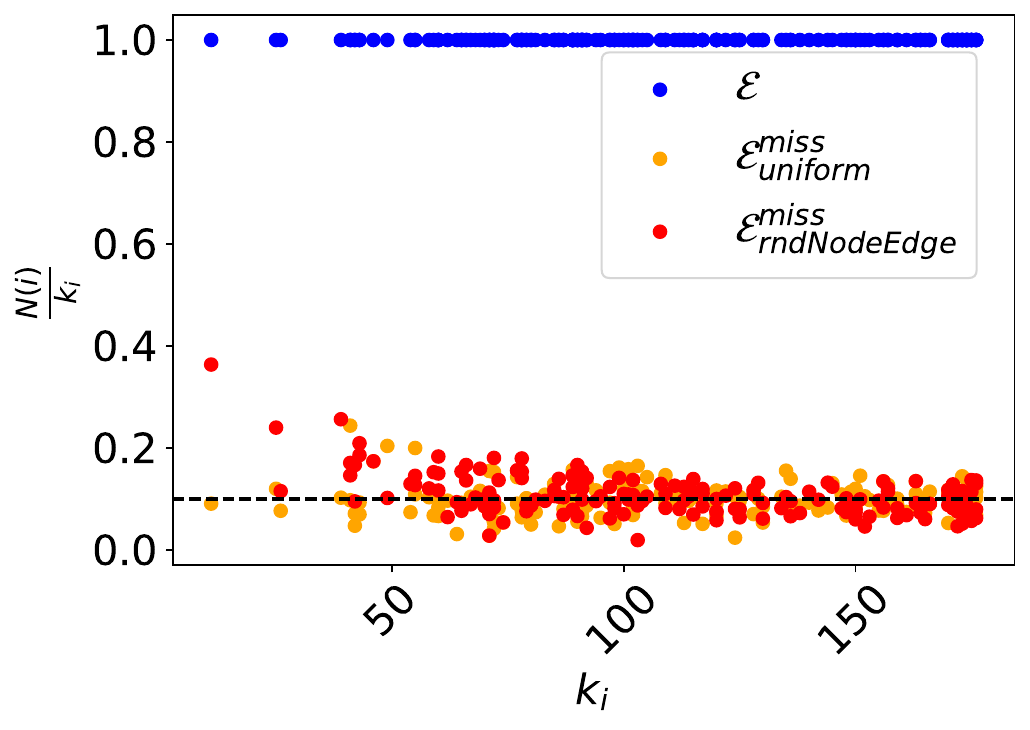}};    
\node[anchor= west] (b) at (fig2.north west) {\textbf{b.1)}};
\node[anchor=north west](fig3) at (fig1.south west) {\includegraphics[width=0.49\textwidth]{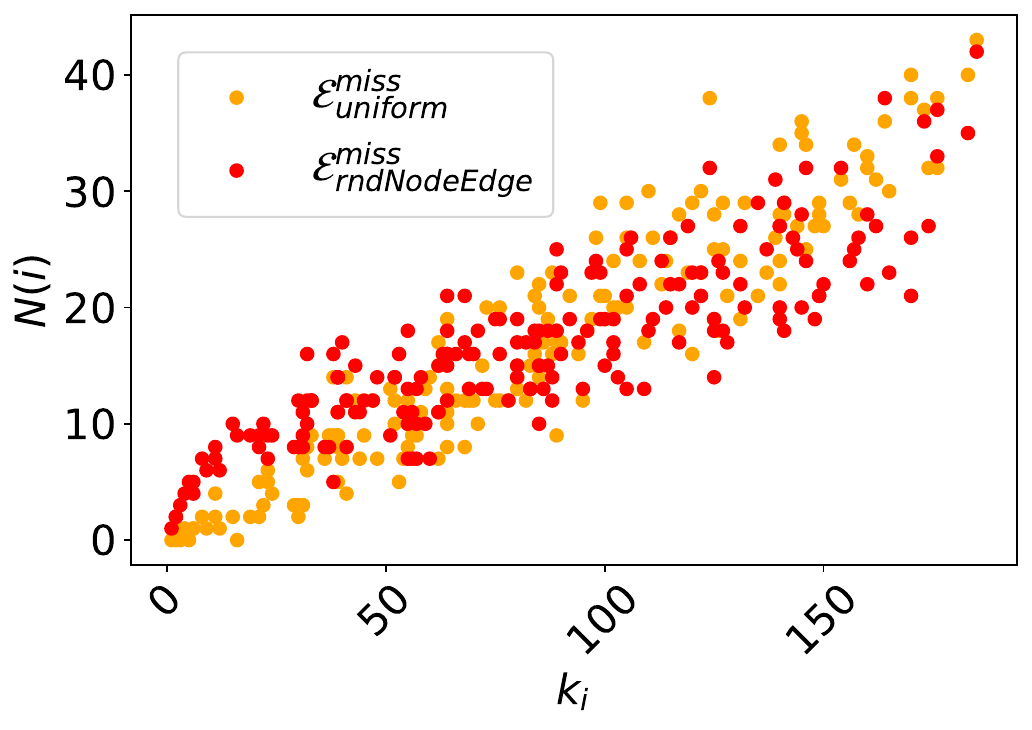}};    
\node[anchor=north west] (c) at (fig3.north west) {\textbf{a.2)}};
\node[anchor=north west](fig4) at (fig3.north east) {\includegraphics[width=0.49\textwidth]{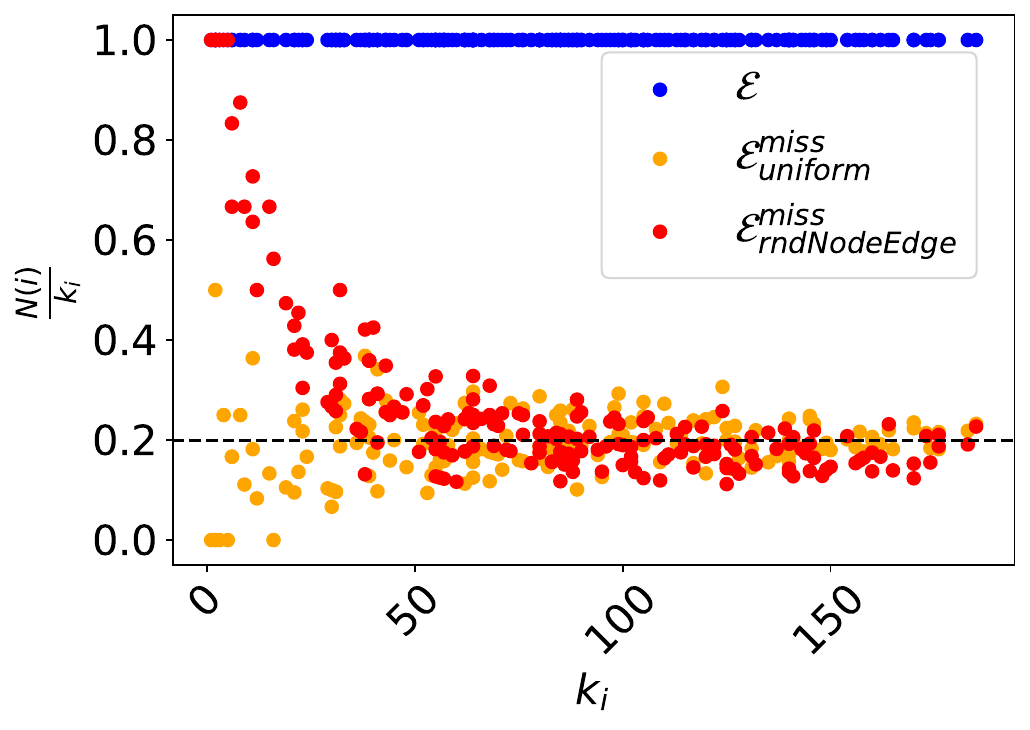}};    
\node[anchor=north west] (d) at (fig4.north west) {\textbf{b.2)}};
\end{tikzpicture}
\caption{\textcolor{black}{Panels \textbf{a.1-a.2}: number of occurrences of node $i$ in $\mathcal{E}^{miss}_{uniform}$, populated with randomly selected links from the year $2000$ of the WTW (panel \textbf{a.1}, $10\%$ of links) and eMID (panel \textbf{a.2}, $20\%$ of links) and $\mathcal{E}^{miss}_{rndNodeEdge}$, populated by implementing the \texttt{Random Node-Edge Sampler} for the year $2000$ of the WTW (panel \textbf{a.1}, $10\%$ of links) and eMID (panel \textbf{a.2}, $20\%$ of links), scattered versus the degree $k_i$. Panels \textbf{b.1-b.2}: number of occurrences of node $i$ in $\mathcal{E}$, $\mathcal{E}^{miss}_{uniform}$ and $\mathcal{E}^{miss}_{rndNodeEdge}$ rescaled by the degree, scattered versus the degree $k_i$.}}
\label{fig:degdistrib}
\end{figure}

{\color{black}
The results that we have shown in the main text have been obtained by populating $\mathcal{E}^{miss}$ with links sampled from $\mathcal{E}$ with a uniform probability~\cite{zhou_predicting_2009,lu_link_2011,parisi_entropybased_2018,mungo_reconstructing_2023}. Since node $i$ appears in $\mathcal{E}$ as a vertex of an undirected edge exactly $k_i$ times, we expect the same dependence on the degree to characterise $\mathcal{E}^{miss}$ as well: in other words, uniformly sampling $\mathcal{E}$ leads to populate $\mathcal{E}^{miss}$ with (links connecting) high-degree nodes. It is, thus, natural to ask how a different sampling scheme would affect the results.

In order to answer such a question, we have replicated the analysis under a couple of `targeted' sampling prescriptions. In the first case, we have removed links by selecting edges with a probability that is inversely proportional to the product of the degrees of the involved vertices: more specifically, at the generic iteration, the link connecting nodes $i$ and $j$ was sampled with a probability reading $p_{ij}\propto 1/k_ik_j$ (see figures~\ref{fig:targeted_WTW} and~\ref{fig:targeted_eMID}). In the second case, we have implemented the so-called \texttt{Random Node-Edge Sampler}~\cite{krishnamurthy_reducing_2005}, first sampling a node uniformly at random and, then, sampling an edge uniformly at random from the ones that are incident to the sampled node - the procedure is repeated until the desired size is reached.

Comparing figures~\ref{fig:targeted_WTW} and~\ref{fig:targeted_eMID} with figures~\ref{fig:ROCcurves} and~\ref{fig:EMID_hbars} reveals that \emph{i)} although removing links randomly leads to remove more links from the more connected nodes, the `targeted' sampling prescriptions considered as alternatives to the uniform one - basically performing the same, as evident upon looking at figure~\ref{fig:degdistrib} - may incur into the same problems; \emph{ii)} the performance of all methods degrades when `targeted' sampling prescriptions are considered - the methods taking as inputs the degrees being more affected than others. We are, thus, led to conclude that existing link prediction algorithms (be they black-box or white-box) perform better for better connected nodes.}

\clearpage

\hypertarget{AppE}{}
\section*{APPENDIX E.\\COMPUTATIONAL COMPLEXITY OF MISSING LINKS PREDICTION ALGORITHMS}\label{AppE}

\setcounter{section}{0}
\renewcommand{\thefigure}{E\arabic{figure}}
\setcounter{figure}{0}
\renewcommand{\thetable}{E\arabic{table}}
\setcounter{table}{0}

{\color{black}
Let us, now, study the computational complexity of the algorithms considered in the present work: the results are reported in tables~\ref{tab:Running_time_wtw} and~\ref{tab:Running_time_emid}. In the case of white-box methods, the total computational time corresponds to the average time needed for training and predicting across training set-test set realisations (i.e. $T^{tr}+T^{te}$): let us observe that four white-box algorithms (CL, CM, FM, GM) neatly beat their black-box counterpart. In the case of black-box methods, instead, it is also necessary to take into account the time needed to tune the hyperparameters (i.e. $T^{tune}$) which, however, does not need to be repeated across training set-test set realisations: still, adding it leads the computational complexity of the black-box methods to significantly exceed the one of white-box methods.}

\begin{table}[t!]
\centering
\begin{tabular}{ll|lr|lr|lr|lr}
\hline
\hline
&& \multicolumn{8}{c}{WTW features} \\
\cline{3-10} &&\multicolumn{2}{c|}{$\omega_i$} & \multicolumn{2}{c|}{$\omega_i,\delta_{ij}$} & \multicolumn{2}{c|}{$k_i$} & \multicolumn{2}{c}{$k_i,\delta_{ij}$} \\
\hline
\hline
\multirow{2}{*}{black-box algorithms}& $T^{tr}+T^{te}$ & \multicolumn{2}{r|}{\textcolor{white}{FM} $753\pm 154$} & \multicolumn{2}{r|}{$813\pm 168$} & \multicolumn{2}{r|}{$203\pm 97$} & \multicolumn{2}{r}{$550\pm 162$} \\
& $T^{tune}$ & \multicolumn{2}{r|}{\textcolor{white}{FM} $(1.23\pm 0.19)\times 10^5$} & \multicolumn{2}{r|}{$(1.30\pm 0.15)\times 10^5$} & \multicolumn{2}{r|}{$(1.41\pm 0.21)\times 10^5$} & \multicolumn{2}{r}{$(1.30\pm 0.18)\times 10^5$} \\
\hline
\multirow{3}{*}{white-box algorithms}&\multirow{3}{*}{$T^{tr}+T^{te}$} & FM & $53\pm 4$ & FMD & $3023\pm 325$ & CM & $34\pm 12$ & CMD & $546\pm 918$ \\
& & & & GM & $62\pm 9$ & CL & $4\pm 1$ & & \\
& & & &&  & fit2SM & $22640\pm 2827$ & & \\
\hline
\hline
\end{tabular}
\caption{\textcolor{black}{Computational complexity of the models considered in the present manuscript (expressed in milliseconds), for all years of the WTW. \textcolor{black}{We have selected the $10\%$ of links $40$ times to populate the test set, on which the prediction is performed after having trained the model on the remaining pairs of nodes.} Values are reported as averages (across training set-test set realisations) plus/minus one standard deviation.}}
\label{tab:Running_time_wtw}
\end{table}

\begin{table}[t!]
\centering
\begin{tabular}{ll|lr|lr}
\hline
\hline
& & \multicolumn{4}{c}{eMID features (yearly)} \\
\cline{3-6}& & \multicolumn{2}{c|}{$s_i$} & \multicolumn{2}{c}{$k_i$} \\
\hline
\hline
\multirow{2}{*}{black-box algorithms}& $T^{tr}+T^{te}$ & \multicolumn{2}{r|}{\textcolor{white}{FM} $695\pm 848$} & \multicolumn{2}{r}{$140\pm 90$} \\
& $T^{tune}$ & \multicolumn{2}{r|}{\textcolor{white}{FM} $(7.94\pm 1.75)\times10^{4}$} & \multicolumn{2}{r}{$(8.32\pm 2.97)\times10^{4}$} \\
\hline
\multirow{3}{*}{white-box algorithms}&\multirow{3}{*}{$T^{tr}+T^{te}$} & FM & $15\pm 10$ & CM & $35\pm 32$ \\
& & & & CL & $3\pm 2$ \\
& & & & fit2SM & $13348\pm 8345$ \\
\hline
\hline
&& \multicolumn{4}{c}{eMID features (monthly)} \\
\cline{3-6}& & \multicolumn{2}{c|}{$s_i$}& \multicolumn{2}{c}{$k_i$} \\
\hline
\hline
\multirow{2}{*}{black-box algorithms}& $T^{tr}+T^{te}$ & \multicolumn{2}{r|}{\textcolor{white}{FM} $283\pm 172$} &\multicolumn{2}{r}{$92\pm 40$} \\
& $T^{tune}$ & \multicolumn{2}{r|}{\textcolor{white}{FM} $(6.26\pm 0.86)\times10^{4}$} &\multicolumn{2}{r}{$(5.94\pm 0.93)\times10^{4}$} \\
\hline
\multirow{3}{*}{white-box algorithms}&\multirow{3}{*}{ $T^{tr}+T^{te}$} & FM & $7\pm 2$ & CM & $17\pm 5$ \\
& & & & CL & $2\pm 0$ \\
& & & & fit2SM & $6906\pm 3637$\\
\hline
\hline
& &\multicolumn{4}{c}{eMID features (daily)} \\
\cline{3-6}& & \multicolumn{2}{c|}{$s_i$}& \multicolumn{2}{c}{$k_i$} \\
\hline
\hline
\multirow{2}{*}{black-box algorithms}& $T^{tr}+T^{te}$ & \multicolumn{2}{c|}{\textcolor{white}{FM} $90\pm 45$} &\multicolumn{2}{r}{$47\pm 26$} \\
&$T^{tune}$ & \multicolumn{2}{c|}{\textcolor{white}{FM} $(5.15\pm 1.05)\times10^{4}$} &\multicolumn{2}{r}{$(4.23\pm 0.91)\times10^{4}$} \\
\hline
\multirow{3}{*}{white-box algorithms}&\multirow{3}{*}{ $T^{tr}+T^{te}$} & FM & $4\pm 1$ & CM &$12\pm 1$ \\
& & & & CL & $2\pm 0$ \\
& & & & fit2SM & $2529\pm 1300$ \\
\hline
\hline
\end{tabular}
\caption{\textcolor{black}{Computational complexity of the models considered in the present manuscript (expressed in milliseconds), for snapshots of eMID at different aggregation levels (top: yearly; middle: monthly; bottom: daily). \textcolor{black}{We have selected the $20\%$ of links $30$ times to populate the test set, on which the prediction is performed after having trained the model on the remaining pairs of nodes.} Values are reported as averages (across training set-test set realisations) plus/minus one standard deviation.}}
\label{tab:Running_time_emid}
\end{table}

\clearpage

\hypertarget{AppF}{}
\section*{APPENDIX F.\\STATISTICAL SIGNIFICANCE OF PREDICTION PERFORMANCES}\label{AppF}

\setcounter{section}{0}
\renewcommand{\thefigure}{F\arabic{figure}}
\setcounter{figure}{0}
\renewcommand{\thetable}{F\arabic{table}}
\setcounter{table}{0}

\begin{figure}[t!]
\begin{tikzpicture}[every node/.style={inner sep=0pt,outer sep=0pt}]
\node[anchor=north west](tab) at (0,0) {\begin{tabular}{c|c|c|c|c}
\hline
\hline
WTW features & White-box algorithms & $\langle\Delta_\text{AUROC}\rangle$ & $t$ & $p$ \\
\hline
\hline 
$\omega_i$ & FM &$ -7.261\times10^{-2}$ & $-174.862$ & $0$ \\
\hline
\multirow{2}{*}{$\omega_i,\delta_{ij}$} & {FMD} & $-7.628\times10^{-2}$ & $-149.266$ & $0$\\
& {GM}&$ -7.148\times10^{-2}$ & $-144.894$ & $0$ \\
\hline
\multirow{3}{*}{$k_i$} & {CM}& $ -7.614\times10^{-4}$ & $-18.083$ & $4.972\times10^{-55}$\\
& {CL} & $ -1.786\times10^{-2}$ & $-119.152$ & $0$\\

& {fit2SM} &$ -7.499\times10^{-4}$ & $-17.741$ & $1.740\times10^{-53}$ \\
\hline
{$k_i,\delta_{ij}$} & {CMD} &$ -3.870\times10^{-3}$ & $-39.052$ & $6.451\times10^{-145}$ \\
\hline
\hline
\end{tabular}
};
\node[anchor=north,shift={(0,-.5)}](FMD) at (tab.south) {\includegraphics[trim= 10 0 10 21, clip,width=0.32\linewidth]{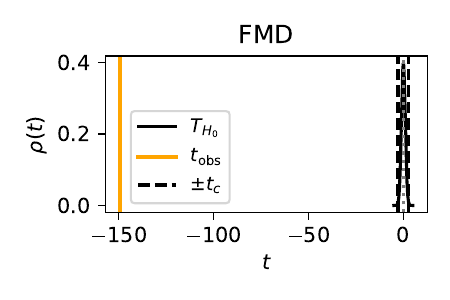}};
\node[anchor=north east] (FM) at (FMD.north west) {\includegraphics[trim= 10 0 10 21, clip,width=0.32\linewidth]{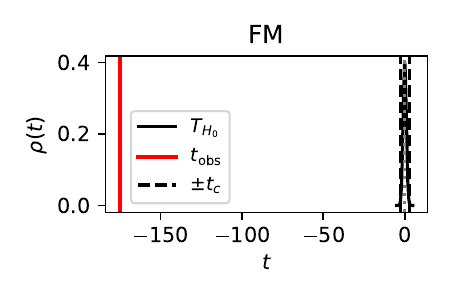}};
\node[anchor=north west](GM) at (FMD.north east) {\includegraphics[trim= 10 0 10 21, clip,width=0.32\linewidth]{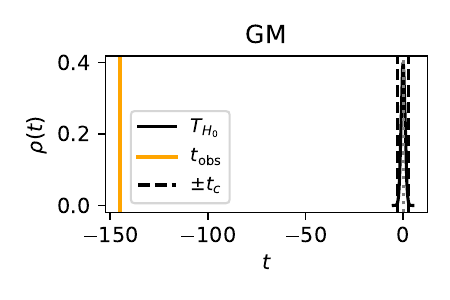}};
\node[anchor=north west](CM) at (FM.south west) {\includegraphics[trim= 10 0 10 21, clip,width=0.32\linewidth]{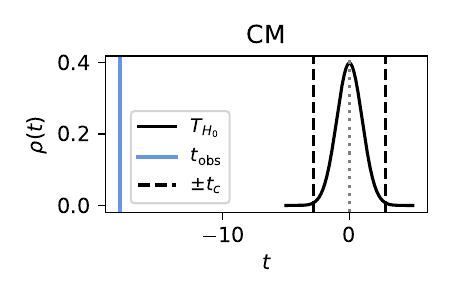}};
\node[anchor=north west](CL) at (CM.north east) {\includegraphics[trim= 10 0 10 21, clip,width=0.32\linewidth]{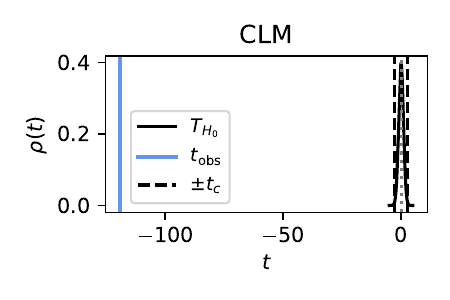}};
\node[anchor=north west](fit2SM) at (CL.north east) {\includegraphics[trim= 10 0 10 21, clip,width=0.32\linewidth]{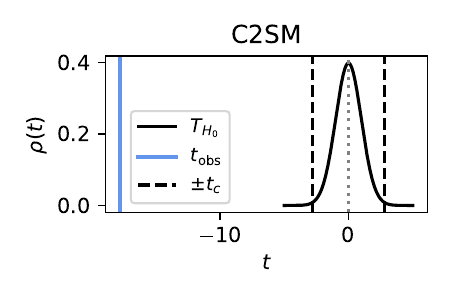}};
\node[anchor=north](CMD) at (CL.south) {\includegraphics[trim= 10 10 10 21, clip,width=0.32\linewidth]{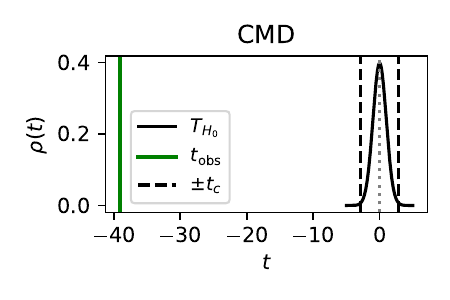}};
\node[anchor= south east] at (tab.north west) {\textbf{a)}};
\node[anchor= south  west] at (FM.north west) {\textbf{b)}};
\node[anchor= south west] at (FM.north) {\textbf{FM}};
\node[anchor=south  west] at (FMD.north) {\textbf{FMD}};
\node[anchor= south west] at (CM.north) {\textbf{CM}};
\node[anchor= south west] at (CL.north) {\textbf{CL}};
\node[anchor= south west] at (fit2SM.north) {\textbf{fit2SM}};
\node[anchor=south  west] at (CMD.north) {\textbf{CMD}};
\node[anchor=south  west] at (GM.north) {\textbf{GM}};
\end{tikzpicture}
\caption{\textcolor{black}{Panel \textbf{a}: statistical significance of the prediction performances of each white-box method versus its black-box counterpart (evaluated in terms of AUROC score), for the WTW when $\mathcal{E}^{miss}$ is populated with the $10\%$ of the existing links. For each pair of methods, the difference of performances is paired across all iterations ($40$) per year ($11$), corresponding to a sample size of $440$ points. The table entries report the (sample) average difference $\overline{\Delta}$, the t-value expressing the distance of the (sample) average difference from $0$, in units of the (sample) variance and the p-value. Panel \textbf{b}: distributions representing our paired samples t-test for the AUROC score and the WTW. Results do not change if the analysis is carried out in an yearly fashion.}}
\label{fig:ttest_WTW01AUROC}
\end{figure}

\begin{figure}[t!]
\begin{tikzpicture}[every node/.style={inner sep=0pt,outer sep=0pt}]
\node[anchor=north west](tab) at (0,0) {\begin{tabular}{c|c|c|c|c}
\hline
\hline
WTW features & White-box algorithms & $\langle\Delta_\text{AUPRC}\rangle$ & $t$ & $p$ \\
\hline
\hline 
$\omega_i$ & FM & $-0.183$ & $-115.614$ & $0$ \\
\hline
\multirow{2}{*}{$\omega_i,\delta_{ij}$} & {FMD} & $-0.178$&$-98.287$&$0$\\
& {GM}&$ -0.184$ & $-98.971$ & $0$ \\
\hline
\multirow{3}{*}{$k_i$} & {CM}& $0.002$&$5.692$&$2.30\times10^{-8}$ \\
& {CL} & $-0.047$ & $-92.973$ & $0$\\

& {fit2SM} &$ 0.002$ & $6.541$ & $1.70 \times10^{-10} $ \\
\hline
{$k_i,\delta_{ij}$} & {CMD} & $3.4\times10^{-5}$&$ 0.082$ & $0.93$ \\
\hline
\hline
\end{tabular}
};
\node[anchor=north,shift={(0,-.5)}](FMD) at (tab.south) {\includegraphics[trim= 10 0 10 21, clip,width=0.32\linewidth]{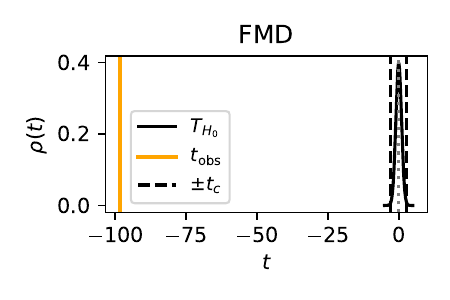}};
\node[anchor=north east](FM) at (FMD.north west) {\includegraphics[trim= 10 0 10 21, clip,width=0.32\linewidth]{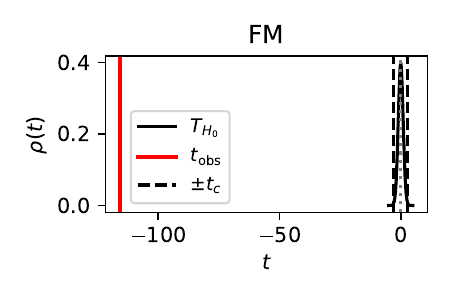}};
\node[anchor=north west](GM) at (FMD.north east) {\includegraphics[trim= 10 0 10 21, clip,width=0.32\linewidth]{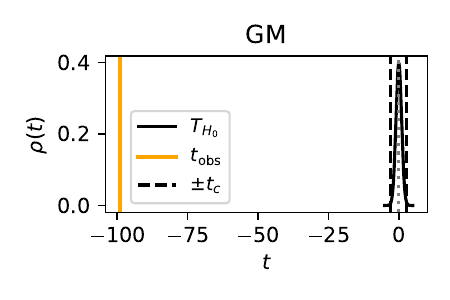}};
\node[anchor=north west](CM) at (FM.south west) {\includegraphics[trim= 10 0 10 21, clip,width=0.32\linewidth]{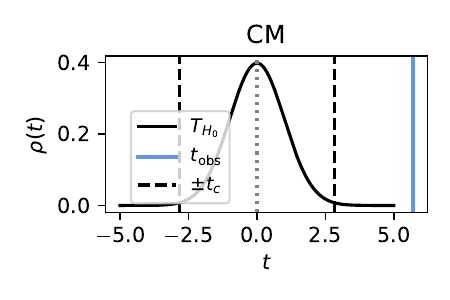}};
\node[anchor=north west](CL) at (CM.north east) {\includegraphics[trim= 10 0 10 21, clip,width=0.32\linewidth]{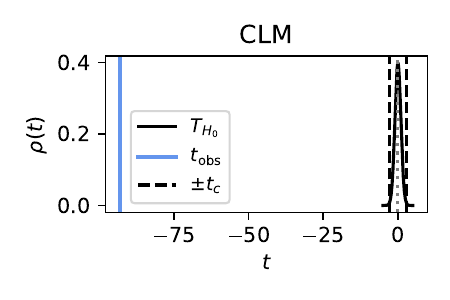}};
\node[anchor=north west](fit2SM) at (CL.north east) {\includegraphics[trim= 10 10 10 21, clip,width=0.32\linewidth]{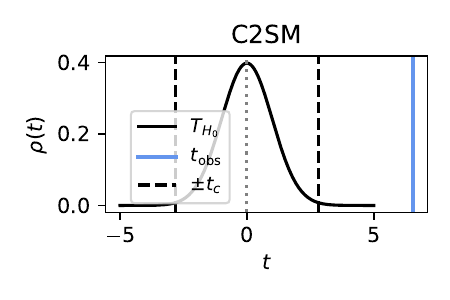}};
\node[anchor=north](CMD) at (CL.south) {\includegraphics[trim= 10 10 10 21, clip,width=0.32\linewidth]{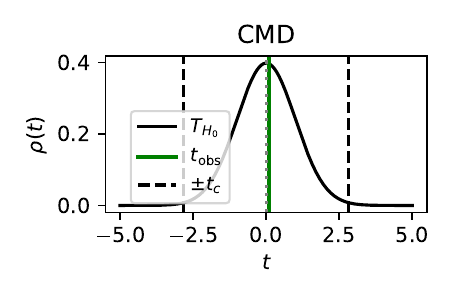}};
\node[anchor= south east] at (tab.north west) {\textbf{a)}};
\node[anchor= west] at (FM.north west) {\textbf{b)}};
\node[anchor= south west] at (FM.north) {\textbf{FM}};
\node[anchor= south west] at (FMD.north) {\textbf{FMD}};
\node[anchor= south west] at (CM.north) {\textbf{CM}};
\node[anchor= south west] at (CL.north) {\textbf{CL}};
\node[anchor= south west] at (fit2SM.north) {\textbf{fit2SM}};
\node[anchor= south west] at (CMD.north) {\textbf{CMD}};
\node[anchor= south west] at (GM.north) {\textbf{GM}};
\end{tikzpicture}
\caption{\textcolor{black}{Panel \textbf{a}: statistical significance of the prediction performances of each white-box method versus its black-box counterpart (evaluated in terms of AUPRC score), for the WTW when $\mathcal{E}^{miss}$ is populated with the $10\%$ of the existing links. For each pair of methods, the difference of performances is paired across all iterations ($40$) per year ($11$), corresponding to a sample size of $440$ points. The table entries report the (sample) average difference $\overline{\Delta}$, the t-value expressing the distance of the (sample) average difference from $0$, in units of the (sample) variance and the p-value. Panel \textbf{b}: distributions representing our paired samples t-test for the AUPRC score and the WTW. Results do not change if the analysis is carried out in an yearly fashion.}}
\label{fig:ttest_WTW01AUPRC}
\end{figure}

\begin{figure}[t!]
\flushleft
\begin{tikzpicture}[every node/.style={inner sep=0pt,outer sep=0pt}]
\node[anchor=north west](tab1) at (0,0) {\begin{tabular}{c|c|c|c|c}
\hline
\hline
eMID features (yearly) & White-box algorithms & $\langle\Delta_\text{AUROC}\rangle$ & $t$ & $p$ \\
\hline
\hline 
$\omega_i$ & FM & $ -7.759\times10^{-2}$ & $-37.787$ & $1.363\times10^{-56}$ \\
\hline
\multirow{3}{*}{$k_i$} & {CM}& $ -3.968\times10^{-3}$ & $-6.929$ & $6.410\times10^{-10}$  \\
& {CL} &$-1.047\times10^{-2}$ & $-17.016$ & $1.010\times10^{-29}$\\
& {fit2SM} &$ -4.035\times10^{-3}$ & $-7.073$ & $3.311\times10^{-10}$ \\
\hline
\hline
\end{tabular}
};
\node[anchor=north,shift={(-0.5,-.5)}](FM1) at (tab1.south west) {\includegraphics[trim= 10 10 10 21, clip,width=0.25\linewidth]{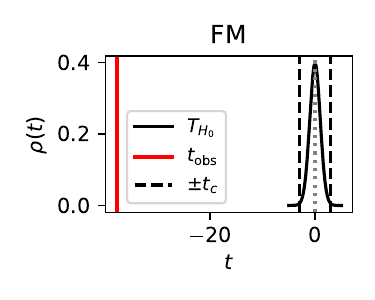}};
\node[anchor=north west](CM1) at (FM1.north east) {\includegraphics[trim= 10 10 10 21, clip,width=0.24\linewidth]{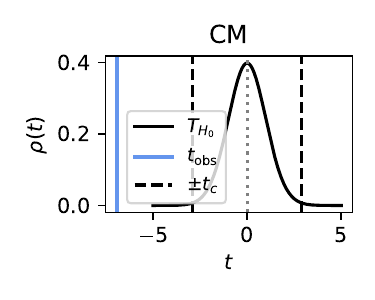}};
\node[anchor=north west](CL1) at (CM1.north east) {\includegraphics[trim= 10 10 10 21, clip,width=0.24\linewidth]{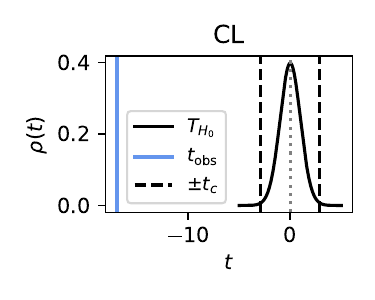}};
\node[anchor=north west](fit2SM1) at (CL1.north east) {\includegraphics[trim= 10 10 10 21, clip,width=0.24\linewidth]{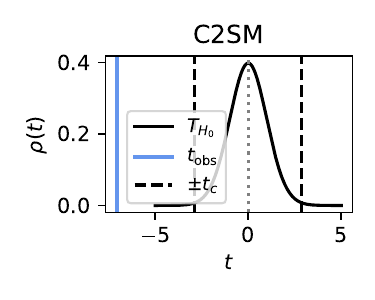}};
\node[anchor= south east] at (tab1.north west) {\textbf{a.1)}};
\node[anchor= south west] at (FM1.north west) {\textbf{a.2)}};
\node[anchor= south west] at (FM1.north) {\textbf{FM}};
\node[anchor= south west] at (CM1.north) {\textbf{CM}};
\node[anchor= south west] at (CL1.north) {\textbf{CL}};
\node[anchor= south west] at (fit2SM1.north) {\textbf{fit2SM}};
\node[anchor=north west,shift={(0.5,-.5)}](tab2) at (FM1.south) {\begin{tabular}{c|c|c|c|c}
\hline
\hline
eMID features (quarterly) & White-box algorithms & $\langle\Delta_\text{AUROC}\rangle$ & $t$ & $p$ \\
\hline
\hline 
$\omega_i$ & FM &$ -6.174\times10^{-2}$ & $-62.250$ & $1.205\times10^{-92}$ \\
\hline
\multirow{3}{*}{$k_i$} & {CM}&$ -2.443\times10^{-3}$ & $-10.308$ & $3.415\times10^{-18}$ \\
& {CL} &$ -9.205\times10^{-3}$ & $-22.093$ & $6.323\times10^{-44}$\\

& {fit2SM} &$ -2.650\times10^{-3}$ & $-11.014$ & $7.054\times10^{-20}$ \\
\hline
\hline
\end{tabular}
};
\node[anchor=north,shift={(-0.5,-.5)}](FM2) at (tab2.south west) {\includegraphics[trim= 10 10 10 21, clip,width=0.25\linewidth]{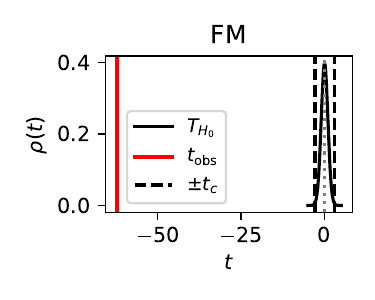}};
\node[anchor=north west](CM2) at (FM2.north east) {\includegraphics[trim= 10 10 10 21, clip,width=0.24\linewidth]{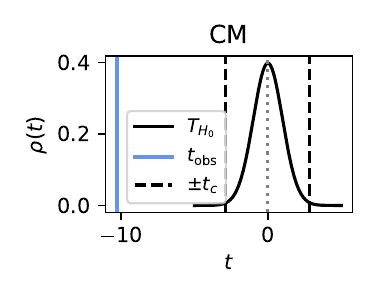}};
\node[anchor=north west](CL2) at (CM2.north east) {\includegraphics[trim= 10 10 10 21, clip,width=0.24\linewidth]{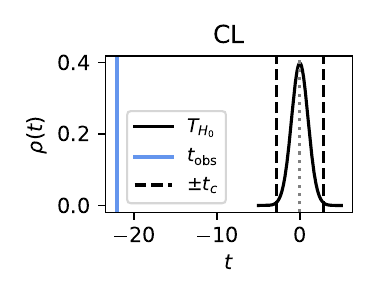}};
\node[anchor=north west](fit2SM2) at (CL2.north east) {\includegraphics[trim= 10 10 10 21, clip,width=0.24\linewidth]{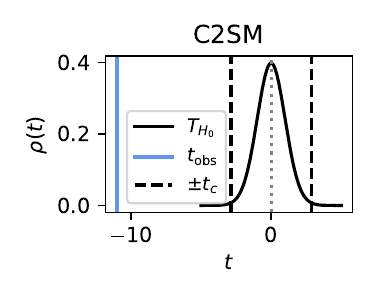}};
\node[anchor= south east] at (tab2.north west) {\textbf{b.1)}};
\node[anchor= south west] at (FM2.north west) {\textbf{b.2)}};
\node[anchor= south west] at (FM2.north) {\textbf{FM}};
\node[anchor= south west] at (CM2.north) {\textbf{CM}};
\node[anchor= south west] at (CL2.north) {\textbf{CL}};
\node[anchor= south west] at (fit2SM2.north) {\textbf{fit2SM}};
\node[anchor=north west,shift={(0.5,-.5)}](tab3) at (FM2.south)  {\begin{tabular}{c|c|c|c|c}
\hline
\hline
eMID features (weekly) & White-box algorithms & $\langle\Delta_\text{AUROC}\rangle$ & $t$ & $p$ \\
\hline
\hline 
$\omega_i$ & FM &$ -2.249\times10^{-2}$ & $-19.311$ & $1.774\times10^{-38}$ \\
\hline
\multirow{3}{*}{$k_i$} & {CM}& $ 2.151\times10^{-4}$ & $0.247$ & $8.050\times10^{-1}$ \\
& {CL} & $ -3.830\times10^{-3}$ & $-3.843$ & $1.966\times10^{-4}$\\
& {fit2SM} &$ -1.045\times10^{-3}$ & $-1.141$ & $2.564\times10^{-1}$ \\
\hline
\hline
\end{tabular}
};
\node[anchor=north,shift={(-0.5
,-.5)}](FM3) at (tab3.south west) {\includegraphics[trim= 10 10 10 21, clip,width=0.24\linewidth]{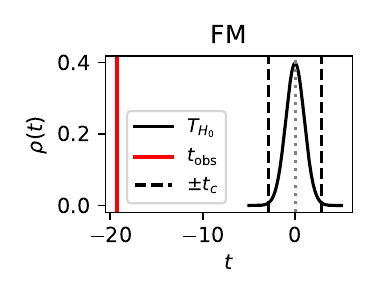}};
\node[anchor=north west](CM3) at (FM3.north east) {\includegraphics[trim= 10 10 10 21, clip,width=0.24\linewidth]{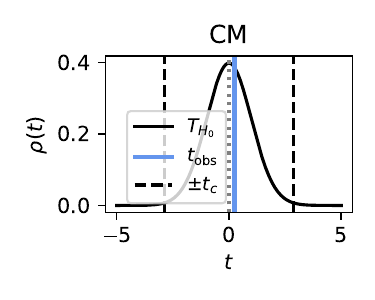}};
\node[anchor=north west](CL3) at (CM3.north east) {\includegraphics[trim= 10 10 10 21, clip,width=0.24\linewidth]{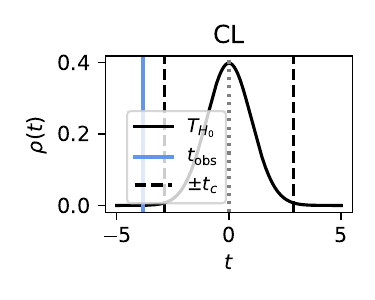}};
\node[anchor=north west](fit2SM3) at (CL3.north east) {\includegraphics[trim= 10 10 10 21, clip,width=0.24\linewidth]{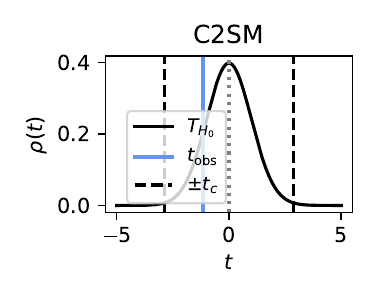}};
\node[anchor= south east] at (tab3.north west) {\textbf{c.1)}};
\node[anchor= south west] at (FM3.north west) {\textbf{c.2)}};
\node[anchor= south west] at (FM3.north) {\textbf{FM}};
\node[anchor= south west] at (CM3.north) {\textbf{CM}};
\node[anchor= south west] at (CL3.north) {\textbf{CL}};
\node[anchor= south west] at (fit2SM3.north) {\textbf{fit2SM}};
\end{tikzpicture}
\caption{\textcolor{black}{Panels \textbf{a.1,b.1,c.1}: statistical significance of the prediction performances of each white-box method versus its black-box counterpart (evaluated in terms of AUROC score), for eMID when $\mathcal{E}^{miss}$ is populated with the $20\%$ of the existing links. For each pair of methods, the difference of performances is paired across all iterations for all snapshots at the \textbf{a)} yearly ($3$ snapshots with $30$ iterations each), \textbf{b)} quarterly ($4$ snapshots with $30$ iterations each) and \textbf{c)} weekly ($4$ snapshots with $30$ iterations each) aggregation levels, corresponding to a sample size of $90$ points (yearly) and $120$ points (quarterly and weekly). The table entries report the (sample) average difference $\overline{\Delta}$, the t-value expressing the distance of the (sample) average difference from $0$, in units of the (sample) variance and the p-value. Panels \textbf{a.2,b.2,c.2}: distributions representing our paired samples t-test for the AUROC score and eMID, at the corresponding aggregation level.}}
\label{fig:ttest_eMID02AUROC}
\end{figure}

\begin{figure}[t!]
\flushleft
\begin{tikzpicture}[every node/.style={inner sep=0pt,outer sep=0pt}]
\node[anchor=north west](tab1) at (0,0) {\begin{tabular}{c|c|c|c|c}
\hline
\hline
eMID features (yearly) & White-box algorithms & $\langle\Delta_\text{AUPRC}\rangle$ & $t$ & $p$ \\
\hline
\hline 
$\omega_i$ & FM & $ -1.893\times10^{-1}$ & $-38.767$ & $1.588	\times 10^{-57}$ \\
\hline
\multirow{3}{*}{$k_i$} & {CM}& $ 5.891\times10^{-3}$ & $4.872$ & $4.772\times10^{-6}$ \\
& {CL} &$ -8.339\times10^{-3}$ & $-4.716$ & $8.843\times10^{-6}$\\

& {fit2SM} &$ 6.095\times10^{-3}$ & $4.990$ & $2.967\times10^{-6}$ \\
\hline
\hline
\end{tabular}
};
\node[anchor=north,shift={(-0.5,-.5)}](FM1) at (tab1.south west) {\includegraphics[trim= 10 10 10 21, clip,width=0.25\linewidth]{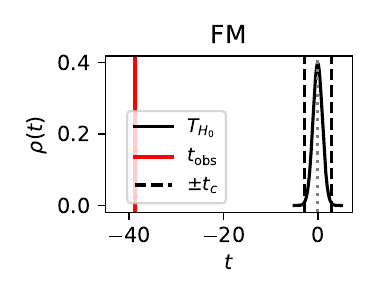}};
\node[anchor=north west](CM1) at (FM1.north east) {\includegraphics[trim= 10 10 10 21, clip,width=0.24\linewidth]{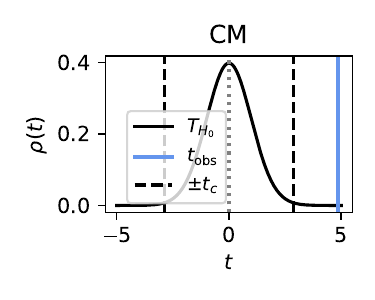}};
\node[anchor=north west](CL1) at (CM1.north east) {\includegraphics[trim= 10 10 10 21, clip,width=0.24\linewidth]{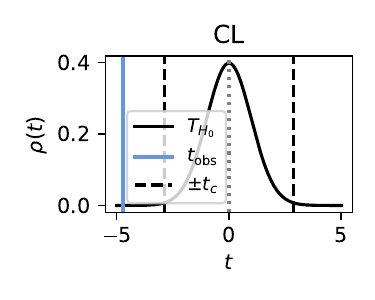}};
\node[anchor=north west](fit2SM1) at (CL1.north east) {\includegraphics[trim= 10 10 10 21, clip,width=0.24\linewidth]{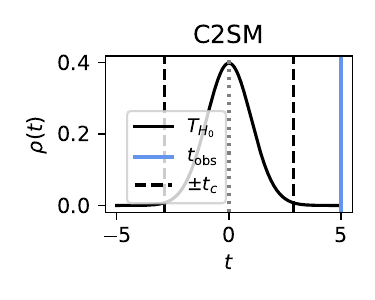}};
\node[anchor= south east] at (tab1.north west) {\textbf{a.1)}};
\node[anchor= south west] at (FM1.north west) {\textbf{a.2)}};
\node[anchor= south west] at (FM1.north) {\textbf{FM}};
\node[anchor= south west] at (CM1.north) {\textbf{CM}};
\node[anchor= south west] at (CL1.north) {\textbf{CL}};
\node[anchor= south west] at (fit2SM1.north) {\textbf{fit2SM}};
\node[anchor=north west,shift={(0.5,-.5)}](tab2) at (FM1.south){\begin{tabular}{c|c|c|c|c}
\hline
\hline
eMID features (quarterly) & White-box algorithms & $\langle\Delta_\text{AUPRC}\rangle$ & $t$ & $p$ \\
\hline
\hline 
$\omega_i$ & FM & $ -9.318\times10^{-2}$ & $-58.908$ & $6.972\times10^{-90}$ \\
\hline
\multirow{3}{*}{$k_i$} & {CM}& $ -3.339\times10^{-3}$ & $-3.274$ & $1.388\times10^{-3}$ \\
& {CL} &$ -1.535\times10^{-2}$ & $-12.219$ & $9.558\times10^{-23}$\\

& {fit2SM} &$ -3.180\times10^{-3}$ & $-3.094$ & $2.464\times10^{-3}$ \\
\hline
\hline
\end{tabular}
};
\node[anchor=north,shift={(-0.5,-.5)}](FM2) at (tab2.south west) {\includegraphics[trim= 10 10 10 21, clip,width=0.25\linewidth]{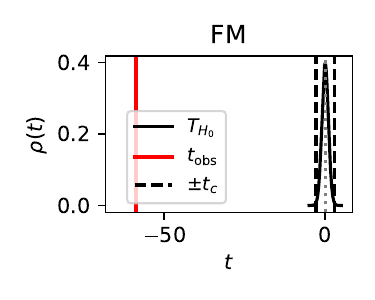}};
\node[anchor=north west](CM2) at (FM2.north east) {\includegraphics[trim= 10 10 10 21, clip,width=0.24\linewidth]{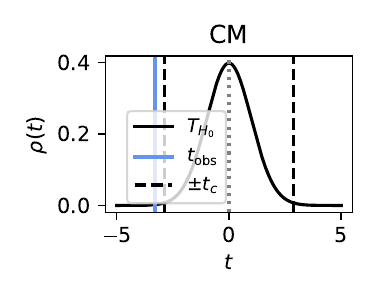}};
\node[anchor=north west](CL2) at (CM2.north east) {\includegraphics[trim= 10 10 10 21, clip,width=0.24\linewidth]{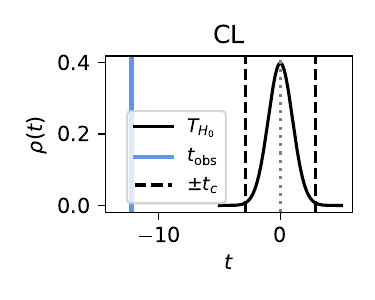}};
\node[anchor=north west](fit2SM2) at (CL2.north east) {\includegraphics[trim= 10 10 10 21, clip,width=0.24\linewidth]{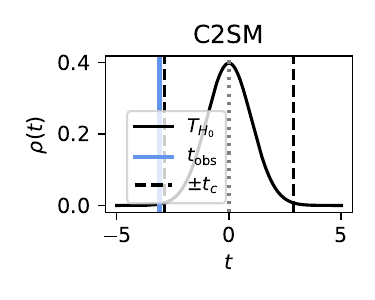}};
\node[anchor= south east] at (tab2.north west) {\textbf{b.1)}};
\node[anchor= south west] at (FM2.north west) {\textbf{b.2)}};
\node[anchor= south west] at (FM2.north) {\textbf{FM}};
\node[anchor= south west] at (CM2.north) {\textbf{CM}};
\node[anchor= south west] at (CL2.north) {\textbf{CL}};
\node[anchor= south west] at (fit2SM2.north) {\textbf{fit2SM}};
\node[anchor=north west,shift={(0.5,-.5)}](tab3) at (FM2.south) {\begin{tabular}{c|c|c|c|c}
\hline
\hline
eMID features (weekly) & White-box algorithms & $\langle\Delta_\text{AUPRC}\rangle$ & $t$ & $p$ \\
\hline
\hline 
$\omega_i$ & FM &$ -8.639\times10^{-3}$ & $-10.299$ & $3.583\times10^{-18}$ \\
\hline
\multirow{3}{*}{$k_i$} & {CM}& $ 4.178\times10^{-3}$ & $5.438$ & $2.913\times10^{-7}$ \\
& {CL} & $ 1.391\times10^{-3}$ & $1.791$ & $7.576\times10^{-2}$\\
& {fit2SM} &$ 3.880\times10^{-3}$ & $5.074$ & $1.449\times10^{-6}$ \\
\hline
\hline
\end{tabular}
};
\node[anchor=north,shift={(-0.5,-.5)}](FM3) at (tab3.south west)  {\includegraphics[trim= 10 10 10 21, clip,width=0.24\linewidth]{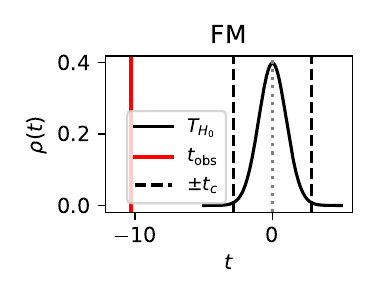}};
\node[anchor=north west](CM3) at (FM3.north east) {\includegraphics[trim= 10 10 10 21, clip,width=0.24\linewidth]{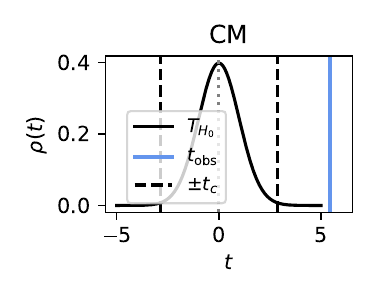}};
\node[anchor=north west](CL3) at (CM3.north east) {\includegraphics[trim= 10 10 10 21, clip,width=0.24\linewidth]{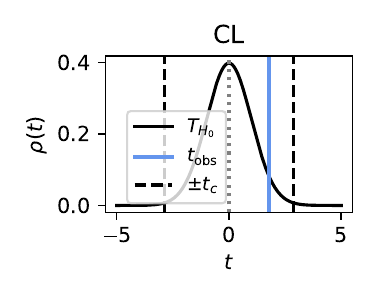}};
\node[anchor=north west](fit2SM3) at (CL3.north east) {\includegraphics[trim= 10 10 10 21, clip,width=0.24\linewidth]{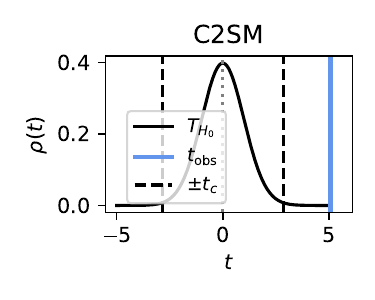}};
\node[anchor= south east] at (tab3.north west) {\textbf{c.1)}};
\node[anchor= south west] at (FM3.north west) {\textbf{c.2)}};
\node[anchor= south west] at (FM3.north) {\textbf{FM}};
\node[anchor= south west] at (CM3.north) {\textbf{CM}};
\node[anchor= south west] at (CL3.north) {\textbf{CL}};
\node[anchor= south west] at (fit2SM3.north) {\textbf{fit2SM}};
\end{tikzpicture}
\caption{\textcolor{black}{Panels \textbf{a.1,b.1,c.1}: statistical significance of the prediction performances of each white-box method versus its black-box counterpart (evaluated in terms of AUPRC score), for eMID when $\mathcal{E}^{miss}$ is populated with the $20\%$ of the existing links. For each pair of methods, the difference of performances is paired across all iterations for all snapshots at the \textbf{a)} yearly ($3$ snapshots with $30$ iterations each), \textbf{b)} quarterly ($4$ snapshots with $30$ iterations each) and \textbf{c)} weekly ($4$ snapshots with $30$ iterations each) aggregation levels, corresponding to a sample size of $90$ points (yearly) and $120$ points (quarterly and weekly). The table entries report the (sample) average difference $\overline{\Delta}$, the t-value expressing the distance of the (sample) average difference from $0$, in units of the (sample) variance and the p-value. Panels \textbf{a.2,b.2,c.2}: distributions representing our paired samples t-test for the AUPRC score and eMID, at the corresponding aggregation level.}}
\label{fig:ttest_eMID02AUPRC}
\end{figure}

\begin{figure}[t!]
\begin{tikzpicture}
\node[anchor=north west](sign) at (FM.south west) {\includegraphics[trim= 10 10 10 0, clip,width=\linewidth]{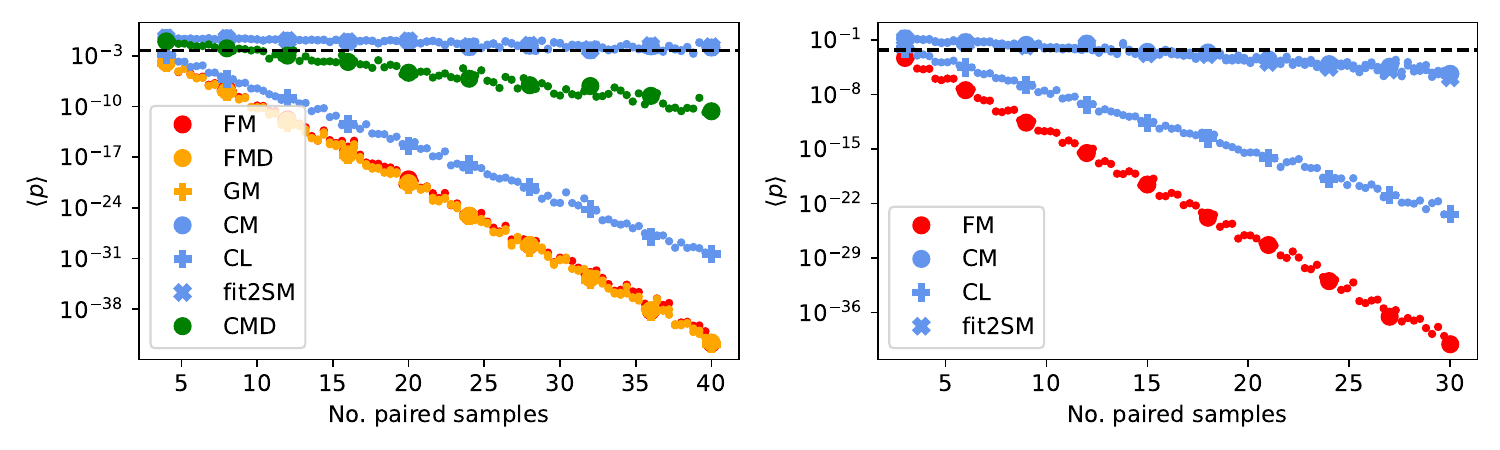}};
\node[anchor= north west] at (sign.north west) {\textbf{a)}};
\node[anchor= north west] at (sign.north) {\textbf{b)}};
\end{tikzpicture}
\caption{\textcolor{black}{Panel \textbf{a}: sensitivity analysis of the results concerning the statistical significance of the prediction performances of our algorithms for the year 2000 of the WTW (AUROC score). $\mathcal{E}^{miss}$ is populated with the $10\%$ of the existing links. For each pair of methods, the difference of performances is paired across the number of samples indicated on the x-axis. Panel \textbf{b}: sensitivity analysis of the results concerning the statistical significance of the prediction performances of our algorithms for the year 2000 of eMID (AUROC score). $\mathcal{E}^{miss}$ is populated with the $20\%$ of the existing links. For each pair of methods, the difference of performances is paired across the number of samples indicated on the x-axis. Our results are robust across the cardinality of samples.}}
\label{fig:sensitivity}
\end{figure}

{\color{black}
Let us, now, check if the prediction performances of the algorithms considered in the present work are significantly different or not. To this aim, let us consider the score difference

\begin{equation}
\Delta_\text{Score}=\text{Score}_\text{wb}-\text{Score}_\text{bb}
\end{equation}
between the white-box methods considered here and their black-box counterpart(s) and ask if it is significantly different from zero. More formally, we carry out a number of paired samples t-tests: under the null hypothesis $\Delta_\text{Score}=0$, the test statistics

\begin{equation}
t=\frac{\overline{\Delta}-0}{s/\sqrt{n}},
\end{equation}
with

\begin{equation}
\overline{\Delta}=\frac{\sum_{i=1}^n\Delta_i}{n}
\end{equation}
and

\begin{equation}
s=\sqrt{\frac{\sum_{i=1}^n(\Delta_i-\overline{\Delta})^2}{n-1}},
\end{equation}
obeys a Student's t-distribution with $n-1$ degrees of freedom - for all practical purposes, a Normal distribution (i.e. a Gaussian distribution with $\mu=0$ and $\sigma^2=1$).

For what concerns the WTW, the paired samples t-test is carried out by pooling together the $40$ observations (i.e. training set-test set splits) per year, amounting at $440$ observations in total. While figure~\ref{fig:ttest_WTW01AUROC} suggests that black-box methods always perform significantly better than their white-box counterparts, figure~\ref{fig:ttest_WTW01AUPRC} reveals the picture to be more nuanced: in this second case, in fact, a clear winner cannot be unambiguously identified as \emph{i)} black-box models perform better than the white-box counterparts not taking structural information as input; \emph{ii)} the former and the latter perform in a statistically undistinguishable way when `mixed' structural information is taken as input; \emph{iii)} white-box models (typically) perform better than black-box counterparts when either the plain degree sequence or non-linear structural information is taken as input.

For what concerns eMID, instead, the paired samples t-test is carried out by pooling together the $40$ observations (i.e. training set-test set splits) per year, amounting at $440$ observations in total. As figures~\ref{fig:ttest_eMID02AUROC} and~\ref{fig:ttest_eMID02AUPRC} suggest, the presence of a winner depends on the level of aggregation as well, the finer the aggregation the higher the chance that white-box models are favoured.

Finally, as the sensitivity analysis depicted in figure~\ref{fig:sensitivity} confirms, our results are quite robust across the cardinality of samples.}

\clearpage

\clearpage

\clearpage

\clearpage

\clearpage

\end{document}